\documentclass[twocolumn,times,tighten]{aastex63}
\received{} \revised{J} \accepted{}
\shorttitle{The Central 300\,pc of the Galaxy probed by H$_3^+$
  and CO} \shortauthors{Oka et al.}
\begin{document}

\title{ The Central 300\,pc of the Galaxy probed by infrared
  spectra of H$_3^+$ and CO: part I.  \\ Predominance of warm
  and diffuse gas and high H$_2$ ionization rate}

%------------------------------------------------------------
% \correspondingauthor{Takeshi Oka}
\author{Takeshi Oka}
%\author[0000-0000-0000-0000]{Takeshi Oka}
\affiliation{Department of Astronomy and Astrophysics and
  Department of Chemistry, the Enrico Fermi Institute,
  University of Chicago, Chicago, IL 60637 USA;
  t-oka@uchicago.edu}
% \email{t-oka@uchicago.edu}

\author{T. R. Geballe} \affiliation{Gemini Observatory, Hilo, HI
  96720 USA}

\author{Miwa Goto} \affiliation{Universit\"ats-Sternwarte
  M\"unchen, D-81679, Munich, Germany}

\author{Tomonori Usuda} \affiliation{National Astronomical
  Observatory of Japan, Tokyo, Japan}
  
\author{Benjamin, J. McCall} 
\altaffiliation{Current address: Hanley Sustainability
  Institute, Department \\ of Physics, and Department of Chemistry,
  University of\\  Dayton, Dayton, OH, 45469 USA}
\affiliation{Department of
  Chemistry, Department of Physics, and Department of Astronomy,
  University of Illinois, Urbana, IL 61801 USA}

\author{Nick Indriolo}
\affiliation{National Astronomical
  Observatory of Japan, Tokyo, Japan}

%============================================================
\vspace{4mm}
\begin{abstract}

The molecular gas in the Central Molecular Zone (CMZ) of the
Galaxy has been studied using infrared absorption spectra of
H$_3^+$ lines at 3.5--4.0 $\mu$m and CO lines near 2.34
$\mu$m. In addition to the previously reported spectra of these
lines toward 8 stars located within 30\,pc of Sgr~A$^\ast$,
there are now spectra toward $\sim$30 bright stars located from
140\,pc west to 120\,pc east of Sgr~A$^\ast$. The spectra show
the presence of warm ($T\sim 200$\,K) and diffuse ($n <
100$\,cm$^{-3}$) gas with $N$(H$_3^+$) $\sim 3 \times
10^{15}$\,cm$^{-2}$ on majority of sightlines. Instead of our
previous analysis in which only electrons from photoionization
of carbon atoms were considered, we have developed a simple
model calculation in which the cosmic ray ionization of H$_2$
and H are also taken into account. We conclude: (1) Warm and
diffuse gas dominates the volume of the CMZ. The volume filling 
factor of dense gas must be much less than 0.1 and the CMZ is
not as opaque as previously considered.  The X-ray emitting
ultra-hot 10$^8$\,K plasma, which some thought to dominate the
CMZ, does not exist over extended regions. (2) The cosmic ray
ionization rate is $\zeta \sim 2 \times 10^{-14}$\,s$^{-1}$,
higher than in Galactic dense clouds and diffuse clouds by
factors of $\sim$1000 and $\sim$100, respectively. If the
equipartition law stands, this suggests a pervading magnetic
field on the order of $\sim$100\,$\mu$G.

\end{abstract}

%============================================================
\section{Introduction} 

The Galactic center (GC) is rich in extraordinary objects such
as the central supermassive black hole, Sgr~A$^\ast$, three
massive star clusters --- the Central, the Quintuplet, and the
Arches --- gigantic non-thermal radio filaments, giant molecular
clouds, giant \ion{H}{2} regions such as Sgr A, B, C, D, E, and
others. As viewed from earth the GC is a strong emitter of
radio, far-infrared, infrared, X-rays, and $\gamma$-rays, as was
noted by the early broadband observations in those wavelength
regions (Reber 1944; Piddington \& Minnett 1951 in the radio
continuum, Hildebrand et al. 1978 in the sub-millimeter,
Hoffmann \& Frederick 1969 in the far-infrared, Becklin \&
Neugebauer 1968, 1969 and Low et al. 1969 in the mid- and
near-infrared, Bowyer et al. 1965 and Kellogg et al. 1971 in
X-rays and Clark et al. 1968 in $\gamma$-rays). Both line
emission and line absorption also are prevalent in the GC, with
pioneering detections made by Rougoor \& Oort (1960) of the
21\,cm \ion{H}{1} line emission, Bolton et al. (1964a) and
others that same year of OH absorption, Cheung et al. (1968) and
Penzias et al. (1971) of NH$_3$ and CO emission, respectively,
Aitken et al (1974) of atomic fine structure line emission, and
Gatley et al. (1984) of H$_2$ line emission.
%------------------------------------------------------------
\subsection{The Central Molecular Zone: previous observations}

The central region of the GC, with a radius of $\sim$150\,pc (we
use a GC distance of 8\,kpc) and a thickness of a few tens of
parsecs, was designated the Central Molecular Zone (CMZ) by
Morris and Serabyn (1996) because of its high concentration of
molecules. This region which constitutes a mere $\sim$10$^{-5}$
of the volume of the Galaxy was estimated by them to contain
$\sim$10\% of all Galactic molecules. The high abundances of
many molecular species in the CMZ has allowed the study of the
physics and chemistry, and the morphology of molecular gas in
the CMZ extensively by radio spectroscopy, particularly via the
centimeter absorption spectrum of OH (e.g. Robinson \& McGee
1970; Cohen \& Few 1976; Boyce \& Cohen 1994) and H$_2$CO
(e.g. Scoville et al. 1972; Whiteoak \& Gardener 1979; Cohen \&
Few 1981; Zylka et al. 1992; Ao et al. 2013; Ginsburg et
al. 2016). The inversion lines of NH$_3$ (e.g. G\"usten et
al. 1981; Morris et al. 1983; H\"uttemeister et al. 1993;
Nagayama et al. 2007; Mills \& Morris 2013; Krieger et
  al. 2017) are unique in that their frequencies are nearly
independent of rotational quantum numbers $J$ and $K$; in
addition, $J = K$ levels are metastable, thus allowing
observations of very high rotational levels (see Appendix
C). Millimeter wave line emission from HCN (e.g. Fukui et
al. 1977; Lee 1996; Jackson et al. 1996), CS (e.g. Bally et
al. 1988; Tsuboi et al. 1999; Lang et al. 2002), and SiO
(e.g. H\"uttemeister et al. 1998; Riquelme et al. 2010),
molecules with large dipole moments (2.985\,D, 1.958\,D, and
3.098\,D, respectively) have fast spontaneous emission and thus
high critical densities, and have been useful for observing
dense regions ($n > 10^4$\,cm$^{-3}$).

Millimeter emission lines of $^{12}$C$^{16}$O and its isotopic
species $^{13}$C$^{16}$O and $^{12}$C$^{18}$O (e.g. Bania 1977;
Liszt \& Burton 1978; Heiligman 1987; Bally et al. 1988; Sofue
1995; Oka et al. 1998b; Sawada et al. 2001; Martin et al. 2004)
have provided by far the most detailed and extensive information
on the molecular content of the CMZ. The high abundance of CO,
second only to H$_2$, makes CO a general probe of molecular
regions having a wide range of temperatures, number densities,
and column densities. At shorter wavelengths, far- to
mid-infrared rotational emission spectral lines of H$_2$ have
provided information on warm molecular gas
(Rodr\'iguez-Fern\'andez, et al. 2001) in the CMZ. The near
infrared vibrational emission spectrum of this molecule gives
information on UV excitation there (Pak et al. 1996).

In addition to the above molecules, which contain at most two
heavy ($M > 10$\,u) atoms, many molecules and molecular ions
with more than two heavy atoms have been detected in the GC
(Requena-Torres et al. 2006, 2008) and some have been used for
mapping of the CMZ (e.g. Jones et al. 2012, 2013).
The highest mass molecule detected in the CMZ was HC$_5$N
  (Avery et al. 1976, Armijos-Abenda\~no et al. 2015), until
  HC$_7$N was observed more than 40 years later (Zeng et
  al. 2018).

%------------------------------------------------------------
% Section 1.2
\subsection{Filling factor of dense molecular gas in the CMZ}

The majority of the radio and millimeter molecular line emission
observed in the CMZ has been interpreted as arising in dense ($n
\geq 3 \times 10^3$\,cm$^{-3}$) gas (e.g. Mills et
al. 2018). The ubiquity of this emission has led to a high
estimated volume filling factor, $f\geq 0.1$, for such dense
molecular gas in the CMZ (Morris \& Serabyn 1996). For an
average value $n = 10^4$\,cm$^{-3}$, the lower limit of
that filling factor corresponds to an average H$_2$ column
density of $N$(H$_2$)$\sim 5 \times 10^{23}$\,cm$^{-2}$ over the
radial distance of 150\,pc and an average visual extinction of
$A_V \geq 500$ in the CMZ's dense gas using the standard ratio
of dust to gas (Bohlin et al. 1978; Predehl \& Schmitt
1995). Such high average column density and associated
extinction appear to conflict with values derived from
observations of the thermal emission from dust at sub-millimeter
and millimeter wavelengths by Bally et al. (2010). From their
data (see, e.g., their Table~3) we estimate that average values
of $N$(H$_2$) and $A_V$ in the CMZ are roughly 30 times less.

The estimate also appears to conflict with measurements of
visual extinction toward many objects located within 40\,pc of
the very center, derived from infrared observations, which is
well known to be $\sim$30\,mag (e.g., Cotera et al. 2000), also
more than an order of magnitude less than the Morris \& Serabyn
lower limit. Larger area infrared studies, covering the entire
CMZ (Schultheis et al. 2009), give fairly similar values,
although somewhat higher values in some regions. In addition, it
has been estimated that one-third of the extinction to objects
close to Sgr~A$^\ast$ arises in molecular gas in foreground
spiral arms (Whittet et al. 1997).

Brightness-limited infrared studies, such as that of Cotera et
al. (2000) and the study described here, will tend to select
objects in front of densest and highest extinction CMZ
clouds. Nevertheless, it seems clear from the above
near-infrared and far infrared--millimeter extinction results,
and from the spectra that we present and discuss in this paper,
that the previously estimated $f\geq 0.1$ for the volume filling
factor of $n \geq 10^4$\,cm$^{-3}$ gas in the CMZ is far too
high.

It has been pointed out by the referee that lower filling
factors than this have been implicit in some previous papers on
dense gas (e.g. Launhardt et al. 2002; Molinari et al. 2011;
Sormani et al. 2018; Kruijssen et al. 2019).

%------------------------------------------------------------
% Section 1.3
\subsection{Previous observations of diffuse molecular gas in the CMZ}

Although a large majority of papers reporting radio emission
from molecules in the CMZ deduce high H$_2$ number densities,
$n$(H$_2$)$\geq 3 \times 10^3$\,cm$^{-3}$, there are a few
exceptions in which molecular emission lines at radio and
millimeter wavelengths were interpreted to be from clouds of
lower density than dense clouds. Oka et al. (1998a; Tomoharu Oka
of Keio University, not to be confused with the first author of
the present paper) concluded ``the total CO emission from the
Galactic center is dominated by the emission from low-density
[$n$(H$_2$)$\approx 10^{2.5}$\,cm$^{-3}$] gas''. Independently,
Dahmen et al. (1998) proposed a somewhat higher average density
[$n$(H$_2$)$\approx 10^3$\,cm$^{-3}$] based on observations of
C$^{18}$O line emission.  They also report $\approx
  10^2$\,cm$^{-3}$ as a minor component. In a noteworthy
but little referenced paper, Magnani et al. (2006) observed
3335\,MHz CH emission, which traces low density molecular gas,
in a 30\arcmin $\times$ 30\arcmin~region and found broad line
profiles that are quite different from those of CO emission
lines, which arise predominantly in dense clouds.

Although not mentioned explicitly in their publications, the
absorption spectra of OH (Goldstein et al. 1964; Bolton et
al. 1964b; Robinson \& McGee 1970) and H$_2$CO (Scoville et
al. 1972) in the CMZ are likely to arise in low-density gas,
some of which may be the gas with $n < 100$\,cm$^{-3}$ most
effectively probed by H$_3^+$ that we have observed. Recently
Herschel observations of rotational absorption spectra of
molecular ions such as OH$^+$ and H$_2$O$^+$ (e.g. Goicoechea et
al. 2013; Indriolo et al. 2015), CH$^+$ and SH$^+$ (Godard et
al. 2012), and ArH$^+$ (Schilke et al. 2014) also indicate the
presence of low-density regions. These molecular ions are
rapidly destroyed by H$_2$ at the Langevin rate and thus can
exist only in diffuse regions with low fractions of molecular
hydrogen $f$(H$_2$). So far, they have been observed only toward
Sgr~A and Sgr~B which contain dense giant molecular cloud
complexes, but these molecular ions must reside in diffuse
clouds in the outskirts of the complexes. Even the CMZ's HF and
H$_2$O, which mainly exist in dense clouds, were interpreted to
reside partly in the low-density clouds by Sonnentrucker et
al. (2013) who conclude that a ``very dilute phase seems
pervasive around the Galactic center.''

%------------------------------------------------------------
\subsection{Temperatures}

Kinetic temperatures in the CMZ have been found to be
significantly higher than the few tens of Kelvins typically
found in the Galactic disk's interstellar medium. Early
far-infrared studies found dust temperatures, $T_d$, less than
30\,K (Hildebrand et al. 1978; Odenwald \& Fazio 1984) and more
definitively later (e.g. $T_d =15-22$\,K by Lis et
al. 2001). However, many studies of molecular line emission have
reported kinetic temperatures, $T_{\rm k}$, of $25-80$\,K,
indicating dynamical heating of the gas  (Ao et al. 2013;
  Ginsburg et al. 2016; Krieger et al. 2017). Observations of
rotational emission lines of H$_2$ demonstrate that gas with
$T_{\rm k} = 150$\,K is widespread (Rodr\'iguez-Fern\'andez, et
al. 2001). Although H$_2$ coexists with CO at high densities
($10^{3.5-4.0}$\,cm$^{-3}$), most of the H$_2$ line emission may
well originate in surrounding lower density regions. Hot NH$_3$
is also observed in the CMZ, implying gas temperatures of a
  few hundred Kelvins (Mauersberger et al. 1986). This is
discussed in Appendix C.

%------------------------------------------------------------
\subsection{A brief history of interstellar H$_3^+$}

Trihydrium, a term recommended by J. K. G. Watson (private
communication) and used here for the first time, otherwise known
as H$_3^+$, in which a proton is attached to H$_2$ in an
equilateral triangular configuration, was discovered by
J. J. Thomson (1911). In H$_2$-dominated interstellar gases this
simplest polyatomic molecule is the most abundant molecular
ion. Martin et al. (1961) first proposed detecting interstellar
H$_3^+$. Observing interstellar H$_3^+$ attained great
importance when Herbst \& Klemperer (1973) and Watson (1973)
deduced that most interstellar molecules are produced by
ion-neutral reactions and that H$_3^+$ plays the central role as
the universal proton donor (acid) to initiate chains of such
reactions. The infrared spectrum of the fundamental $\nu_2$ band
with band origin at 3.97\,$\mu$m), was observed in the
laboratory by Oka (1980). Following unsuccessful searches (Oka
1981; Geballe \& Oka 1989), detections of absorption lines of
interstellar H$_3^+$ in two dense interstellar clouds were made
by Geballe \& Oka (1996). The H$_3^+$ column densities observed
in many dense clouds by them and others (McCall et al 1999;
Kulesa 2003; Brittain et al. 2004; Gibb et al. 2010) are
approximately as expected from the theory. Readers are referred
to a recent review (Oka 2013) for more details of the history,
astronomy, physics, and chemistry of this fundamental molecular
ion.

%------------------------------------------------------------
Once detected, H$_3^+$ has been observed in numerous
interstellar molecular environments. Two big surprises were
found soon after its discovery. First, absorption lines of
H$_3^+$ with similar strengths as those in dense clouds were
discovered in diffuse clouds (McCall et al. 1998a; Geballe et
al. 1999; McCall et al. 2002), implying H$_3^+$ column densities
in diffuse clouds are comparable to those in dense clouds,
despite the much higher abundance in diffuse clouds of
H$_3^+$-destroying free electrons created by the ionization of
carbon atoms. Extinctions through diffuse clouds are typically
an order of magnitude less than through dense clouds, indicating
that the fraction of H$_3^+$, $X$(H$_3^+$)$\equiv
n$(H$_3^+$)/$n_{\rm H}$, is $\sim$10 times higher in diffuse
clouds than in dense clouds. Because of this higher abundance,
H$_3^+$ has emerged as a valuable probe of physical conditions
in diffuse clouds. Further studies have led to the important
conclusion that in the Galactic disk, on average the rate of
cosmic ray ionization of H$_2$ in Galactic diffuse clouds is 10
times higher than in dense clouds (McCall et al. 2003; Indriolo
et al. 2007; Indriolo \& McCall 2012), contradicting the
previous notion that the rate is more or less uniform in the
Galaxy.

Second, H$_3^+$ column densities more than 10 times greater than
the highest ones observed in the Galactic disk were found toward
bright, dust-embedded infrared stars in the CMZ, GCS\,3-2 in the
Quintuplet Cluster and GCIRS~3 in the Central Cluster (Geballe
et al. 1999) and subsequently toward many other stars in the
CMZ, as discussed below. This finding initiated our long term
systematic study of the CMZ using H$_3^+$ as a probe, which
continues to this day. H$_3^+$ has been observed also in two
extragalactic objects the ultraluminous galaxy
IRAS~08572+3915\,NW (Geballe et al. 2006) and type II Seyfert
galaxy NGC~1068 (Geballe et al. 2015).

%============================================================
% Figure 1 % section 1.6
%------------------------------------------------------------
\setcounter{figure}{0}
\begin{figure}
\includegraphics[angle=-90,width=0.5\textwidth]{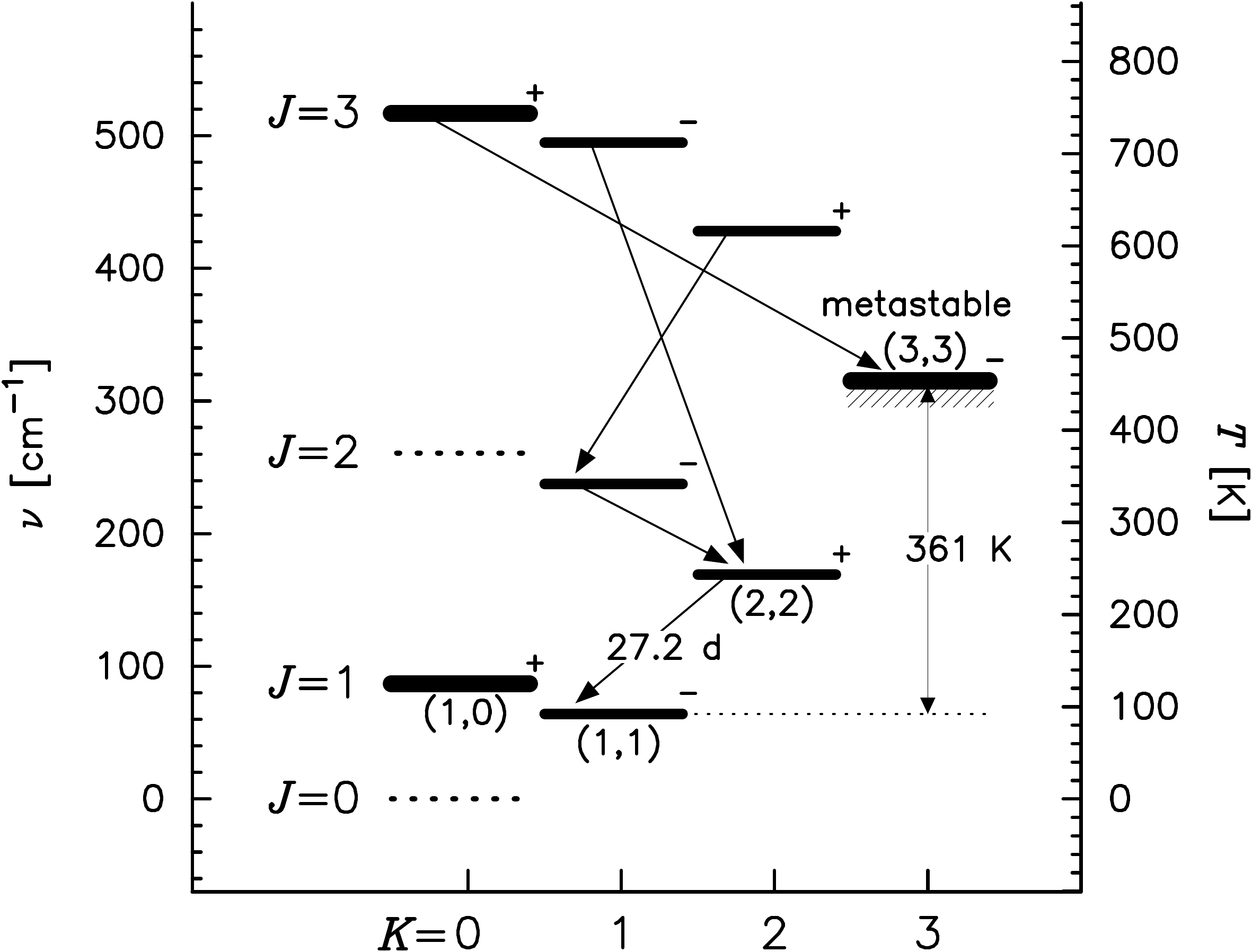}
\caption{Lower rotational levels in the ground vibrational state
  of H$_3^+$, composed of ortho ($K=3n$, bold line) and para
  ($K=3n\pm1$, thin line) nuclear spin states. The $+$ and $-$
  signs indicate the parity $= (-1)^K$. The (0,0) and (2,0)
  levels are forbidden due to nuclear spin statistics (as for
  the symmetric inversion levels of NH$_3$). Downward arrows
  indicate spontaneous emission due to spontaneous breakdown of
  symmetry. The energy gap between the lowest level and the
  (3,3) level and the lifetime of the (2,2) level for
  spontaneous emission are shown.  In the warm diffuse molecular
  gas of the CMZ, because of rapid spontaneous emission only
  three levels (1,1), (1,0), and (3,3) are significantly
  populated; in denser warm molecular gas the (2,2) level is
  also significantly populated.}
\end{figure}
%------------------------------------------------------------
% Section 1.6 
\subsection{ H$_3^+$ in the Central Molecular Zone: warm and diffuse gas}

The kinetic temperatures of both dense and diffuse molecular
clouds in the Galactic disk are typically a few tens of
Kelvins. Toward stars in the Central and Quintuplet Clusters,
however, large amounts of H$_3^+$ in the metastable ($J$, $K$) =
(3,3) level, 361\,K above the lowest (1,1) level (See Figure~1),
have been found via detection of absorption in the $R$(3,3)$^l$
line at 3.53\,$\mu$m (Goto et al. 2002). The relative
populations in those levels imply that H$_3^+$ in the CMZ exists
in gas at temperatures of a few hundred Kelvins (Oka et
al. 2005). A significant population of H$_3^+$ in the (3,3)
level has become the fingerprint of diffuse gas in the
CMZ. Attempts to detect H$_3^+$ in the lower lying (2,2) level,
which is only 150\,K above the (1,1) level, in the CMZ, via the
$R$(2,2)$^l$ line at 3.62\,$\mu$m have largely failed (e.g. Goto
et al. 2002; Oka et al. 2005). The extreme population inversion
of the (3,3) and (2,2) levels is due to spontaneous emission
from (2,2) to (1,1) a consequence of the spontaneous breakdown
of symmetry (SBS, see Appendix C). A lifetime of 27 days of the
(2,2) level against this emission has been determined from high
quality ab initio theory (Neale et al. 1996) and corresponds to
a critical density of 200\,cm$^{-3}$ assuming the Langevin rate
constant of $2\times 10^{-9}$\,cm$^3$\,s$^{-1}$ for
H$_3^+$-H$_2$ collisions. The decisive population inversion
between these levels is a clear demonstration that the density
of the warm gas is much less than 200\,cm$^{-3}$. The rotational
levels of H$_3^+$, shown in Figure~1, are ideally arranged to
study this warm, low density environment.

A more quantitative treatment of thermalization, in which many
SBS emissions are treated accurately and collisional transitions
approximately, based only on the principle of detailed balance
(Oka \& Epp 2004), has led to the finding of two-dimensional
relationships between $T_{\rm k}$ and $n$(H$_2$) and between the
column density ratios $N$(3,3)/$N$(1,1) and
$N$(3,3)/$N$(2,2). These relations and their inverses together
with the observed H$_3^+$ column densities allowed H$_3^+$ in
the CMZ to be used as a thermometer and densitometer for
GCS\,3-2 (likely the brightest infrared star in the CMZ, based
on its infrared magnitudes and colors), and for seven other
stars located from Sgr~A$^\ast$ to 30\,pc east of it (Oka et
al. 2005; Goto et al. 2008). Unlike the physical collisions
between NH$_3$ and H$_2$ in which the ortho/para spin states
conserve and the $\Delta k = \pm 3$ rule holds (Oka 1973),
collisions between H$_3^+$ and H$_2$ are chemical collisions in
which the protons are exchanged between H$_3^+$ and H$_2$,
spin states may change, and thus the $\Delta k$ rule is violated
(Appendix C). The rotational distribution of H$_3^+$ is
subthermal and our attempts to detect higher metastable levels
(4,4), (5,5), and (6,6) have not been successful.

In this paper, we focus our analysis and discussion on H$_3^+$
in diffuse molecular CMZ gas, which, as we describe in
subsequent sections, appears to exist in similar conditions
throughout the CMZ and thus can be treated generally. In
contrast, the contributions to the spectra on these sightlines
by H$_3^+$ in dense clouds in the CMZ differ for each sightline,
reflecting local conditions in those clouds, and are better
dealt with individually. Observations and analyses of H$_3^+$ in
dense clouds on several sightlines toward stars in the CMZ have
been previously published (Goto et al. 2011, 2013, 2014).

%------------------------------------------------------------
% Section 1.7 
\subsection{Production and destruction of H$_3^+$ in diffuse gas}

In interstellar molecular gas H$_3^+$ is mainly produced by
cosmic ray ionization of H$_2$ to H$_2^+$ at a slow rate $\zeta$
followed by the very rapid Langevin reaction,

\begin{equation}
  {\rm H_2 + H_2^+ \rightarrow H_3^+ + H},
\end{equation}
\noindent
in which a proton hops from H$_2^+$ to H$_2$ (Section 2.2.1. of
Oka 2013). In diffuse clouds, H$_3^+$ is mainly destroyed by
dissociative recombination with electrons,

\begin{equation}
  {\rm H_3^+ + e^- \rightarrow H + H + H, ~~H_2 + H}
\end{equation}
\noindent
whose rate constant is $\sim$100 times larger than the Langevin
rate constant (Section 2.3.3. of Oka 2013).

For H$_3^+$ in the diffuse molecular gas of the CMZ this simple
chemistry leads to a linear relation between the product of
$\zeta$ and column length, $L$, of the absorbing diffuse gas and
the total H$_3^+$ column density $N$(H$_3^+$),
\begin{equation}
  \zeta L = 2 k_{\rm e} N({\rm H_3^+}) (n_{\rm C}/n_{\rm
    H})_{\rm SV} R / f({\rm H_2}),
\end{equation}
\noindent
on the assumption that all electrons are from photoionization of
C atoms (Oka et al. 2005). In the above relation $k_{\rm e}$ is
the rate constant for dissociative recombination (McCall et
al. 2004), $(n_{\rm C}/n_{\rm H})_{\rm SV}$ is the carbon to
hydrogen ratio in the solar vicinity (Sofia et al. 2004), $R$ is
the increase of metallicity from the solar vicinity to the GC,
and $f({\rm H_2}) = 2 n({\rm H_2})/[n({\rm H}) + 2n({\rm H_2})]$
is the fraction of molecular hydrogen. The observed large
H$_3^+$ column densities in the CMZ's warm and diffuse gas,
$N({\rm H_3^+})\sim 3 \times 10^{15}$\,cm$^{-2}$ (Oka et
al. 2005; Goto et al. 2008), yield high values of $\zeta L$ and,
given reasonable constraints on $L$, likely cosmic ray
ionization rates of H$_2$ on the order of $\zeta \sim 3 \times
10^{-15}$\,s$^{-1}$ an order of magnitude higher than in
Galactic diffuse clouds and two orders of magnitude higher than
in dense clouds. The above equation and resulting estimate for
$\zeta$ (which was obtained on the assumptions that $f({\rm
  H_2}) \sim 1$ and $R \sim 3$), are modified in the present
paper in a major way as discussed below and in Section~5.2.

Recently, Le Petit et al. (2016) in the Laboratoire d'\'Etudes
du Rayonnement et de la Mati\`ere en Astrophysique et
Atmosph\`eres (LERMA) Meudon analyzed the H$_3^+$ column
densities in the CMZ observed by us (Oka et al. 2005; Goto et
al. 2008, 2011). Hereafter we refer to their analysis as the
Meudon analysis. Instead of the simple chemistry based only on
equations (1) and (2), they used the Meudon PDR code (Le Petit
et al. 2006) in which a great many species (165) and chemical
reactions (2850) are simultaneously considered. They also used
Herschel observations of HF, OH$^+$, H$_2$O$^+$, and H$_3$O$^+$
toward Sgr~B2 to constrain the physical and chemical parameters
of the CMZ. They extended the calculations to high values of
$\zeta$ where the linear relation between $\zeta$ and
$N$(H$_3^+$) of Eq. (3) no longer holds, and obtained values for
$\zeta$ of $(1-11) \times 10^{-14}$\,s$^{-1}$, considerably
larger than those derived by Oka et al. (2005) and Goto et
al. (2008). They also used a more detailed calculation of
thermalization by G\'omez-Carrasco et al. (2012), which is based
on the statistical theory of Park \& Light (2007a), to determine
temperature and density. The Meudon analysis is discussed in
Section 5.1.2. and 5.2.6. For a review of variations in the
derived values of $\zeta$ over the years, see Oka (2019).

%------------------------------------------------------------
\subsection{The $v = 2 \leftarrow 0$ first overtone infrared spectrum of CO}

Unlike H$_3^+$, which is abundant both in diffuse and dense
clouds, CO abundances and column densities in diffuse clouds are
much lower than in dense clouds, due to diffuse clouds being
permeated by CO - dissociating ultraviolet photons. Because of
this difference, observations of H$_3^+$ combined with properly
designed observations of CO can discriminate between H$_3^+$ in
dense gas and diffuse gas. On sightlines to the CMZ,
discrimination is made easier by the diffuse and dense gas
having different radial velocities and line widths. In
particular, the molecular clouds in the three foreground spiral
arms produce narrow absorption features at well-known radial
velocities, which usually can be easily separated from the
absorption by the CMZ's diffuse gas.

The rotationally resolved infrared spectrum of $^{12}$CO in the
CMZ was initially observed in the $v = 1 \leftarrow 0$
fundamental band near 4.7\,$\mu$m (Geballe, et al. 1989) and
later by others (see Moultaka et al. 2019 and references
therein).  Because lines of this band that originate in the most
populated levels are heavily saturated, in our studies we have
used lines of the $v = 2 \leftarrow 0$ first overtone band near
2.34\,$\mu$m. The transition dipole moment of this band $\langle
2|\mu|0 \rangle = (0.006518\pm 0.000028)$\,D (Zou \& Varanasi
2002) is 16 times lower than that of the fundamental band
$\langle 1|\mu|0 \rangle = 0.1055$\,D; that is, its absorption
is 262 times weaker. Because of the very low abundance of CO in
diffuse clouds, first overtone absorption lines there are
undetectable or very weak. However, in dense clouds, where CO
has a high abundance, the overtone CO lines are much stronger
and easily detectable. They are seldom saturated, however,
allowing one to determine reliable CO column densities. Like the
lines of the fundamental vibration-rotation band, the low-lying
transitions of the pure rotational spectrum of $^{12}$CO
(permanent dipole moment $\mu = 0.1098$\,D; Muenter 1975) at
millimeter and sub-millimeter wavelengths are highly saturated.

%------------------------------------------------------------
\subsection{Observable sightlines in the CMZ and outline of the paper}

Until 2008, spectra of H$_3^+$ in the CMZ were confined to
sightlines toward previously known bright dust-embedded stars in
the Central Cluster of massive stars surrounding Sgr~A$^\ast$
(Becklin \& Neugebauer 1975), similar stars in the Quintuplet
Cluster (Nagata et al. 1990; Okuda et al. 1990), and the Nagata,
Hyland, Straw (NHS) stars reported by Nagata et
al. (1993). Stars in the latter two groups are located no more
than 30\,pc to the east of Sgr~A$^\ast$ and thus collectively
they and the Central Cluster sightlines sample only a small
fraction of the longitudinal extent of the CMZ. In 2008, we
began to search for bright stars suitable for high-resolution
spectroscopy of the relatively weak lines of H$_3^+$, across the
entire $r \sim 150$\,pc CMZ. This search is outlined in Section
2.2.  A paper containing spectra of H$_3^+$ and CO toward two
such stars in the Sgr~B and Sgr~E complexes (Geballe \& Oka
2010) and a subsequent paper (Goto et al. 2011) both reported
the discovery of H$_3^+$ in warm diffuse molecular gas and dense
molecular gas on these sightlines. The presence of similar gas
far from the warm diffuse gas seen toward stars near the center
of the CMZ, suggested that the newly found gas is widespread in
the CMZ. However, observations on additional sightlines that are
both more widely and more uniformly spread across the CMZ are
required to more stringently test this possibility. This paper
reports and discusses such observations.

We describe the observational considerations and the
observations themselves in Section~2. The spectra are presented
in Section~3. The basic interpretation of the spectra and
determination of temperatures and column densities of the
diffuse and dense CMZ gas are given in Section~4. In Section~5
the new analysis method is presented and compared with the
Meudon analysis, and values of the cosmic ray ionization rate,
the column length of warm diffuse gas, and particle densities
are estimated. The implications of these results are discussed
in Section~6 and a summary and main conclusions are presented in
Section 7. Readers may understand the essence of this paper by
reading the last section alone. Some details of our analysis are
discussed separately in appendices to this paper. More detailed
information on and discussion of the morphology and dynamics of
the gas obtained from velocity profiles of the lines will be
given in a separate paper as Part II of this series.

%============================================================
% Table 1 % section 2.1
%============================================================
\setcounter{table}{0} \begin{deluxetable}{rccc}
\tablecaption{H$_3^+$ and CO transitions\label{t1}}
% \tableline \tableline
\tablehead{
  \colhead{Transitions}        &
  \colhead{$\nu$ [cm$^{-1}$]}  &
  \colhead{$\lambda$ [$\mu$m]} &
  \colhead{$|\mu|^2$ [$D^2$]}  
%  \colhead{Interference}\\
%   \colhead{\multicolumn{5}{c}{H$_3^+$ }} \\
  }
% \tableline
% \tableline
\startdata
\multicolumn{4}{c}{H$_3^+$} \\
\cline{1-4}
$Q$(1,0)           & 2529.724 & 3.95300 & 0.0254 \\% 3
$Q$(1,1)           & 2545.420 & 3.92862 & 0.0128 \\% 2
$\ast R$(1,1)$^l$  & 2691.443 & 3.71548 & 0.0141 \\% 1
$\ast R$(1,0)      & 2725.898 & 3.66852 & 0.0259 \\% 3
$\ast R$(1,1)$^u$  & 2726.220 & 3.66808 & 0.0158 \\% 3
$\ast R$(2,2)$^l$  & 2762.070 & 3.62047 & 0.0177 \\% 1
$\ast R$(3,3)$^l$  & 2829.925 & 3.53366 & 0.0191 \\% 5
$\ast R$(4,4)$^l$  & 2894.488 & 3.45484 & 0.0197 \\% 1
% \tableline \tableline
\cline{1-4}
\multicolumn{4}{c}{CO} \\
\cline{1-4}
% \tableline
$R$(0)             & 4263.837 & 2.34531 & 0.0000425  \\
$R$(1)             & 4267.542 & 2.34327 & 0.0000283  \\
$R$(2)             & 4271.177 & 2.34128 & 0.0000255  \\
$R$(3)             & 4274.741 & 2.33932 & 0.0000243  \\
$R$(4)             & 4278.234 & 2.33741 & 0.0000236  \\
$R$(5)             & 4281.657 & 2.33554 & 0.0000232  
\enddata
% \tableline \tableline
\tablecomments{Wavenumbers $\nu$, vacuum wavelengths $\lambda$,
  and transition strengths $|\mu|^2$ of the H$_3^+$ $\nu_2$
  fundamental band and CO first overtone band $v=2\rightarrow 0$
  used in this paper.
  %The last column gives degree of   atmospheric interference of %the H$_3^+$ line.  See text for   the interference.
 Excerpts from Table 2 of Oka (2013).}
% \end{deluxetable}
\end{deluxetable}

%------------------------------------------------------------
%============================================================
\section{Observation}
%------------------------------------------------------------
% 2.1
\subsection{Spectral Lines and Observing Sites}

The $R$ and $Q$ branch lines of the fundamental $\nu_2$
vibration-rotation band of H$_3^+$ that are of particular
astrophysical interest in studying the CMZ occur in the $L$
atmospheric window from 3.45 to 3.95\,$\mu$m, a region
relatively free from atmospheric interference (see Figure~1 of
McCall et al. 1998b). That wavelength range is well above the
2.73\,$\mu$m $\nu_1$ and 2.66\,$\mu$m $\nu_3$ bands of H$_2$O
and well below that molecule's 6.27\,$\mu$m $\nu_2$ band. The
principal interference is caused by $P$ branch lines of the
3.31\,$\mu$m $\nu_3$ band of CH$_4$ and by the 3.17\,$\mu$m
2$\nu_2$ band of H$_2$O. Particularly annoying is the strong
$5_{05} \leftarrow 6_{34}$ transition of the latter band, whose
frequency, 2830.008\,cm$^{-1}$, is only 0.083\,cm$^{-1}$ higher
than that of the $R$(3,3)$^l$ line of H$_3^+$. Observations of
that H$_3^+$ line are thus very sensitive to variations in the
column density of water vapor above the observatory, and it is
generally more difficult to obtain reliable line profiles of it
at lower altitude sites such as Cerro Pach\'on (2715\,m) and
Cerro Paranal (2635\,m), than on Maunakea (4200\,m).

Relevant H$_3^+$ transitions with frequencies and (vacuum)
wavelengths taken from Oka (2013) and squares of dipole moments
$|\mu|^2$ are listed in Table~1. From the observed equivalent
widths $W_\lambda$, the H$_3^+$ column densities in the lower
levels can be calculated by $N({\rm H_3^+})_{\rm level} =
(3hc)/(8 \pi^3 \lambda) W_\lambda / |\mu|^2$. Because of the
rapid SBS spontaneous emissions and the efficient chemical
collisions with H$_2$, only four levels (1,1), (1,0), (2,2), and
(3,3) in the ground vibrational state are observably populated
in the CMZ. Therefore the $\nu_2$ transitions that we have
primarily observed are $R$(1,1)$^l$, $R$(3,3)$^l$, the close
doublet of $R$(1,0) and $R$(1,1)$^u$ (velocity separation =
35\,km\,s$^{-1}$), and $R$(2,2)$^l$. The $R$(1,1)$^l$ and
$R$(3,3)$^l$ lines have been detected toward all stars in the
CMZ while $R$(2,2)$^l$ has been detected only toward a few stars
that lie behind or inside dense clouds within the CMZ. The
notation $R$(1,1)$^l$ signifies the transition ($v_2=1$, $l=1$
lower, $J^\prime=2$, $K^\prime=2$) $\leftarrow$ ($v_2=l=0$,
$J=1$, $K=1$). See Lindsay and McCall (2001) for more details on
the nomenclature. Other transitions in Table~1 such as $Q$(1,0)
and $Q$(1,1) have been observed toward a limited number of
stars.

CO has been observed mostly through its $K$-band $v = 2
\leftarrow 0$ vibration-rotation band. The relevant parameters
for the observed lines are given in Table~1.  The low $J$ lines
in this band are sharp toward the CMZ and are only moderately
affected by atmospheric absorption. For some stars that are too
faint in the $K$ band ($K > 11.5$\,mag) for sufficiently
accurate high-resolution absorption spectroscopy, the
fundamental bands of $^{13}$CO and C$^{18}$O have been observed
instead.

%------------------------------------------------------------
% 2.2
\subsection{Selection of background infrared sources}

High-resolution infrared absorption spectroscopy of H$_3^+$ and
CO at high S/N can be conducted only toward background objects
with bright 2--5\,$\mu$m infrared continua; such objects are
almost entirely stars. Bright stars with complex photospheric
spectra (e.g. Ridgway et al. 1984) cannot be used, however, as
their spectra are superimposed on the lines arising in the
interstellar gas, making analysis of the interstellar lines
difficult or impossible. This eliminates as probes most
infrared-bright stars, which are red giants.  Consequently, only
luminous hot stars whose spectra contain only lines of a few
elements (e.g., hydrogen, helium, carbon) and luminous stars
that are deeply embedded in warm opaque dust shells are suitable
background sources. This is a serious limitation, when compared
to observations of CO radio wavelength emission which can be
conducted toward any molecular cloud. However, absorption
spectroscopy has the advantages that (a) the location of the
absorbing gas is known to be in front of the continuum source
hence it is clear whether the absorbing gas is moving towards or
away from the star, (b) the optical depths of lines of the
H$_3^+$ fundamental and CO first overtone bands are almost
always low ($\tau < 0.15$) and thus column densities are
reliably determined, (c) the beam size is very small, on the
order of 0\farcs00001, and (d) the column density can be simply
calculated from the equivalent width without the need to
consider collisions. The large emission optical depths of the
highly saturated low $J$ millimeter transitions of CO emission
and observations of those lines with the large beam sizes of
single dish radio telescopes (10\arcsec-- 20\arcsec) both tend
to give the impression that the CMZ is filled with high density
gas.  Thus, infrared absorption spectroscopy of CO and H$_3^+$
toward point sources provides more reliable information on the
amounts of both dense and diffuse gas.

%============================================================
% Figure 2 starchart T section 2.2
%------------------------------------------------------------
% \setcounter{figure}{1}
\begin{figure*}
\includegraphics[width=1.\textwidth]{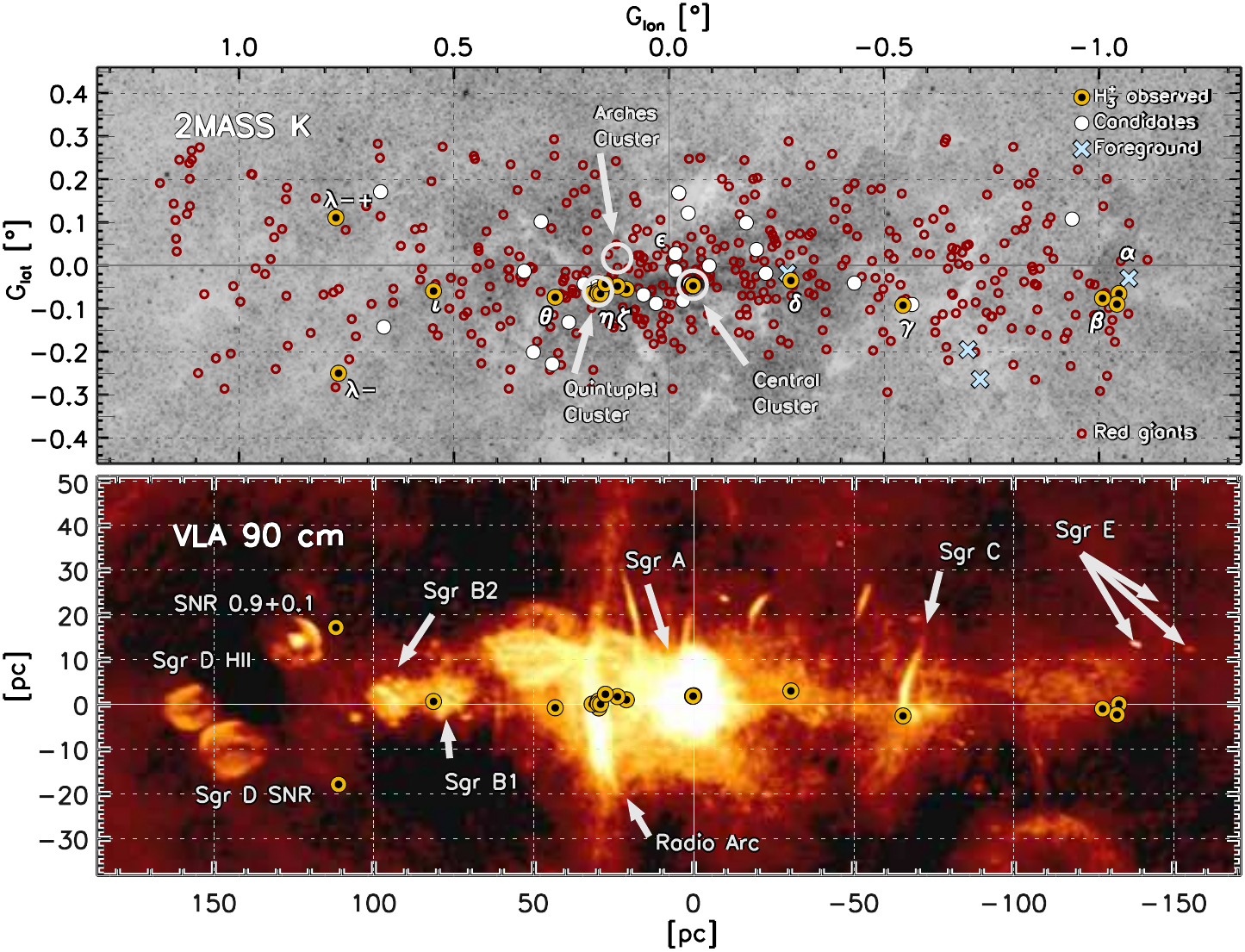}
\caption{Top figure: Locations of stars observed in the
  program. Symbols are as follows: yellow circles with dots --
  objects for which a set of H$_3^+$ and CO spectra have been
  observed; white circles -- objects with clean continuum for
  which part of H$_3^+$ and CO spectra have been observed or yet
  to be observed; red circles -- red giants which have CO
  bandhead photospheric absorptions at 2.295\,$\mu$m in the
  $K$-band medium resolution spectroscopic survey (Geballe et
  al. 2019) and not suitable for the H$_3^+$ spectroscopy;
  crosses -- objects in front of the CMZ.  All of the objects
  with yellow circles with dots and a few of the objects with
  white circles are listed in Table~2.  The three supermassive
  star clusters are indicated.  Several stars in each of the
  Central Cluster and the Quintuplet Cluster are overlapped and
  expressed with one circle.  The background is the 2MASS image
  at $K_S$-band from Skrutskie et al. (2006).  Bottom figure:
  Locations of the stars on which a set of H$_3^+$ and CO
  spectra have been observed overlaid on the 90\,cm radio
  continuum map taken from LaRosa et al. (2000). The major radio
  sources are marked and indicated. The projected physical
  distances are shown in grids centered at Sgr~A$^\ast$,
  assuming the distance to the Galactic center of 8\,kpc.}
\end{figure*}
%------------------------------------------------------------

Until very recently almost no background sources suitable for
spectroscopy of interstellar H$_3^+$ and CO on lines of sight
toward the CMZ were known further than $\sim$30\,pc from the
center of the CMZ. In 2008 we initiated a search for bright,
dust-embedded stars within the entire CMZ, a 10 times larger
region. We first used the catalogue of bright YSO candidates in
the ISOGAL survey (Felli et al. 2002) which yielded one star
suitable for spectroscopy of H$_3^+$, but which turned out to be
in the foreground (see Section~3 and Figure~4). We then
initiated a much more extensive star search using the Spitzer
Space Telescope Galactic Legacy Infrared Mid-Plane Survey
Extraordinaire ({\it GLIMPSE}) Catalog (Ram\'irez et al. 2008)
which includes photometry from the Two Micron All Sky Survey
({\it 2MASS}) Point Source Catalog (Skrutskie et al. 2006). Out
of the over million objects in the catalogue, 6000 stars with
$L\leq 8$\,mag are located in the central 2.39\degr $\times$
0.6\degr ($\sim 340 \times 85$\,pc) of the Galaxy, corresponding
roughly to the dimensions of the CMZ. We dropped obvious
late-type and/or low extinction stars based on the {\it 2MASS}
and {\it GLIMPSE} photometry. The remaining $\sim$500 stars have
been examined during the last eight years by medium resolution
$K$-band spectroscopy, which can distinguish dust-embedded stars
from other stars with spectral features characteristic of red
giants. For details of the star search and the $K$-band
  spectra of stars suitable for H$_3^+$ spectroscopy, see Geballe
  et al. (2019).

Plots of the survey highlighting the stars suitable for H$_3^+$
spectroscopy are shown in Figure~2. We have obtained spectra of
$\sim$40 of these stars, 29 of which show absorption by warm
diffuse CMZ gas, and are listed in Table 2. A very few clearly
do not; these are likely to be foreground stars. For several the
spectra are inconclusive, mainly due to low S/N.  The newly
found stars are all in the {\it 2MASS} catalogue. For
convenience, in discussions and compilations we have also
designated these stars using the Greek alphabet from alpha to
lambda (from west to east), and by adding $+$ and $-$
to the Greek notation as additional nearby stars are found.

%============================================================
% Table 2 % section 2.2
%============================================================
\setcounter{table}{1} \begin{deluxetable*}{ll rr rr c}
% \tablewidth{0pt}
\tablecaption{Targets toward which H$_3^+$ in the CMZ has been observed.\label{t2}}
% \tableline \tableline
\tablehead{
  \colhead{2MASS}&
  \colhead{Star}&
  \colhead{$l$}&
  \colhead{$b$}&
  \colhead{$K$}&
  \colhead{$L$}& 
  \colhead{$N({\rm H_3^+})_{\rm diffuse}$\tablenotemark{a}}  \\
   \colhead{}&
   \colhead{}&
   \colhead{[\degr]}&
   \colhead{[\degr]}&
   \colhead{[mag]}  &
   \colhead{[mag]}  &
   \colhead{[$10^{15}$\,cm$^{-2}$]}
}

\startdata
17432173$-$2951430 & $\alpha$    & $-$1.0463 & $-$0.0651 & 6.484\phantom{0}  & 3.792\phantom{0}&   1.57\\
17432823$-$2952159 & $\alpha+$   & $-$1.0417 & $-$0.0899 & 10.104\phantom{0} & 7.02\phantom{00}&   4.81\\
17432988$-$2950074 & $\beta$     & $-$1.0082 & $-$0.0762 & 8.823\phantom{0}  & 4.531\phantom{0}&   1.53\\
17443734$-$2927557 & $\gamma-$   & $-$0.5651 & $-$0.0905 & 10.289\phantom{0} & 7.032\phantom{0}&        \\
17444083$-$2926550 & $\gamma$    & $-$0.5442 & $-$0.0925 & 9.404\phantom{0}  & 6.406\phantom{0}&   2.87\\
17445895$-$2923259 & $\gamma+$   & $-$0.4604 & $-$0.1183 & 11.145\phantom{0} & 7.431\phantom{0}&        \\
17450483$-$2911464 & $\delta$    & $-$0.2834 & $-$0.0350 & 9.044\phantom{0}  & 6.507\phantom{0}&   \\
                & GCIRS~8$^\ast $& $-$0.0486 & $-$0.0422 & 13.3\phantom{000} &                 &   \\
                   & GCIRS~10W   & $-$0.0536 & $-$0.0469 & 11.170\phantom{0} &                 &   \\
                   & GCIRS~3     & $-$0.0576 & $-$0.0461 & 10.64\phantom{00} & 4.84\phantom{00}&   \\
                   & GCIRS~21    & $-$0.0561 & $-$0.0471 & 10.4\phantom{000} & 6.29\phantom{00}&   \\
                   & GCIRS~16~NE & $-$0.0552 & $-$0.0467 & 9.18\phantom{00}  & 7.16\phantom{00}&   \\
                   & GCIRS~1W    & $-$0.0549 & $-$0.0473 & 8.9\phantom{000}  & 4.92\phantom{00}&   \\
17452861$-$2856049 & $\epsilon$  & $-$0.0150 &    0.0274 & 9.218\phantom{0}  & 6.580\phantom{0}&   \\
17460215$-$2857235 & $\epsilon+$ &    0.0300 & $-$0.0882 & 8.076\phantom{0}  & 4.197\phantom{0}&    \\
17460433$-$2852492 & NHS~21      &  0.0992   & $-$0.0553 & 7.558\phantom{0}  & 4.604\phantom{0}&   1.03\\
17460562$-$2851319 & NHS~22      &  0.1200   & $-$0.0482 & 7.462\phantom{0}  & 6.369\phantom{0}&   \\
17460825$-$2849545 & NHS~42      &  0.1479   & $-$0.0426 & 8.325\phantom{0}  & 6.613\phantom{0}&   3.03\\
17461524$-$2850035 & NHS~25      &  0.1591   & $-$0.0654 & 7.291\phantom{0}  & 5.629\phantom{0}&   \\
17461783$-$2850074 & $\eta$      &  0.1631   & $-$0.0742 & 7.84\phantom{00}  & 5.519\phantom{0}&   \\
17461412$-$2849366 & GCS~3-4     &  0.1634   & $-$0.0581 & 7.717\phantom{0}  & 3.903\phantom{0}&   \\
17461471$-$2849409 & GCS~3-2     &  0.1635   & $-$0.0605 & 7.293\phantom{0}  & 3.162\phantom{0}&   3.48\\
17461586$-$2849456 & GCS~4       &  0.1646   & $-$0.0647 & 7.236\phantom{0}  & 3.527\phantom{0}&   \\
17461481$-$2849343 & GCS~3-1     &  0.1652   & $-$0.0598 & 7.52\phantom{00}  & 4.693\phantom{0}&   \\
17461798$-$2849034 & FMM~362     &  0.1787   & $-$0.0653 & 7.094\phantom{0}  & 6.41\phantom{00}&   2.63\\
17463219$-$2844546 & $\theta$    &  0.2647   & $-$0.0738 & 9.209\phantom{0}  & 6.381\phantom{0}&   3.37\\
17470898$-$2829561 & $\iota$     &  0.5477   & $-$0.0593 & 10.445\phantom{0} & 6.579\phantom{0}&   \\
17482472$-$2824313 & $\lambda-$  &  0.7685   & $-$0.2501 & 9.539\phantom{0}  & 6.721\phantom{0}&   2.85\\
17470137$-$2813006 & $\lambda-+$ &  0.7746   &    0.1108 & 9.978\phantom{0}  & 7.062\phantom{0}&   \\
\enddata
\tablenotetext{a}{$N({\rm H_3^+})_{\rm diffuse}$ is the total
  column density in $10^{15}$\,cm$^{-2}$ of H$_3^+$ in the
  diffuse gas of the CMZ. In addition to sightlines listed in
  this table, following sightlines have been observed but
  H$_3^+$ in the CMZ have not been definitely detected:
  17431001$-$2951460 ($\alpha-$), 17445538$-$2941284 ($\beta+$),
  17444319$-$2937526 ($\beta++$), 17443734$-$2927557
  ($\gamma-$), 17451618$-$2903156 (ISOGAL), 17452861$-$2856049
  ($\epsilon$), 17460215$-$2857235 ($\epsilon+$),
  17474486$-$2826365 ($\kappa$), 17473680$-$2816005
  ($\lambda$). Some of these are believed to be foreground stars.}
\end{deluxetable*}

%------------------------------------------------------------
%----------------------------------------------------------
% Section  2.3
\subsection{Spectrometers and Telescopes}

Five infrared spectrographs at five different telescopes have
been used to obtain the spectra presented here, three in Hawaii
and two in Chile. Cold Grating Spectrometer 4 (CGS4) with a
spectral resolution of $R\sim 40,000$ was used on the 3.8-m
United Kingdom Infrared Telescope (UKIRT) from 1997 to 2004. The
8.2-m Subaru Telescope's Infrared Camera and Spectrograph (IRCS)
with $R \sim 20,000$ was used from 2001 to 2015. The Phoenix
spectrometer with $R \sim 60,000$ was employed from 2003 to 2010
and 2016 to 2017 at the 8.1-m Gemini South telescope. The
Cryogenic Infrared Echelle Spectrograph (CRIRES) with $R \sim
50,000-100,000$ was used in 2007 on the 8.2-m Very Large
Telescope (VLT) of the Paranal Observatory. The Gemini
Near-Infrared Spectrograph (GNIRS) with $R \sim 20,000$ was used
from 2011 to 2014 at the 8.1-m Frederick C. Gillett Gemini North
telescope. Details of the operation of these spectrometers and
data reduction procedures are given in Geballe et al. (1999) for
CGS4, Goto et al. (2002) for IRCS, McCall et al. (2002) for
Phoenix, Goto et al. (2014) for CRIRES, and Geballe et
al. (2015) for GNIRS.

The IRCS, which contains an echelle and a cross-dispersion
grating, allows multiple H$_3^+$ lines to be observed
simultaneously (see Goto et al. 2002). For the other
spectrometers the key lines can only be made one line at a
  time except for the $R$(1,0)-$R$(1,1)$^u$ pair
  which are separated by 0.322\,cm$^{-1}$ and for GNIRS the
  pair and the $R$(1,1)$^l$ line, which are separated by
  35\,cm$^{-1}$.
In contrast all strong lines
of the CO overtone $R$ branch are closely spaced in wavelength
and could be observed simultaneously by all
spectrographs. Absorption lines of H$_3^+$ in diffuse clouds
have depths on the order of 0.5--15\% requiring integration
times sufficient for the signal-to-noise ratios of up to few
hundred on the continua of the background stars. Depending on
the spectrometer and the brightness of stars this required
exposure times of 10 minutes to a few hours. Altogether,
$\sim$100 lines of H$_3^+$ have been observed, amounting to a
total integration time of several hundred hours. The CO lines
typically have higher absorption depths than the H$_3^+$ lines
and for them signal-to-noise ratios of $\sim$30 were sufficient.
%============================================================
\section{Description of Spectra}
Spectra of many of the suitable background stars in lines of
H$_3^+$and CO are shown in Figures 3--8. The spectra display a
wide variety of profiles and are grouped so as to illustrate key
characteristics. Each figure is described briefly below and is
discussed in detail in Section~4.

%============================================================
% Figure 3
%------------------------------------------------------------
\setcounter{figure}{2}
\begin{figure}
 \includegraphics[width=0.48\textwidth]{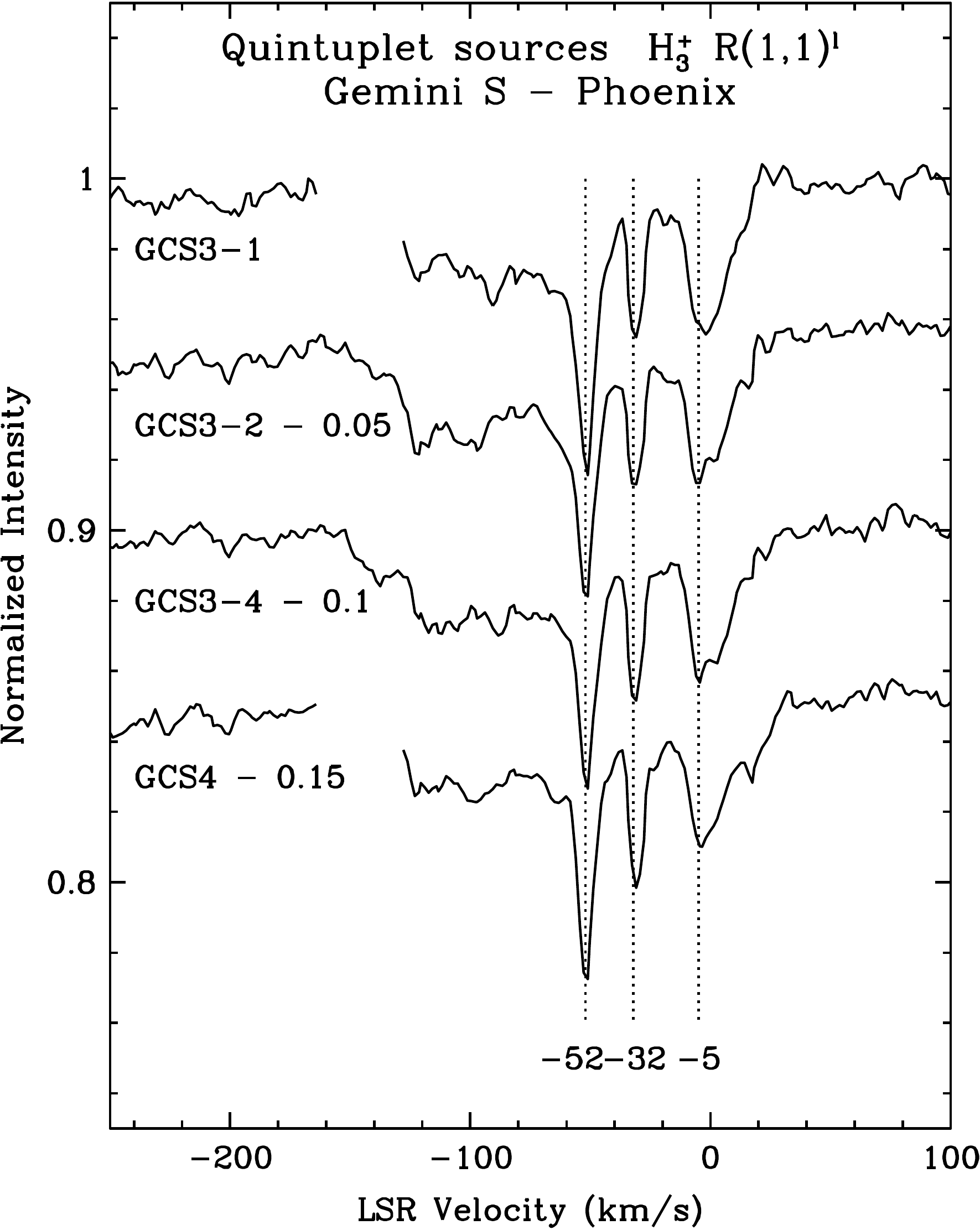}
  \caption{The H$_3^+$ $R$(1,1)$^l$ absorption toward four
    brighter stars of the Quintuplet Cluster observed by the
    Phoenix Spectrometer at the Gemini South Observatory. The 3
    sharp absorptions at velocities of $-$52, $-$32, and
    $-$5\,km\,s$^{-1}$ arise in the three foreground spiral
    arms. Note the presence of broad absorption troughs, due to
    H$_3^+$ in the CMZ, in addition to the narrow features.}
\end{figure}
%----------------------------------------------------------
%============================================================
% Figure 4
%------------------------------------------------------------
\setcounter{figure}{3}
\begin{figure}
 \includegraphics[width=0.48\textwidth]{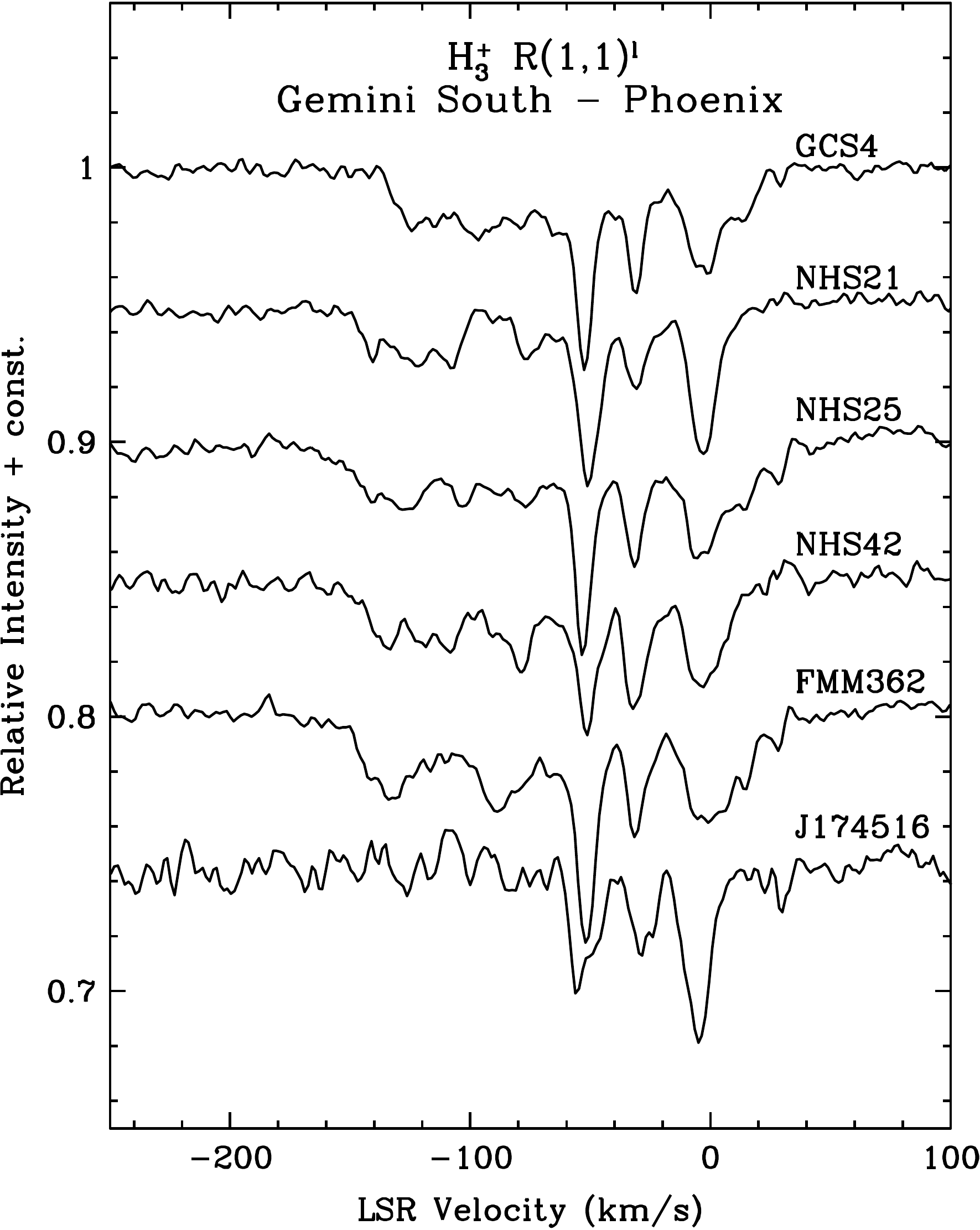}
  \caption{Spectra of the H$_3^+$ $R$(1,1)$^l$ line toward six
    stars near the Quintuplet Cluster. The three sharp
    absorptions at $\sim$ $-$52, $-$32, and $-$5\,km\,s$^{-1}$
    are due to H$_3^+$ in the foreground three spiral arms. The
    spectra of GCS\,4, NHS\,25, NHS\,42, and FMM\,362 show
    troughs from $-$150 to 0\,km\,s$^{-1}$ with similar
    equivalent widths, while the spectrum of NHS\,21 shows a
    weaker trough with a gap between $-$100 and
    $-$80\,km\,s$^{-1}$, The bottom spectrum toward the
    2MASS/ISOGAL star J17451618$-$2903156 has no absorption
    trough, only the narrow features arising in the spiral arms
    indicating that this star lies between the CMZ and the
    3\,kpc arm.}
\end{figure}
%----------------------------------------------------------

%============================================================
% Figure 5
%------------------------------------------------------------
\begin{figure*}
\begin{center}
\includegraphics[width=0.31\textwidth]{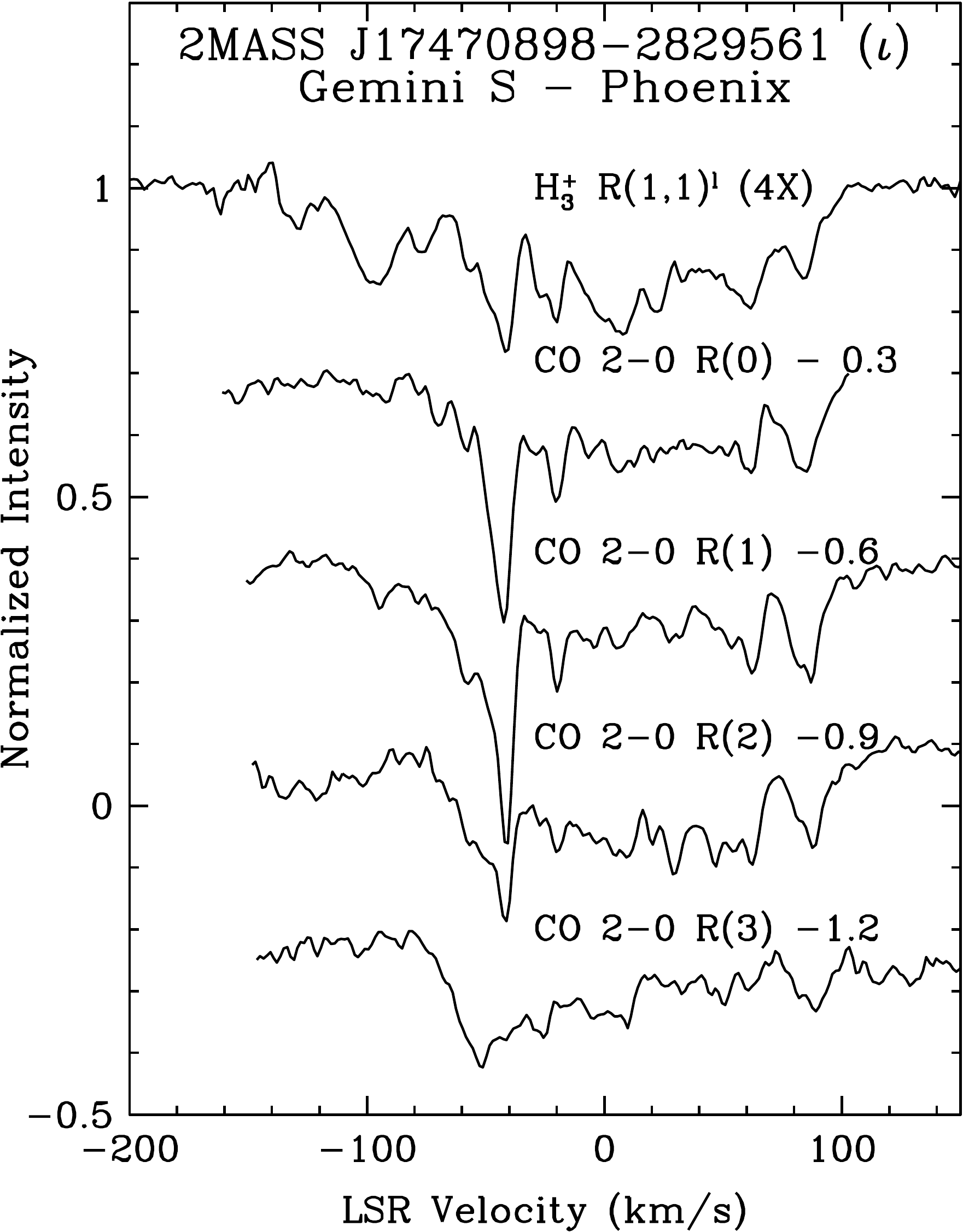}
\includegraphics[width=0.32\textwidth]{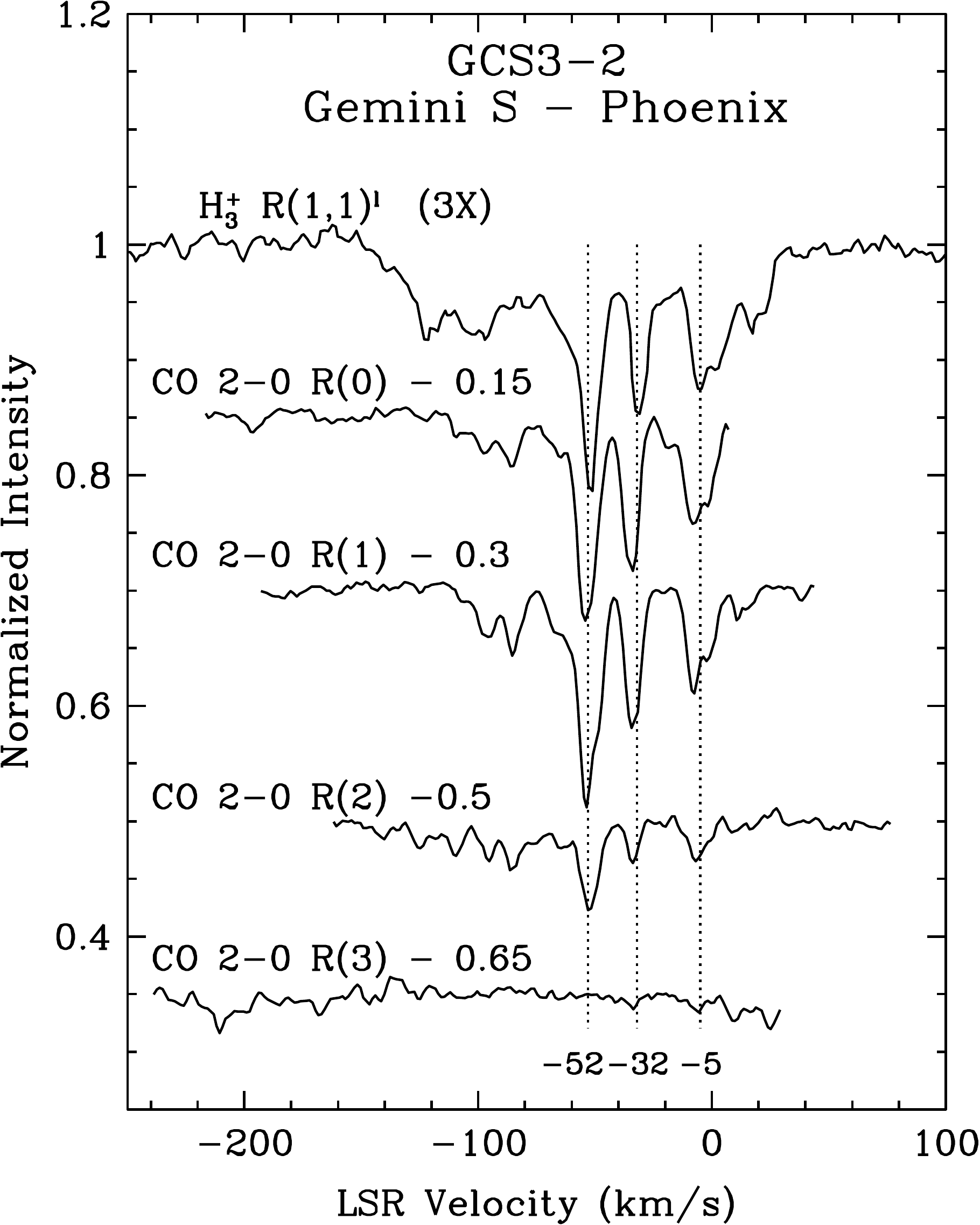}
\includegraphics[width=0.32\textwidth]{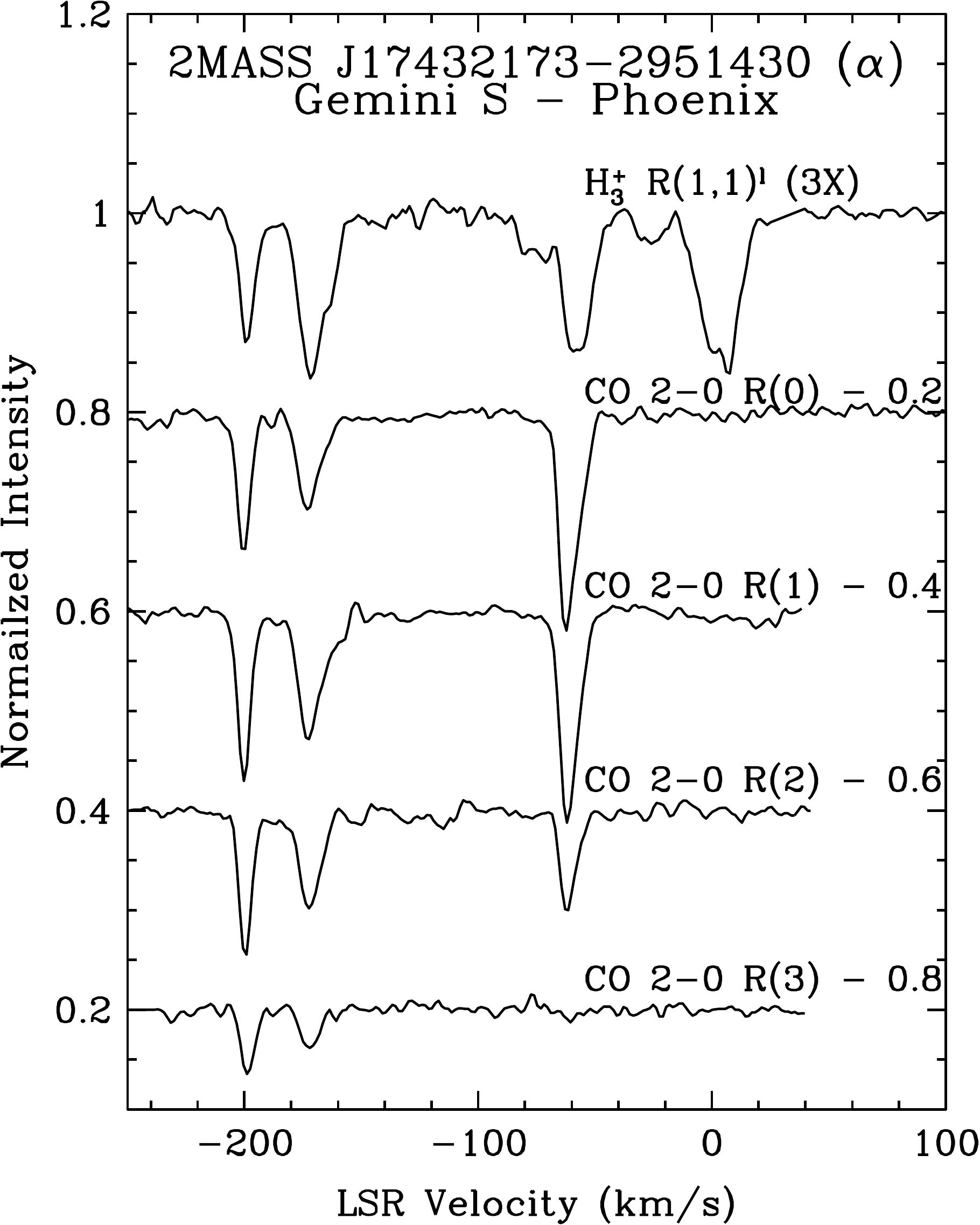}
\end{center}
\caption{$R$(1,1)$^l$ absorption spectra of H$_3^+$ (top traces)
  and CO overtone band absorption lines toward three stars
  located near the western edge (right panel), center (middle
  panel) and between Sgr\,B1 and B2, 85\,pc to the east of
  center (left panel).}
\end{figure*}

%------------------------------------------------------------

Figure~3 contains spectra, obtained with Phoenix in 2003, of the
$R$(1,1)$^l$ line of H$_3^+$ toward four of the five bright
infrared Quintuplet sources located close to the center of the
Quintuplet Cluster. The spectra are nearly identical, which is
perhaps not surprising given the proximities of the stars to one
another. Noteworthy features of the spectra are the three sharp
absorptions at $-$52, $-$32, and $-$5\,km\,s$^{-1}$ (LSR) due to
H$_3^+$ in the three foreground spiral arms, the 3\,kpc (Rougoor
\& Oort 1960; Oort 1977), 4.5\,kpc or Norma (Menon \& Ciotti
1970), and Local arms, respectively, and a broad absorption
trough extending from negative velocities as high as
$-$150\,km\,s$^{-1}$ to 0 or slightly positive velocities. The
same features with similar depths are seen toward sources in the
Central Cluster (e.g., Goto et al. 2014, Figure~3). As discussed
later in this paper, and as argued in previous papers (e.g., Oka
et al. 2005) this high dispersion gas is located within the
CMZ. The almost entirely negative absorption velocities imply
that the vast majority of the gas is moving outward from the
central region of the CMZ.

Figure~4 contains spectra of the same line of H$_3^+$, obtained
in 2008 towards GCS\,4, one of the Quintuplet stars observed in
2003, and five other stars near the Quintuplet Cluster, all
observed using Phoenix. The spectra of GCS\,4 in this figure and
in Figure~3, obtained five years apart, are essentially
identical. The spectra of three of the five other stars show
both the same narrow galactic spiral arms absorption features
and the same broad absorption trough as the Quintuplet
stars. The general similarity of these troughs to those seen
toward the Quintuplet and Central clusters indicates that the
outward-moving, high dispersion gas is not localized to a small
region, but is seen on sightlines covering an extended region of
the CMZ near its center. Toward one of the stars, NHS\,21, a
portion of the trough, between $-$100 and $-$80\,km\,s$^{-1}$,
is missing and the trough between $-$80 and 0\,km\,s$^{-1}$ is
weaker than the previously described stars. The spectrum
observed toward 2MASS\,J17451618$-$2903156 (bottom of the
figure) differs greatly from all of the others in that the
trough is completely absent. Our interpretations of the spectra
of NHS\,21 and J17451618 are given in Section~4.1.

Figure~5 contains spectra of the Quintuplet source GCS\,3-2,
which can be regarded as prototypical of stars located in the
central part of the CMZ, and of two stars located near the
western and eastern edges of the CMZ, J17432173$-$2951430
($\alpha$) and J17470898$-$2829561 ($\iota$). The spectra shown
for each of these stars are of the H$_3^+$ $R$(1,1)$^l$ line and
the four lowest lying $R$-branch lines of CO. J17432173
($\alpha$) is situated near the radio source Sgr~E, while
J17470898 ($\iota$) is located 85\,pc west of Sgr~A$^\ast$, in
the Sgr~B molecular cloud complex between Sgr~B1 and Sgr~B2. All
of the spectra were obtained with Phoenix at Gemini South and
have been published earlier (Oka et al. 2005; Geballe \& Oka
2010). They illustrate the most extreme differences in velocity
profiles that we have found for sightlines in the CMZ.

Toward GCS\,3-2 the CO absorptions consist almost entirely of
the three narrow spiral arm features, with no evidence for the
broad trough that is present in the H$_3^+$ line along with the
narrow features. The spectra towards the other two stars in
Figure~5 are strikingly different from both each other and the
spectra of GCS\,3-2 and other stars in the central region of the
CMZ (e.g., those with spectra shown in Figures~3 and 4).
J17432173 ($\alpha$) exhibits three narrow blue-shifted
absorption features in both H$_3^+$and CO, only one of which,
near $-$60\,km\,s$^{-1}$, corresponds to a foreground spiral
arm.  Its spectra also contain a prominent absorption feature in
H$_3^+$ near 0\,km\,s$^{-1}$, a velocity that approximately
matches that expected from absorption by the local arm, but with
no counterpart absorption at that velocity in CO, unlike the
Quintuplet and Central Cluster sources. In contrast, the spectra
of J17470898 ($\iota$) have much more complex absorption
profiles extending over $\sim$200\,km\,s$^{-1}$ in both
species. Some narrow absorption features, e.g. those at $-$45
and $+$80\,km\,s$^{-1}$ are present in both species, while
others, e.g. at $-$130 and $-$ 100\,km\,s$^{-1}$, are only
present in H$_3^+$. These spectra are discussed in Sections
4.2.1.--4.2.3.

Figure~6 contains spectra of the four lowest-lying $R$-branch
transitions of the $v = 2 \leftarrow 0$ band of CO toward 9
stars scattered across the CMZ, obtained with Phoenix at Gemini
South. As discussed in Section~4.1., the strengths of H$_3^+$
absorption features due to gas in the CMZ indicate that all are
deeply embedded in the CMZ. However, with the exceptions of
J17432173 ($\alpha$) at top left and J17470898 ($\iota$) at
bottom right these CO spectra are dominated by narrow absorption
features whose velocities match those of the foreground spiral
arms. In dense clouds, CO produces strong infrared absorption
features, yet no additional strong absorption features
attributable to dense clouds in the CMZ are seen toward the
other 10 stars. The interpretations of these spectra are given
in Section~4.1.

Figure~7 contains velocity profiles of the H$_3^+$ $R$(1,1)$^l$
and $R$(3,3)$^l$ lines toward 9 stars in the CMZ, as observed by
CGS4 at UKIRT, Phoenix at Gemini South, IRCS at Subaru, GNIRS at
Gemini North, and CRIRES at the VLT.  As explained in
Section~1.6., the presence of the $R$(3,3)$^l$ line is a
hallmark of warm gas. To date that line has been detected
nowhere else in the Galaxy except in the CMZ, whereas the
$R$(1,1)$^l$ absorption line, which originates from the lowest
energy level of H$_3^+$, is observed both in the CMZ and in cold
foreground clouds. The presence of the $R$(3,3)$^l$ line on
nearly all sightlines to sources in the direction of the CMZ
that have been observed to date indicates that the surface
filling factor of the warm gas in the CMZ is near unity. As
discussed in Section~5., it is highly likely that this warm gas
fills a large fraction of the front half of the CMZ.

Figure~8 is a comparison of spectra, obtained at Subaru, of the
$R$(1,1)$^l$ and $R$(2,2)$^l$ lines toward nine stars located
within the CMZ. As explained in Section~1.6, in warm diffuse gas
the strength of the $R$(2,2)$^l$ line relative to $R$(1,1)$^l$
is highly sensitive to density. In the CMZ the $R$(2,2)$^l$ line
has been detected only toward J17470898 ($\iota$) in the
  Sgr~B complex (as shown in the figure) and toward GCIRS~3 in
the Central Cluster whose sightline intersect dense gas in the
2-pc diameter circumnuclear ring (Goto et al. 2014). The
non-detection of this line on all other CMZ sightlines known to
contain warm gas (by virtue of the presence of absorption by the
$R$(3,3)$^l$ line) implies that the portions of those sightlines
passing through the CMZ cross low density gas only.

%============================================================
\section{Interpretation}

Here we address, in considerably more detail than in the
previous largely descriptive section, numerous issues regarding
the interpretation of the spectra, presented here and elsewhere,
of H$_3^+$ and CO toward stars within the CMZ.

%------------------------------------------------------------
% Section 4.1
\subsection{How deeply are the survey stars embedded in the CMZ?}

A major question in interpreting infrared absorption
spectroscopy of an object located within a region with high
extinction is the depth of the object within the obscuring
material. If the extinction within the CMZ is sufficiently
large, then a brightness-limited CMZ survey such as ours will
tend to select sources that are more shallowly embedded in the
CMZ than average. There has been some thought that this
selection effect is quite significant for the objects whose
absorption spectra we have been measuring and that the stars we
have observed are in the outer part of the CMZ, close to its
front surface.

Towards the Central Cluster, a region of diameter $\sim$3 pc
containing many luminous stars, and centered on Sgr~A$^\ast$ and
thus arguably at the center of the CMZ, this possibility has
been decisively negated by observations of many bright infrared
stars of the Central Cluster (e.g. Viehmann et al. 2005). As
discussed in Section 1.2., these stars suffer extinctions of
only $\sim$30 visual magnitudes, corresponding to $\sim$3 and
$\sim$1 magnitudes in the $K$ and $L$ bands, respectively (Rieke
\& Lebofsky 1985), of which some occurs outside of the CMZ
(e.g., in the foreground spiral arms). Our survey stars range in
$L$ magnitude from 3 to 8, suggesting a wide range of intrinsic
stellar brightness and that extinction influencing the selection
of targets is a minor issue, at least to the depth of the
Central Cluster.  Our spectral lines of the CO $v = 2 \leftarrow
0$ band toward these sources show little or no absorption due to
dense clouds, apart from those in the three foreground spiral
arms (e.g., see Figure~6 and Goto et al. 2014,
Figure~3). Clearly, giant molecular clouds do not exist in the
CMZ in front of the Central Cluster; any such clouds that exist
on that general sightline must be located behind the Central
Cluster.

The Quintuplet Cluster is also thought to be located deep within
the CMZ and close to the Central Cluster, as is the Arches
Cluster, with all three clusters having formed in the last few
to several million years (e.g. Schneider et al. 2014; Espinoza
et al. 2009). As in the case of the Central Cluster, our spectra
of CO $v = 2 \leftarrow 0$ band lines (e.g., GCS\,3-2 in
Figures~5 and 6) toward Quintuplet Cluster stars show no
evidence for large dense clouds in the front half of the CMZ
on these sightlines. In addition, the high total column
densities of H$_3^+$ in the absorption troughs, on the order of
$3 \times 10^{15}$\,cm$^{-2}$, observed toward GCS\,3-2 (Oka et
al 2005) the other GCS stars (see Figure~3), are comparable to
those toward the stars in the Central Cluster (Goto et al. 2008,
2013, 2014), which is an argument for the Quintuplet Cluster
being as deeply embedded as the Central Cluster. Finally, the
extinctions to stars in those two clusters (Liermann et
al. 2010) are comparable. Thus, the evidence is overwhelming
that the Quintuplet Cluster is as deeply embedded in the CMZ as
the Central Cluster.  The extinction to the Arches Cluster
(Habibi et al 2013) is similar to that to the Quintuplet
Cluster, indicating that it is as deeply embedded in the CMZ as
the other two clusters. However, the stars of the Arches Cluster
are too faint for high-resolution spectroscopy of H$_3^+$, as
they are still on the main sequence and less luminous than the
more evolved stars in the Quintuplet Cluster.

The depths in the CMZ of the other stars that we have observed
are not known. However, the observed column densities of warm
H$_3^+$ in Table~2 give an approximate idea of their depths
inside the CMZ. This is because the ubiquity of the $R$(3,3)$^l$
line (Figure~7) indicates that the surface filling factor of the
CMZ's warm diffuse gas approaches 100\,\%. Its volume filling
factor is likely more than 50\,\%, at least for the front
hemisphere of the CMZ, as estimated in Section 5.  This,
together with the constancy of the H$_3^+$ number density (see
Figure~3 of Oka 2013), makes the column density of warm H$_3^+$
in diffuse clouds an approximate yardstick for measuring the
depths of stars in the CMZ.

The absorption profiles of the H$_3^+$ $R$(1,1)$^l$ line toward
six stars near the Quintuplet Cluster, shown in Figure~4,
illustrate this. After excluding the three sharp absorption
components due to H$_3^+$ in dense clouds in the three
foreground spiral arms, there remain broad blue-shifted
absorption ``troughs'' extending from $\sim -150$\,km\,s$^{-1}$
to near 0\,km\,s$^{-1}$, similar to those seen in the Quintuplet
and Central Clusters. Similar to the stars in the Quintuplet and
Central Clusters, the troughs are due to warm diffuse gas. The
large equivalent widths, comparable to those of the cluster
stars, toward four of these stars, GCS\,4 (which is a member of
the Quintuplet Cluster) NHS\,25, NHS\,42, and FMM\,362, indicate
that they are as deeply embedded in the CMZ as the Central
Cluster. The spectrum of NHS\,21 is somewhat different; the
trough is not as deep and has a gap from $-$100 to
$-$80\,km\,s$^{-1}$. That suggests that NHS\,21 is located
somewhat in front of the aforementioned four stars, although
still within the CMZ and much of the outwardly moving diffuse
gas.  The difference between the spectra of these five stars and
that of J174516 (ISOGAL bottom spectrum) is much more
pronounced, as a trough in the latter is completely absent. The
narrow spiral arm features are still present in its spectrum,
however, implying that J174516 lies between the outer (front)
edge of the CMZ and the innermost (3\,kpc) arm.

Some of the other stars that have been observed in H$_3^+$ also
appear to be more shallowly embedded in the CMZ. Figure~7
contains spectra of one example, J17444083$-$2926550 ($\gamma$),
which is located roughly midway between the center of the CMZ
and its western edge. Its spectra show strong absorption by warm
diffuse gas at high negative velocities, but little or no
absorption by warm diffuse gas at less negative velocities where
the trough is prominent toward stars located near the center of
the CMZ. Like the spectrum of this star, the spectra of other
stars on similarly situated sightlines tend to show their
strongest absorption by warm diffuse gas at the high negative
velocity end of the absorption trough.
%------------------------------------------------------------
\subsection{Differentiating between dense and diffuse gas}

The state of carbon changes from C$^+$ to C to CO as cloud
density changes from diffuse to dense (e.g. Figure~1 of Snow \&
McCall 2006). On the other hand, H$_3^+$ exists in any
interstellar environment as long as H$_2$ abounds; that is, in
diffuse molecular clouds as well as in dense clouds. Thus, as
discussed in Section~1.8, high-resolution spectra of infrared CO
and H$_3^+$ absorption lines on the same sightline can allow one
to discriminate between H$_3^+$ in diffuse molecular clouds and
in dense clouds. The spectra in Figure~5 demonstrate three
strikingly different examples of the use of this property.

%------------------------------------------------------------
% Section 4.2.1
\subsubsection{GCS\,3-2: Diffuse blue-shifted gas}

The Quintuplet Cluster star GCS\,3-2, the brightest infrared
star in the CMZ that we have employed as a background source ($L
= 3.16$\,mag), is located 30.7\,pc to the east of Sgr~A$^\ast$
and 2.0\,pc below the Galactic plane. Its CO overtone spectrum
(middle panel of Figure~5) is dominated by CO in dense clouds in
the three foreground spiral arms at $-$52, $-$32, and
$-$5\,km\,s$^{-1}$. The only additional absorptions due to CO,
at $-$123, $-$97, and $+$19\,km\,s$^{-1}$, are weak, indicating
that there is little dense gas within the CMZ on this sightline.
As pointed out earlier, this paucity of dense CMZ gas is
generally found towards all of the stars observed by us except
those near Sgr~E (Section~4.2.2.) and Sgr~B (4.2.3.).  The
H$_3^+$ $R$(1,1)$^l$ spectrum in the same figure contains
absorption features at the above six velocities, indicating that
the absorbing H$_3^+$ responsible for those features must be in
the same dense clouds. Removing them (by chopping them off as in
Figure~4 of Oka et al. 2005) leaves a wide trough of absorption
by H$_3^+$, from $-$150 to $+$32\,km\,s$^{-1}$ arising in
diffuse molecular clouds in the CMZ. Also as pointed out
earlier, this gas is much warmer than diffuse gas in the
Galactic disk.  Blueshifted H$_3^+$ absorption troughs of
similar depth and velocity extent are present toward many stars
located in the central regions of the CMZ (e.g., Figure~4). This
suggests that the blue shifted warm diffuse gas occupies a
region of large and continuous or near-continuous radial extent
in the front half of the CMZ, presumably with the highest
velocities closest to the front edge and the lowest velocities
near the center. The extensions of the troughs to slightly
red-shifted velocities observed on sightlines to GCS\,3-2 and
some of the stars in Figures~3 and 4 could mean that those stars
are located slightly behind the region from which gas is being
or has been ejected.

%------------------------------------------------------------
\subsubsection{J17432173$-$2951430 ($\alpha$): Diffuse gas moving perpendicular to sightline}

This bright star ($L = 3.79$\,mag), located 139\,pc to the west
of Sgr~A$^\ast$ and 2.7\,pc below the Galactic plane, near Sgr~E
and the western edge of the CMZ, most simply demonstrates the
use of the CO overtone spectrum for discriminating between
the H$_3^+$ in diffuse and dense clouds. The H$_3^+$ spectrum
(right in Figure~5) is composed of four prominent absorptions at
LSR velocities $-$200\,km\,s$^{-1}$ (depth 4.4\,\%),
$-$172\,km\,s$^{-1}$ (5.6\,\%), $-$60\,km\,s$^{-1}$ (4.6\,\%),
and a blend at 0\,km\,s$^{-1}$ and $+$8\,km\,s$^{-1}$ (5.4\,\%),
together with two much weaker absorptions at $-$27\,km\,s$^{-1}$
(0.9\,\%) and a doublet centered at $-$75\,km\,s$^{-1}$
(1.5\,\%). Three of the four deep H$_3^+$ absorptions are
accompanied by CO absorption at the same velocities; thus we
interpret them as arising in dense gas. Of these, the two
highest negative velocities agree with the velocities of $\sim
-$200\,km\,s$^{-1}$ observed toward the Sgr~E complex in
$^{13}$CO and the H70$\alpha$ radio recombination line by Liszt
(1992) and more precisely with $-$207\,km\,s$^{-1}$ and
$-$174\,km\,s$^{-1}$ of the CO cloud used for observations of
N$^{+}$ and C$^{+}$ by Langer et al. (2015). Thus, these two
dense clouds lie within the CMZ. They are fairly compact, as
they are only seen over a small region in the CMZ.  Where they
are present, their spectral profiles vary drastically with the
Galactic coordinates as can be seen by comparing the H$_3^+$
spectrum of $\alpha$ with the H$_3^+$ spectra of nearby stars
$\alpha+$, and $\beta$ (Figure~7).
The source-to-source variations must be due to the peculiar
morphology and dynamics of the dense clouds in this part of the
CMZ and are not of interest in this paper.
The third
deep absorption at $-$60\,km\,s$^{-1}$ is caused by H$_3^+$ in
dense clouds of the foreground 3\,kpc spiral arm. Dense clouds
in the other two spiral arms, near $-$30 and 0\,km\,s$^{-1}$, do
not exist on this sightline.

The deep blend of absorptions by H$_3^+$ at 0\,km\,s$^{-1}$ and
$+8$\,km\,s$^{-1}$ and the weaker features at $-27$ and
$-75$\,km\,s$^{-1}$ are also due to H$_3^+$ in diffuse molecular
clouds. It is possible that some of the H$_3^+$ absorption near
zero velocity is due to cold foreground diffuse gas, as such
gas, which could be associated with the local arm, does not
produce detectable CO absorption.  However, the presence of
strong absorption in the $R$(3,3)$^l$ line of H$_3^+$ at these
low velocities (Figure~7, top left panel) implies that much, if
not all of the absorbing H$_3^+$ is both diffuse and warm and
therefore located within the CMZ.

Although the near zero radial velocity of warm H$_3^+$ on this
sightline near the western edge of the CMZ could be due to the
absorbing gas traveling at low speed, in view of the rapid
radial outward motions observed closer to the center of the CMZ,
it is much more natural to interpret it as the same rapid gas
motion (velocities up to 150\,km\,s$^{-1}$) as seen toward the
center, but perpendicular to the line of sight.  In other
  words, the overall motion of the warm diffuse gas is most
  naturally interpreted as a roughly radial expansion
  originating near the center of the CMZ. 
This low radial velocity absorption by warm diffuse gas is also
seen in the other stars located near the western edge of the
CMZ, $\alpha+$ and $\beta$ (Figure~7), again demonstrating the
large dimension of the warm, diffuse gas. We interpret the
variation of their H$_3^+$ column densities from sightline to
sightline as due to variations of these stars' depths in the
CMZ. The weaker absorption at $-$27\,km\,s$^{-1}$ and a doublet
centered at $-$75\,km\,s$^{-1}$, which are not accompanied by CO
absorptions, are also due to H$_3^+$ in diffuse clouds. We
interpret them as due to small stray clouds.

%------------------------------------------------------------
% Section 4.2.3
\vspace*{-2mm}
\subsubsection{17470898-2829561($\iota$): The only sightline with CO dominated
by CMZ absorption}

This star, 84.5\,pc east of Sgr~A$^\ast$ (about 60\,\% of the
distance from Sgr~A$^\ast$ to the eastern edge of the CMZ) and
1.8\,pc below the Galactic plane is located between Sgr~B1 and
B2, a region of dense giant molecular clouds undergoing massive
star formation. This is the only sightline in our survey whose
CO absorption is dominated by high dispersion CMZ gas (Figure~5
left). The H$_3^+$ absorption component at $-$98\,km\,s$^{-1}$
is the only clear feature not seen in CO absorption, and thus it
is the only absorption definitely arising in diffuse gas. The
$R$(3,3)$^l$ spectrum (Figure~7) also has a strong absorption at
this velocity, further attesting to this H$_3^+$ absorption
component arising in the warm diffuse gas of the CMZ. The
observed H$_3^+$ column density in the $-$98\,km\,s$^{-1}$
component indicates that this star is embedded in the CMZ at
least a few tens of parsec from the front of the expanding
diffuse gas. The remainder of the H$_3^+$ $R$(1,1)$^l$
absorption, whose depth is approximately 4\,\% continuously from
$v \sim -90$\,km\,s$^{-1}$ to $+100$\,km\,s$^{-1}$, is
accompanied by CO absorption and hence arises in dense gas of
the giant molecular cloud complex. As seen in the figure, the $v
= 2 \leftarrow 0$ absorption lines of CO toward this star are
strong out to the most highly excited line shown [$R$(3)],
unlike the CO spectra toward other stars in which the $R$(3)
absorption is absent or very weak (see Figure~6). Indeed,
significant absorption is present in J17470898 ($\iota$) out to
$R$(5). This indicates high kinetic temperatures in this dense
molecular cloud complex. Apart from the $-$98\,km\,s$^{-1}$
feature, analysis of the complex spectra of this star is beyond
the scope of this paper.

%------------------------------------------------------------
% 4.3
\subsection{Column Densities of Dense Gas in the Foreground and CMZ}

Although the large beam size of single dish radio observations
HPBW $\sim$ 16\arcsec~(0.62\,pc) (Oka et al. 1998b) and the
highly-saturated $J = 1 \rightarrow 0$ CO emission found toward
the CMZ might give the impression that the CMZ is packed with
dense clouds, pencil beam spectroscopy of lines of the first
vibrational overtone of CO toward our survey stars tells quite a
different story. As seen in Figure~6 in 9 stars located from
140\,pc west to 85\,pc east of Sgr~A$^\ast$, the vast majority
of spectra show very little absorption by CO, apart from
features associated with foreground spiral arms, and hence very
little dense gas resident in the CMZ on their sightlines. All of
these stars are deeply embedded in the CMZ as demonstrated by
the observed H$_3^+$ total column densities of $N({\rm H_3^+}) >
2 \times 10^{15}$\,cm$^{-2}$ and in many cases by their inferred
locations relative to the Central Cluster. The only sightlines
where prominent absorption by CO within the CMZ is observed are
toward several stars near Sgr~E on the western extremity of the
CMZ (e.g., J17432173$-$2951430 ($\alpha$), Figure~6 top left),
toward J17470898$-$2829561 ($\iota$) in the Sgr~B molecular
cloud complex (Figure~6 bottom right), and toward several stars
in the Central Cluster.

%============================================================
% Table 3 % 4.3 Column Densities of Dense Gas in the Foreground 
%============================================================
% \documentclass[11pt,letterpaper]{aastex}
% \setcounter{table}{3}
% \documentclass[8pt]{aastex}
% \begin{document}
\tabletypesize{\scriptsize}
\tablewidth{0pt}
\startlongtable
\begin{deluxetable*}{cc cc cccc ccc}
\tablecaption{CO column densities toward stars in the CMZ.\label{t3}}
% \tablenum{3a}
\tablehead{
\colhead{Star     } &
\colhead{location}  &
\colhead{$v$}       &
\colhead{$\Delta$v }&
\colhead{$N$(0)}    &
\colhead{$N$(1)}    &
\colhead{$N$(2)}    &
\colhead{$N$(3)}    &
\colhead{$N{\rm (CO)_{cloud}}$} &
\colhead{$N{\rm (CO)_{total}}$} &
\colhead{$A_V$} \\
\colhead{         } &
\colhead{         } &
\colhead{[km\,s$^{-1}$]}  &
\colhead{[km\,s$^{-1}$]      }  &
\colhead{[$10^{17}$\,cm$^{-2}$]} &
\colhead{[$10^{17}$\,cm$^{-2}$]} &
\colhead{[$10^{17}$\,cm$^{-2}$]} &
\colhead{[$10^{17}$\,cm$^{-2}$]} &
\colhead{[$10^{17}$\,cm$^{-2}$]} &
\colhead{[$10^{17}$\,cm$^{-2}$]} &
\colhead{[mag]} 
}
\startdata
           &    arm 1  & $-$62     &   26  & 4.22  & 5.94  & 2.81  & 0.25  & 12.7  &     arm 12.7  &       arm 13\\
 $\alpha$  &      CMZ  & $-$173    &   26  & 2.17  & 5.00  & 3.62  & 1.18  & 12.0  &     CMZ 22.0  &       CMZ 22\\
           &      CMZ  & $-$200    &   19  & 1.83  & 3.67  & 3.02  & 1.48  & 10.0  &               &             \\ \hline
           &    arm 3  &  $+$3     &   15  & 1.33  & 2.14  & 2.05  &    -  &  5.5  &    arms 10.4  &     arms 10 \\
  $\beta$  &    arm 1  & $-$58bl   &   24  & 1.61  & 1.92  & 1.33  &    -  &  4.9  &               &             \\
           &      CMZ  & $-$193    &   17  & 0.82  & 1.38  & 1.89  &    -  &  4.1  &      CMZ 4.1  &         CMZ 4\\ \hline
           &    arm 3  &     $-$1  &   22  & 1.83  & 3.66  & 2.01  & 0.76  &  8.3  &    arms 24.0  &     arms 24 \\
 $\gamma$  &    arm 2  &    $-$32  &   14  & 1.57  & 2.28  & 1.09  &    -  &  3.9  &               &             \\
           &    arm 1  &    $-$54  &   18  & 3.66  & 4.67  & 1.61  & 0.84  & 10.8  &               &             \\
           &      CMZ  &   $-$136  &   15  & 0.87  & 1.85  & 2.61  & 2.91  &  8.2  &      CMZ 8.2  &        CMZ 8 \\ \hline
           &      CMZ  & $+$13     &   19  & 1.35  & 2.82  & 2.85  & 1.27  &  8.3  &               &             \\
           &    arm 3  & $-$6bl    &   18  & 1.59  & 2.68  & 2.77  & 0.63  &  7.7  &               &             \\
 $\delta$  &    arm 2  & $-$30     &   18  & 2.29  & 3.55  & 2.05  &    -  &  7.9  &    arms 23.5  &     arms 24 \\
           &    arm 1  & $-$53     &   18  & 2.75  & 4.09  & 1.09  &    -  &  7.9  &               &             \\
           &      CMZ  & $-$143    &   14  & 0.69  & 2.10  & 1.61  & 0.68  &  5.1  &     CMZ 13.4  &       CMZ 13 \\ \hline
           &      CMZ  &    $+$57  &   13  & 0.65  &    -  & 1.41  &    -  &  2.1  &      CMZ 4.8  &        CMZ 5\\
           &      CMZ  &    $+$44  &   13  &    -  &    -  & 1.49  & 1.22  &  2.7  &               &             \\
GCIRS\,21  &    arm 3  &     $-$2  &   27  & 7.86  & 10.17 & 7.92  & 4.81  & 30.8  &               &             \\
           &    arm 2  &    $-$31  &   12  & 1.64  & 1.99  & 1.01  &    -  &  4.6  &    arms 45.9  &     arms 46 \\
%           &    arm 1  &    $-$55  &   20  & 4.07  & 4.24  & 2.21  &    -  & 10.5  &               &             \\ \tablebreak
           &    arm 1  &    $-$55  &   20  & 4.07  & 4.24  & 2.21  &    -  & 10.5  &               &             \\ \hline
           &      CMZ  &    $+$68  &   34  &    -  &    -  &    -  & 5.99  &  6.0  &               &             \\
           &      CMZ  &    $+$44  &   13  & 1.35  &    -  & 2.45  & 1.06  &  4.9  &     CMZ 24.9  &      CMZ 25 \\
           &      CMZ  &    $+$25  &   26  & 2.72  & 5.18  & 3.02  & 1.10  & 12.0  &               &             \\
 GCIRS\,3  &    arm 3  &     $-$2  &   27  & 7.95  &    -  &    -  & 1.86  &  9.8  &               &             \\
           &    arm 2  &    $-$31  &   12  & 2.51  & 3.00  &    -  &    -  &  5.5  &    arms 23.5  &     arms 24 \\
           &    arm 1  &    $-$55  &   20  & 3.64  & 3.58  & 1.01  &    -  &  8.2  &               &             \\
           &      CMZ  &    $-$74  &   17  & 0.70  & 0.43  & 0.88  &    -  &  2.0  &               &             \\ \hline
           &      CMZ  &    $+$57  &   13  & 0.22  & 1.45  & 1.29  &    -  &  3.0  &               &             \\
           &      CMZ  &    $+$44  &   13  & 0.24  & 1.67  & 1.37  &    -  &  3.3  &     CMZ 14.8  &      CMZ 15 \\
           &      CMZ  &    $+$25  &   26  &    -  & 1.67  & 3.38  & 1.90  &  7.0  &               &             \\
GCIRS\,16NE&    arm 3  &     $-$2  &   27  & 5.78  & 9.41  & 8.44  & 2.79  & 26.4  &               &             \\
           &    arm 2  &    $-$31  &   12  & 1.54  & 1.74  & 1.13  &    -  &  4.4  &    arms 44.0  &     arms 44 \\
           &    arm 1  &    $-$55  &   20  & 3.90  & 6.15  & 2.41  & 0.72  & 13.2  &               &             \\
           &      CMZ  &    $-$74  &   17  & 0.51  & 1.01  &    -  &    -  &  1.5  &               &             \\ \hline
           &      CMZ  &    $+$57  &   13  & 0.63  & 1.19  &    -  & 0.72  &  2.5  &               &             \\
           &      CMZ  &    $+$44  &   13  & 2.39  & 4.53  & 5.67  & 2.74  & 15.3  &     CMZ 27.7  &      CMZ 28 \\
GCIRS\,1W  &      CMZ  &    $+$25  &   26  & 1.88  & 4.63  & 2.01  & 1.39  &  9.9  &               &             \\
           &    arm 3  &     $-$2  &   27  & 7.98  &    -  & 3.22  & 1.94  & 13.1  &               &             \\
           &    arm 2  &    $-$31  &   12  & 2.02  & 1.92  &    -  &    -  &  3.9  &    arms 25.5  &     arms 26 \\
           &    arm 1  &    $-$55  &   20  & 3.57  & 3.55  & 1.41  &    -  &  8.5  &               &             \\ \hline
           &    arm 3  &   $-$5bl  &   18  & 3.37  & 4.45  & 2.05  &    -  &  9.9  &    arms 25.3  &      arms 25\\
  NHS\,21  &    arm 2  &    $-$35  &   12  & 1.08  & 1.41  & 0.56  &    -  &  3.1  &               &             \\
           &    arm 1  &    $-$51  &   27  & 3.78  & 4.81  & 2.77  & 0.89  & 12.3  &        CMZ -  &       CMZ - \\ \hline
%           &    arm 1  &    $-$51  &   27  & 3.78  & 4.81  & 2.77  & 0.89  & 12.3  &        CMZ -  &       CMZ - \\ \tablebreak
           &    arm 3  &   $-$6bl  &   23  & 2.19  & 3.29  & 1.33  &    -  &  6.8  &    arms 27.3  &      arms 27\\
  NHS\,22  &    arm 2  &    $-$38  &   13  & 0.99  & 0.98  & 0.68  &    -  &  2.7  &               &             \\
%           &    arm 1  &    $-$51  &   20  & 5.37  & 6.88  & 4.74  & 0.84  & 17.8  &        CMZ -  &       CMZ - \\ \hline
           &    arm 1  &    $-$51  &   20  & 5.37  & 6.88  & 4.74  & 0.84  & 17.8  &        CMZ -  &       CMZ - \\\tablebreak
           &      CMZ  &     $+$2  &   12  & 0.72  & 1.67  & 1.61  &    -  &  4.0  &      CMZ 7.1  &       CMZ 7 \\
           &    arm 3  &     $-$6  &   16  & 1.49  & 2.86  & 1.25  &    -  &  5.6  &               &             \\
  NHS\,42  &    arm 2  &    $-$34  &   19  & 2.72  & 4.20  & 1.53  &    -  &  8.5  &    arms 25.4  &     arms 25 \\
           &    arm 1  &    $-$51  &   24  & 3.30  & 5.47  & 2.49  &    -  & 11.3  &               &             \\
           &      CMZ  &    $-$86  &   25  & 0.82  & 1.38  & 0.88  &    -  &  3.1  &               &             \\ \hline
           &    arm 3  &   $-$6bl  &   23  & 2.51  & 3.66  & 0.88  &    -  &  7.1  &               &             \\
  NHS\,25  &    arm 2  &    $-$33  &   13  & 1.74  & 2.46  & 0.68  &    -  &  4.9  &    arms 25.0  &     arms 25 \\
           &    arm 1  &    $-$54  &   25  & 4.41  & 6.41  & 3.06  &    -  & 13.9  &               &             \\ \hline
           &    arm 3  &  $-$4     &   21  & 2.39  & 2.90  & 1.89  &    -  &  7.2  &               &             \\
   $\eta$  &    arm 2  & $-$33     &   12  & 2.00  & 2.46  & 0.68  &    -  &  5.1  &    arms 23.9  &     arms 24 \\
           &    arm 1  & $-$50     &   25  & 4.02  & 5.65  & 1.93  &    -  & 11.6  &               &             \\
           &      CMZ  & $-$108    &   18  & 0.60  & 1.38  & 0.96  &    -  &  2.9  &      CMZ 2.9  &       CMZ 3 \\ \hline
           &    arm 3  &   $-$8bl  &   20  & 2.34  & 2.61  & 1.01  &    -  &  6.0  &               &             \\
           &    arm 2  &    $-$36  &   16  & 1.83  & 2.32  & 0.68  &    -  &  4.8  &    arms 21.5  &     arms 22 \\
 GCS\,3$-$2 &   arm 1  &    $5$4   &   20  & 3.16  & 5.43  & 2.09  &    -  & 10.7  &               &             \\
           &      CMZ  &    $-$66  &   12  & 0.46  & 0.58  & 0.56  &    -  &  1.6  &               &             \\
           &      CMZ  &    $-$85  &   13  & 0.67  & 1.09  & 1.09  &    -  &  2.9  &      CMZ 6.6  &       CMZ 7 \\
%           &      CMZ  &    $-$98  &   12  & 0.48  & 0.91  & 0.68  &    -  &  2.1  &               &             \\  \tablebreak
           &      CMZ  &    $-$98  &   12  & 0.48  & 0.91  & 0.68  &    -  &  2.1  &               &             \\  \hline
           &    arm 3  &   $-$9bl  &   29  & 3.37  & 2.86  & 1.69  &    -  &  7.9  &               &             \\
   GCS\,4  &    arm 2  &    $-$34  &   19  & 2.02  & 2.32  & 0.84  &    -  &  5.2  &    arms 25.6  &      arms 6 \\
           &    arm 1  &    $-$54  &   21  & 3.69  & 5.76  & 3.06  &    -  & 12.5  &               &             \\
           &      CMZ  &    $-$66  &   10  & 0.55  & 1.38  & 0.80  &    -  &  2.7  &      CMZ 2.7  &       CMZ 3 \\ \hline
           &      CMZ  & $+$13     &   13  & 0.34  & 0.65  & 0.84  &    -  &  1.8  &               &             \\
           &    arm 3  &     0     &   12  & 0.75  & 1.41  & 0.60  &    -  &  2.8  &               &             \\
           &    arm 3  & $-$11     &    9  & 1.25  & 0.94  & 0.36  &    -  &  2.6  &    arms 28.9  &     arms 29 \\
 FMM\,362  &    arm 2  & $-$34     &   13  & 2.27  & 1.88  & 1.45  &    -  &  5.6  &               &             \\
           &    arm 1  & $-$53     &   19  & 5.40  & 7.28  & 4.86  & 0.34  & 17.9  &               &             \\
           &      CMZ  & $-$70     &   10  & 0.60  & 0.80  & 1.53  & 0.25  &  3.2  &     CMZ 10.9  &      CMZ 11 \\
           &      CMZ  & $-$93bl   &   28  & 1.54  & 2.14  & 2.01  & 0.25  &  5.9  &               &             \\ \hline
           &    arm 3  &     $-$6  &   16  & 2.22  & 3.37  & 2.57  &    -  &  8.2  &               &             \\
 $\theta$  &    arm 2  &    $-$38  &   17  & 4.53  & 8.18  & 4.54  & 1.60  & 18.9  &    arms 55.0  &     arms 55 \\
           &    arm 1  &    $-$52  &   18  & 6.89  & 11.73 & 7.56  & 1.69  & 27.9  &               &             \\
           &      CMZ  &   $-$101  &   14  & 0.72  & 1.01  &    -  &    -  &  1.7  &      CMZ 1.7  &     CMZ 1.7 \\ 
\enddata
\tablecomments{The LSR velocities $v$, velocity widths
  $\Delta v$, the level column densities $N(J)$, the total
  column densities $N$(CO)$_\mathrm{cloud}$ for each cloud,
  total arm column densities $N$(CO)$_\mathrm{arm}$, total CMZ
  column densities $N$(CO)$_\mathrm{CMZ}$, estimated total
  visual extinction of arms ($A_V$)$_{\rm arm}$ and of the CMZ
  ($A_V$)$_{\rm CMZ}$ are given. ``bl'' denotes blended components.
%   The velocities are in km
%   s$^{-1}$ and column densities are in 10$^{17}$ cm$^{-2}$.
  }
\end{deluxetable*}

% \
 \begin{deluxetable}{cc cccc cc}
\tablecaption{CO column densities toward $\iota$ in the CMZ.\label{t4}}
% \tablenum{3b}
\tablehead{
% \colhead{Star     } &
% \colhead{location}  &
% \colhead{$v$}       &
% \colhead{$\Delta$v }&
\colhead{$N$(0)}    &
\colhead{$N$(1)}    &
\colhead{$N$(2)}    &
\colhead{$N$(3)}    &
\colhead{$N$(4)}    &
\colhead{$N$(5)}    &
\colhead{$N{\rm (CO)_{total}}$} &
\colhead{$A_V$} 
}
\startdata
% $\iota$    & total     &
44.3    & 75.7   & 68.7 & 54.9  & 19.9   & 6.4   & 270                    & 270   \\
\enddata
\tablecomments{Same with Table~\ref{t3} but for the star $\iota$ toward
  Sgr\,B, which has a complicated velocity profile with high
  velocity dispersion; only total CO column densities are
  given.}
 \end{deluxetable}

% \
 % IOTA
%------------------------------------------------------------
% \vspace*{-1.2cm}

The CO column densities for each velocity component toward 18
stars are given in Table~3. They were calculated using $N({\rm
  CO})_{\rm level}= (3hc)/(8\pi^3 \lambda) W_\lambda/|\mu|^2$
from measured equivalent widths $W_\lambda = \int [\Delta
  I(\lambda)]/[I(\lambda)] d\lambda$ of the $R$(0), $R$(1),
$R$(2), and $R$(3) absorption lines for levels $J =$ 0, 1, 2,
and 3, respectively, and the transition strength $|\mu|^2 = (J
+1)/(2J +1) \langle 2|\mu|0 \rangle ^2$ with the transition
dipole moment $\langle 2|\mu|0 \rangle = 0.006518$\,D (Zou \&
Varanasi 2002). It should be noted that these values are much
more accurate and reliable than the CO column densities measured
from radio CO emission because (1) the upper level of a
transition is not populated, (2) the absorption is not saturated
because of the small transition dipole moment, and, most
significantly, (3) the large velocity gradient method (Goldreich
\& Kwan 1974) used to overcome the complication of radiation
trapping need not be invoked to determine the column density. As
discussed in the previous section, the sightline toward
J17470898$-$2829561($\iota$) is unique, with multiple blended
velocity components, which we do not attempt to separate; we
list it separately at Table~4.

%============================================================
% Figure 6
%------------------------------------------------------------
\begin{figure*}
\begin{center}
  \includegraphics[width=0.9\textwidth]{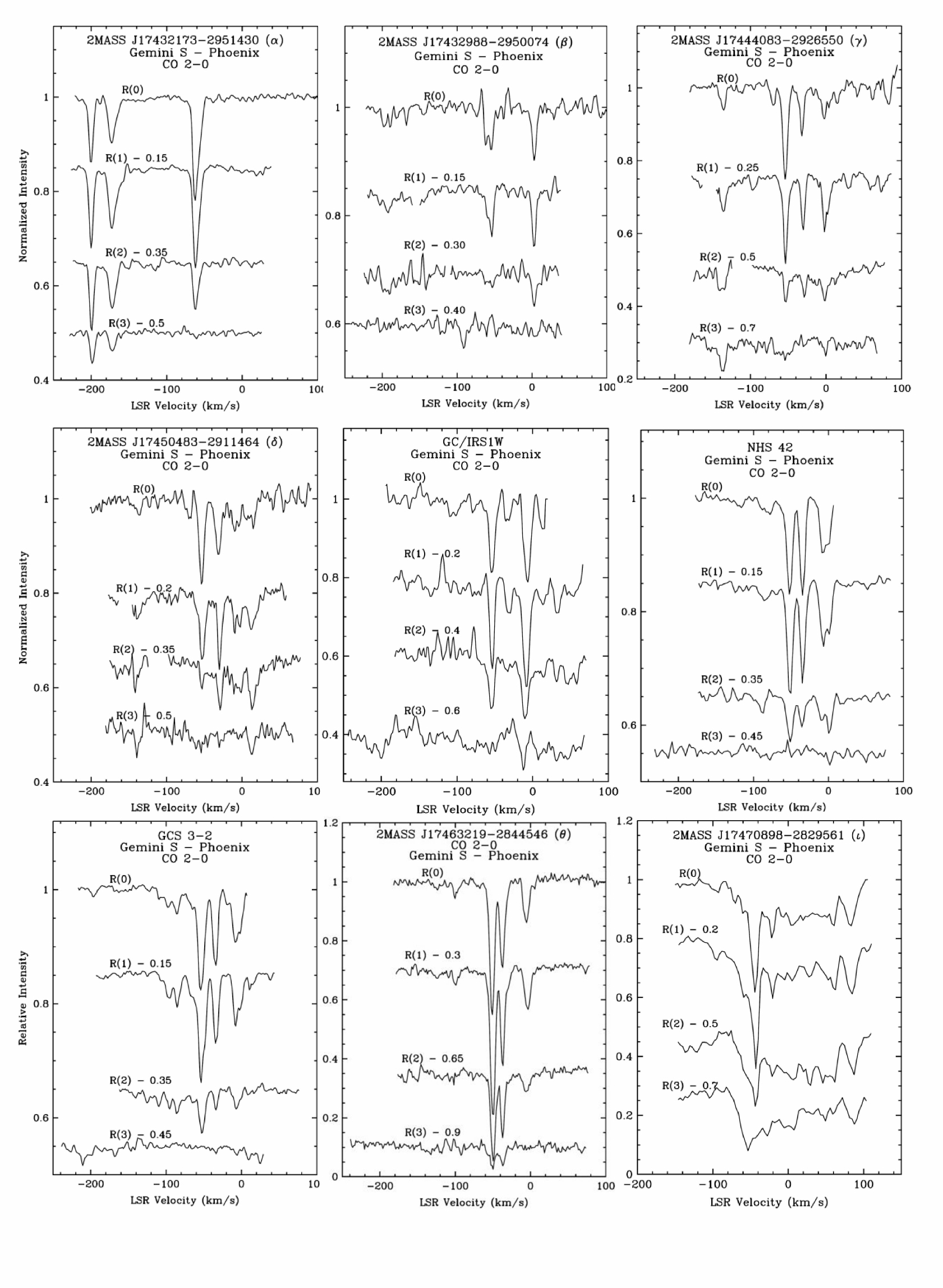}
\end{center}
  \caption{The $\nu = 2 \leftarrow 0$ first overtone absorption
    lines $R$(0), $R$(1), $R$(2), and $R$(3) observed toward
    nine stars in the CMZ using the Phoenix Spectrometer at the
    Gemini South Observatory. The H$_3^+$ absorptions toward
    these stars indicate that the stars are all deeply embedded
    in the CMZ. Note that except for the stars
    J17432173$-$2951430 ($\alpha$) (top left) which is in the
    Sgr\,E complex and J17470898$-$2829561 ($\iota$) (bottom
    right) in the Sgr\,B complex, the CO spectra are dominated
    by absorptions due to the three foreground spiral arms the
    3\,kpc arm, the 4.5\,kpc arm, and the local arm.}
\end{figure*}
%----------------------------------------------------------
% \vspace{-1.4cm}
%------------------------------------------------------------
% S. 4.3.1
\clearpage
\subsubsection{CO in the three spiral arms}

As described earlier, sightlines toward the GC cross three
spiral arms, Norma, Scutum, and Sagittarius (e.g. Figure~2 of
Vall\,ee 2016), which leave their fingerprints on nearly all of
the spectra of H$_3^+$ and CO presented in this paper. Staying
out of the controversy regarding the detailed structure of
spiral arms, and whether or not the Norma arm and the 3\,kpc arm
are identical etc. (e.g. Vall\,ee 2014), we refer to them
hereafter as the 3\,kpc arm (Rougoor \& Oort 1960) (arm 1), the
4.5\,kpc arm (Menon \& Ciotti 1970) (arm 2), and the local arm
(arm 3). Deep absorption due to CO in the 3\,kpc arm is observed
toward all 9 stars in Figure~6 and 18 stars listed in Table~3
and is the strongest of the three spiral arm absorptions toward
most stars. When this absorption feature is not observed toward
a star, we regard the star to be in front of the 3\,kpc arm. The
absorption velocity due to this arm varies from
$-$62\,km\,s$^{-1}$ on the western edge of the CMZ to
$-$50\,km\,s$^{-1}$ on the eastern edge. Such radial-velocity
gradients in Galactic longitude are well known and are used for
studies of kinematics and dynamics of the gas in the GC region
(e.g. Sofue 2006). Weaker absorption lines of CO in the 4.5\,kpc
arm with velocities from $-$32\,km\,s$^{-1}$ (west) to
$-$38\,km\,s$^{-1}$ (east) are observed toward all 20 stars
except two stars in the west end, J17432173$-$2951430 ($\alpha$)
and J17432988$-$2950074 ($\beta$). We interpret this as due to
incidental absence of dense clouds in that arm on those
sightlines. Finally, deep and relatively broad absorption at
$\sim$0\,km\,s$^{-1}$ is due to CO in the pile up of local gas
closest to the sun. This absorption feature is observed toward
all but one star, J17432173$-$2951430 ($\alpha$), whose
sightline must also happen to avoid local dense molecular gas.

The total CO column densities in the three spiral arms, $N({\rm
  CO})_{\rm arms}$, are also listed in Table~3. Fourteen of the
eighteen sightlines have column densities in the range
$(1.04-2.89) \times 10^{18}$\,cm$^{-2}$. Four stars, the
aforementioned J17470898$-$2829561 ($\iota$), GCIRS\,21,
GCIRS\,16NE, and J17473219$-$2844546 ($\theta$) have much higher
values. The higher column densities toward GCIRS\,21 and
GCIRS\,16NE are due to exceptionally high values in the
$\sim$0\,km\,s$^{-1}$ component, where contributions from the
local arm and dense clouds in the CMZ overlap and cannot be
separated. The high value toward Star $\theta$ is due to an
exceptionally high column density of dense molecular gas in the
3\,kpc arm on its sightline. Assuming the ratio of H$_2$ to CO
number densities to be $10^4$, the fourteen ``normal'' values of
$N({\rm CO})_{\rm arms}$ correspond to H$_2$ column densities,
$N({\rm H_2})_{\rm arms} \sim (1-3) \times 10^{22}$\,cm$^{-2}$,
which correspond to $A_V \sim 10-30$\,mag, if the standard
conversion factor from hydrogen column density to color excess
$E(B-V)$ (Bohlin et al. 1978) and $A_V/E(B-V)$ ratio of $\sim$3
(both caused by interstellar dust) are used. The values of $A_V$
for the arms are also given in Table~3. The value of the total
CO column density $N({\rm CO})$ in the units of
$10^{17}$\,cm$^{-2}$ and the visual extinction $A_V$ in
magnitudes happen to be nearly identical using this conversion.

A detailed analysis of CO in the spiral arms is beyond the scope
of this paper and will be published in a separate paper.

%============================================================
% Figure 7
%------------------------------------------------------------
\setcounter{figure}{6}
\begin{figure*}
 \includegraphics[width=\textwidth]{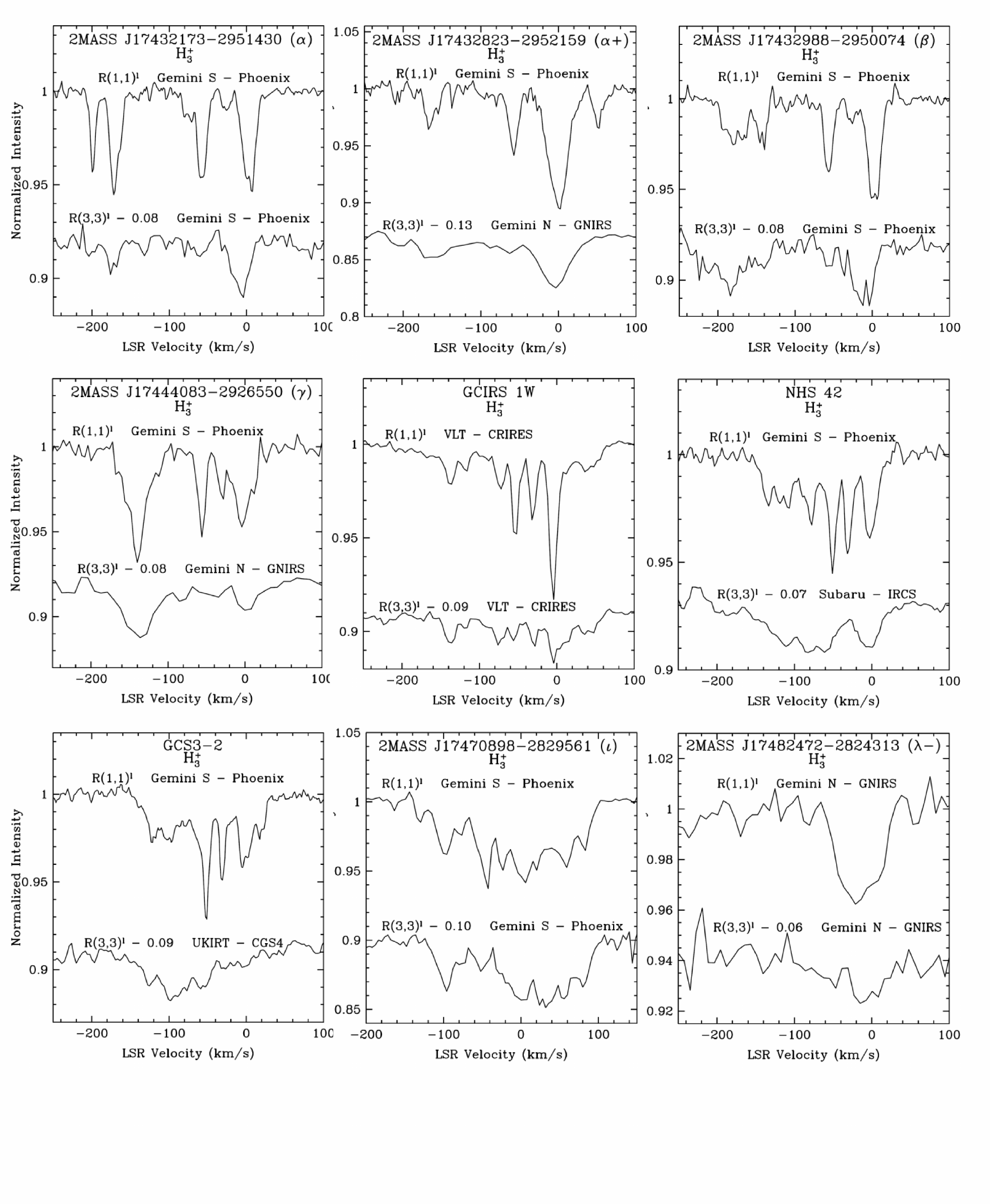}
  \caption{The $R$(1,1)$^l$ (upper trace) and $R$(3,3)$^l$
    (lower trace) absorption lines of H$_3^+$ toward 10 stars in
    the CMZ. The $R$(1,1)$^l$ absorption profiles include
    H$_3^+$ in the both the CMZ and the three foreground spiral
    arms, while the $R$(3,3)$^l$ absorption is entirely from
    H$_3^+$ in the CMZ. }
\end{figure*}
%------------------------------------------------------------
%------------------------------------------------------------
% S. 4.3.2
\subsubsection{ CO in the CMZ}

We have already demonstrated that most of our survey stars are
embedded approximately as deeply in the CMZ as the stars in the
Central Cluster. It is thus apparent, from both Figure~6 and
Table~3, that on most observed sightlines the CO column
densities in the front half of the CMZ are far less than the
column densities in the foreground spiral arms. Out of the 17
stars listed in Table~3, the values of $N({\rm CO})_{\rm CMZ}$
are somewhat higher than $N({\rm CO})_{\rm arms}$ only for 3
stars, $\alpha$ in the Sgr~E complex and GCIRS\,3 and GCIRS\,1W
in the Sgr~A complex. The visual extinctions due to dense clouds
in the CMZ are $A_V~\sim 20-30$\, mag towards the three stars
and $\sim$300\,mag for star $\iota$ listed in Table~4. For the
remaining 14 stars in Table~3, $A_V$ is very much smaller and a
far cry from the mean value $A_V \sim 500$\,mag obtained
assuming that dense ($n \geq 10^4$\,cm$^{-3}$) gas has a volume
filling factor of $f \geq 0.1$ in the CMZ.  In calculating these
extinctions due to dense gas in the CMZ, which are based on our
CO measurements, we have assumed that the dust-to-gas ratio
scales with the carbon abundance, and thus compensates for the
higher than solar value of [C/H] in the CMZ (see Appendix A.2).
This assumption is crudely consistent with the dependency of
dust-to-gas ratio with galactocentric radius found by Giannetti
et al. (2017).

Cotera et al. (2000) and Schultheis et al. (2009) used infrared
photometry of stars in the Galactic center to derive total
visual extinctions (i.e., due to dust in both dense and diffuse
gas). The former authors, who relied on {\it 2MASS} values,
found total extinctions in the range 25-40\,mag for stars in the
CMZ, while the latter authors, who used both {\it 2MASS} and
longer wavelength Spitzer IRAC photometry, which detects more
highly obscured stars than {\it 2MASS}, determined somewhat
higher values.  To compare our values with theirs, we must add
to the extinction that we derive due to dense clouds in the CMZ
and the spiral arms, the extinction by dust in diffuse gas along
the entire sightline. The extinction due to diffuse gas on
sightlines to objects near Sgr~A$^\ast$ has been roughly
estimated by Whittet et al. (1997) to be 20 visual magnitudes,
by subtracting their estimate of 10 visual magnitudes of
extinction due to dust in dense clouds, from the canonical 30
magnitudes of total extinction. Note, however, that almost all
of our estimates of visual extinction by dense clouds on such
sightlines are much higher than 10\,mag (Table~3), suggesting
that the extinction due to dust in diffuse gas may be
significantly less than 20\,mag, or perhaps that the dust-to-gas
ratio in the CMZ derived from $N({\rm CO})$ does not scale with
the carbon abundance. Adding 10--20\,mag to the sum of the
values of $A_V$ due to dense clouds in Table~3 gives total
visual extinctions of 20--70\,mag for most of the sightlines
that we observed, in general agreement with the ranges reported
by Cotera et al. and Schultheis et al.

The case of J17470898$-$2829561 ($\iota$), where the visual
extinction based on the column density of CO is $\sim$300\,mag,
poses a severe problem. The use of a standard extinction
dependency on wavelength (e.g., $\propto \lambda^{-1.7}$) or the
more recently derived $\lambda^{-2.11}$ in the region of the
Central Cluster (Fritz et al. 2011) lead to absurdly high values
for the extinction in the $K$ and $L^\prime$ bands
($\sim$24\,mag and $\sim$10\,mag, respectively for the standard
dependency and $\sim$13\,mag and $\sim$4\,mag, respectively for
the Fritz et al. dependency), given the apparent brightness of
this star ($K_{\rm 2MASS}$=10.4\,mag, $L_{\rm IRAC}$=6.6\,
mag). Either the dust-to-gas ratio in this portion of the Sgr~B
cloud is vastly lower than predicted from the value of [C/H] or
the extinction falloff with increasing wavelength on this
sightline is considerably steeper than the above extinction
curves. It is unclear if either of these possibilities is
manifested in dense clouds elsewhere in the CMZ; if so, the
effect would be to greatly lower the extinctions (reducing the
impact of the selection effect discussed earlier).  Given the
modest dense cloud extinctions found on the other sightlines in
the CMZ, neither of these possibilities affect the major
conclusions this paper. The properties of the interstellar gas
and dust toward star $\iota$ will be presented and discussed in
more detail in a forthcoming paper (Geballe et al., in prep.)

%------------------------------------------------------------
% 4.4
\subsection{H$_3^+$ in the Central Molecular Zone}

As discussed in Section 4.1. and shown in Table~2, column
densities of warm H$_3^+$ on the order of $3 \times
10^{15}$\,cm$^{-2}$ are observed toward stars deeply embedded in
the CMZ. In the following sections, we describe how the
individual column densities in the para-(1,1) ground level, the
ortho-(1,0) level, the unstable para-(2,2) level, and the
metastable ortho-(3,3) level provide information on the
temperature and density of the gas in which H$_3^+$ is
located.

The observed (1,1) and (1,0) column densities include
contributions from the CMZ and the foreground spiral arms. Only
in the high-resolution spectra obtained with UKIRT/CGS4, Gemini
South/Phoenix, and VLT/CRIRES can these contributions be
separated (see Figure~4 of Oka et al. 2005). On the other hand,
the columns densities in the (2,2) and (3,3) levels come
entirely from the CMZ and can be measured with all of the
spectrographs that we have used. Figure 7 shows the $R$(1,1)$^l$
(upper traces) and $R$(3,3)$^l$ (lower traces) absorption
spectra toward nine stars from 139\,pc west of Sgr~A$^\ast$
($\alpha$) to 116\,pc east ($\lambda-$). The column densities
contributed by the CMZ and individual spiral arms, where
separable, are given in Table~5.  For the $R$(1,1)$^l$
absorption different velocity components are separately measured
using high resolution ($R \geq 40,000$)
spectrometers of CGS\,4 at UKIRT, Phoenix at the Gemini
South Observatory, and CRIRES at VLT, in order to 
  discriminate between the H$_3^+$ absorptions in the 3
foreground spiral arms. See Figure~4 of Oka et al. (2005). On
the other hand for the $R$(3,3)$^l$ absorption which does not
exist in the low temperature spiral arms, low resolution
observations suffice.

%------------------------------------------------------------
% 4.4.1
%------------------------------------------------------------
\subsubsection{H$_3^+$ in the metastable (3,3) level: evidence for high temperature}

As indicated in Section~1.6. the ratio of H$_3^+$ column
densities in the metastable (3,3) and the ground (1,1) levels,
separated in energy by 361\,K, is a sensitive measure of the
temperature in diffuse warm gas. Unlike collisions between
NH$_3$ and H$_2$, which are physical and do not convert ortho
and para species (Oka, 1973), collisions between H$_3^+$ with
H$_2$ are chemical, in that protons are exchanged between
H$_3^+$ and H$_2$ and connect the ortho (3,3) level and the para
(1,1) level (Quack 1977; Uy et al. 1997; Oka 2004). Since
ortho-to-para conversion occurs rapidly, the temperature of the
gas can be approximately determined from the excitation
temperature $T_{\rm ex}(3,3/1,1)$ which is calculated from
\begin{equation}
  \frac{N(3,3)}{N(1,1)} = \frac{14}{3}
  \exp\left(-\frac{361\,{\rm K}}{T_{\rm ex}(3,3/1,1)} \right)
\end{equation}
\noindent
The actual kinetic temperature is higher than the excitation
temperature because the rapid symmetry-breaking $\Delta k = \pm
3$ spontaneous emissions cool the H$_3^+$. A more accurate
determination of temperature taking into account this effect
will be discussed in Section~5.1.

%============================================================
% Table 4.4.1 H3+ in metastable (3,3) 
%------------------------------------------------------------
\startlongtable
\begin{deluxetable}{lccccc}
% \def\arraystretch{0.9}
% \tabletypesize{9pt}
\tablewidth{0pt}
% \tablenum{4b}
\tablecaption{Level Column Densities of H$_3^+$ from Spectroscopy at Gemin South, Gemini North and UKIRT\label{t5}}
\tablehead{
  \multicolumn{1}{c}{Star} &
  \colhead{$v_{\rm LSR}$} &
  \colhead{$\Delta v_{\rm LSR}$} &
  \multicolumn{3}{c}{$N$ ($J$,$K$) [10$^{15}$\,cm$^{-2}$]} \\
  \cline{4-6}
  \colhead{} & 
  \colhead{[km\,s$^{-1}$]} &
  \colhead{[km\,s$^{-1}$]} &
  \colhead{(1,1)\tablenotemark{a}}& 
  \colhead{(1,1)\tablenotemark{b}}& 
  \colhead{(3,3)}
}
\startdata
    $\alpha$&$+$2&   35&   0.55\tablenotemark{c}&       &   0.19\tablenotemark{cd}\\
            &$-$26&  17&   0.06\tablenotemark{c}&       &       \\
            &$-$60&  21&   0.39&       &       \\
            &$-$77&  19&   0.12\tablenotemark{c}&       &       \\
            &$-$172& 31&   0.55&       &       \\
            &$-$200& 23&   0.28&       &       \\
       Total&   &    &   1.90&   0.00&   0.19\tablenotemark{c}\\
\hline
  $\alpha +$&$+$51&  30&   0.28\tablenotemark{c}&       &       \\
            &$+$1&   62&   1.82\tablenotemark{c}&       &   0.86\tablenotemark{cd}\\
            &$-$58&  27&   0.48&       &       \\
            &$-$167& 25&   0.32&       &       \\
       Total&   &    &   2.90&   0.00&   0.86\tablenotemark{cd}\\
\hline
     $\beta$&$+$2&   35&   0.58\tablenotemark{c}&       &   0.40\tablenotemark{cd}\\
            &$-$27&  16&   0.09&       &       \\
            &$-$57&  23&   0.30&       &       \\
            &$-$143& 23&   0.22&       &       \\
            &$-$176& 41&   0.43&       &       \\
       Total&   &    &   1.62&   0.00&   0.40\tablenotemark{cd}\\
\hline
    $\gamma$&$-$4&   38&   0.72&       &       \\
            &$-$30&  16&   0.20&       &       \\
            &$-$56&  22&   0.37&       &       \\
            &$-$136& 60&   1.16\tablenotemark{c}&       &   0.64\tablenotemark{c}\\
       Total&   &    &   2.44&       &   0.64\tablenotemark{c}\\
\hline
    $\delta$&$+$0&  10&   0.05&       &       \\
            &$-$8&  18&   0.19&       &       \\
            &$-$29&  20&   0.33&       &       \\
            &$-$54&  28&   0.45&       &       \\
            &$-$126&  30&   0.19&       &       \\
            &$-$155&  18&   0.06&       &       \\
       Total&   &    &   1.27&   0.85\tablenotemark{c}&       \\
\hline
  $$NHS$$ 21&$-$3&  30&   0.34&       &       \\
            &$-$30&  24&   0.14&       &       \\
            &$-$50&  24&   0.32&       &       \\
            &$-$125&  50&   0.45\tablenotemark{c}&       &   0.17\tablenotemark{c}\\
       Total&   &    &   1.25&       &       \\
\hline
  $$NHS$$ 22&$-$9&  29&   0.35&       &       \\
            &$-$38&  17&   0.17&       &       \\
            &$-$57&  22&   0.43&       &       \\
            &$-$83&  24&   0.09&       &       \\
       Total&   &    &   1.04&   0.55\tablenotemark{c}&       \\
% \tablebreak
       \hline
  $$NHS$$ 42&$-$3&  30&   0.14&       &       \\
            &$-$31&  24&   0.19&       &       \\
            &$-$50&  28&   0.25&       &       \\
            &$-$78&  29&   0.13&       &       \\
            &$-$113bl?&  29&   0.11&       &       \\
            &$-$134&  17&   0.06&       &       \\
       Total&   &    &   1.16&   1.24\tablenotemark{c}&       \\
\hline
  $$NHS$$ 25&$+$30&  10&   0.04&       &       \\
            &$+$17&   9&   0.05&       &       \\
            &$-$2bl?&  25&   0.27&       &       \\
            &$-$32&  18&   0.14&       &       \\
            &$-$52&  26&   0.31&       &       \\
            &$-$77&  14&   0.05&       &       \\
            &$-$86&  10&   0.03&       &       \\
            &$-$100&  13&   0.05&       &       \\
            &$-$125&  23&   0.12&       &       \\
            &$-$140?&   8&   0.03&       &       \\
       Total&   &    &   1.09&   1.43\tablenotemark{c}&       \\
\hline
     GCS~3-2&$+$17&  15&   0.10&       &       \\
            &$+$2&  13&   0.14&       &       \\
            &$-$6&  15&   0.16&       &       \\
            &$-$32&  18&   0.21&       &       \\
            &$-$52&  34&   0.43&       &       \\
            &$-$97&  19&   0.08&       &       \\
            &$-$123&  12&   0.03&       &       \\
       Total&   &    &   1.15&   1.45\tablenotemark{cf}&   0.70\tablenotemark{cf}\\
\hline
    FMM\,362&$-$1&  48&   0.55&       &       \\
            &$-$32&  20&   0.20&       &       \\
            &$-$52&  26&   0.41&       &       \\
            &$-$87&  28&   0.18&       &       \\
       Total&   &    &   1.34&   1.13\tablenotemark{c}&       \\
\hline
    $\theta$&$-$1&  24&   0.41&       &       \\
            &$-$32&  10&   0.14&       &       \\
            &$-$51&  32&   0.79&       &   0.36\tablenotemark{cg}\\
            &$-$122&  48&   0.22&       &       \\
       Total&   &    &   1.56&   1.24\tablenotemark{c}&   0.95\tablenotemark{cdg}\\
\hline
     $\iota$&$+$0& 234&   4.32&   0.00&   1.68\tablenotemark{cdg}\\
\hline
 $\lambda -$&$-$19&  89&   1.27\tablenotemark{eg}&       &   0.43\tablenotemark{cdg}\\
\hline
$\lambda -+$&$+$150? &  86&   0.61\tablenotemark{g}&       &       \\
            &$+$50?  &  46&   0.15 &       &       \\
            &$-$30bl & 133&   2.19\tablenotemark{ceg}&       &       \\
       Total&        &    &   2.94 &       &      \\  
\enddata
\tablecomments{Observed peak velocities, absorption widths, the
H$_3^+$ column densities for the ground level for each peaks,
for the trough, and $N$(3,3) for the metastable level.
Most lines have been observed in the Gemini South Observatory
using the Phoenix spectrometer, unless otherwise noted.}
\tablenotetext{a}{Sharp absorption lines with peaks.}
\tablenotetext{b}{Trough absorption component}
\tablenotetext{c}{Isolated diffuse gas.}
\tablenotetext{d}{$N$(3,3) are contaminated by atmospheric water absorption.}
\tablenotetext{e}{$N$(1,1) valuesare blended with the absorptions in spiral arms.}
\tablenotetext{f}{Observed at the UKIRT using the CGS4 spectrometer (Oka et al. 2005)}
\tablenotetext{g}{Observed at the Gemini North Observatory using the GNIRS spectrometer.}
\end{deluxetable}
% \end{splitdeluxetable}
 % H3+
%------------------------------------------------------------
Toward three stars, J17432173$-$2951430 ($\alpha$),
J17432823-2952159 ($\alpha+$), and J17432988-2950074 ($\beta$),
located near Sgr~E and the western edge of the CMZ ($G_{\rm lon}
= -1.0463\degr, -1.0417\degr, -1.0082\degr$, respectively), the
$R$(3,3)$^l$ absorption occurs near 0\,km\,s$^{-1}$ radial
velocity (Figure~7, top three spectra). As discussed in
Section~4.2.2. for the case of the $R$(1,1)$^l$ line, the radial
velocity at these locations is most simply interpreted as gas
expanding perpendicular to the line of sight.

Toward the star J17444083$-$2926550 ($\gamma$) ($G_{\rm lon} =
-0.5442\degr$) near the Sgr~C complex the $R$(3,3)$^l$ spectrum
(Figure~7, middle left) shows a prominent absorption extending
from $-$170 to $-$110\,km\,s$^{-1}$ centered at
$-$136\,km\,s$^{-1}$. As the depth of this star in the CMZ is
unknown, it is not clear if this rather narrow blue-shifted
absorption range and lack of a trough is due to this star's
shallow depth within the CMZ or if it is more deeply embedded
but its sightline somehow avoids gas at lower blue-shifted
velocities.

For the three stars in the central 30\,pc, GCIRS\,1W ($G_{\rm
  lon} = -0.0553\degr$), NHS\,42 ($G_{\rm lon} = 0.1479\degr$),
and GCS\,3-2 ($G_{\rm lon} = 0.1635\degr$), all deeply embedded
in the CMZ, the $R$(3,3)$^l$ spectra show a broad absorption
feature extending from $\sim -$160\,km\,s$^{-1}$ all the way to
$\sim$0\,km\,s$^{-1}$ (Figure~7, middle center and right, and
bottom left panels). In Section~5 we show that this is best
interpreted as originating in a large volume of warm gas
occupying most of the sightline within the front half of the
CMZ. This absorption feature is also seen in the $R$(1,1)$^l$
line as a ``trough'' superimposed on the sharp absorption
features due to the foreground spiral arms (e.g., Figures~3 and
4; see also Figure~4 of Oka et al. 2005).

The sightline toward J17470898$-$2829561 ($\iota$) ($G_{\rm lon}
= 0.5477\degr$, Figure~7, bottom middle panel), located in the
direction of the Sgr~B complex is as unique in its H$_3^+$
spectra as it is in its CO spectra discussed previously
(Section~4.2.3.). The broad $R$(1,1)$^l$ and $R$(3,3)$^l$
absorption features centered at $-$98\,km\,s$^{-1}$ represent
the only clear diffuse cloud component (as there is no CO
absorption feature at this velocity; see Figure~6) . The deep,
broad, and structured $R$(1,1)$^l$ and $R$(3,3)$^l$ profiles
from $\sim -$70\,km\,s$^{-1}$ to $+$100\,km\,s$^{-1}$ are
produced by H$_3^+$ in a conglomeration of warm dense clouds on
the line of sight. Although this gas is denser than the diffuse
gas seen on all other observed sightline in the CMZ, it is not
very dense since the $R$(2,2)$^l$ absorption (discussed in the
next section) is not as strong as it would be if the (2,2) level
were populated in LTE at the temperature derived from equation
(4). The absorption profiles of H$_3^+$ and CO toward this star
are discussed in some detail by Goto et al. (2011), and are
beyond the scope of this paper.

The star J17482472$-$2824313 ($\lambda-$) ($G_{\rm lon} =
0.7745\degr$ Figure~7 bottom right) is the easternmost star
(116\,pc from Sgr~A$^\ast$) for which both the $R$(1,1)$^l$ and
$R$(3,3)$^l$ line profiles have been observed. In both lines the
only velocity component present is centered near
$-$20\,km\,s$^{-1}$. We interpret this rather low velocity in
the same way as we did for the westernmost stars, $\alpha$,
$\alpha$+, and $\beta$, in Section~4.2.2, as a radial expansion
nearly perpendicular to the line of sight.

%============================================================
% Figure 8
%------------------------------------------------------------
\setcounter{figure}{7}
\begin{figure*}
\includegraphics[width=\textwidth]{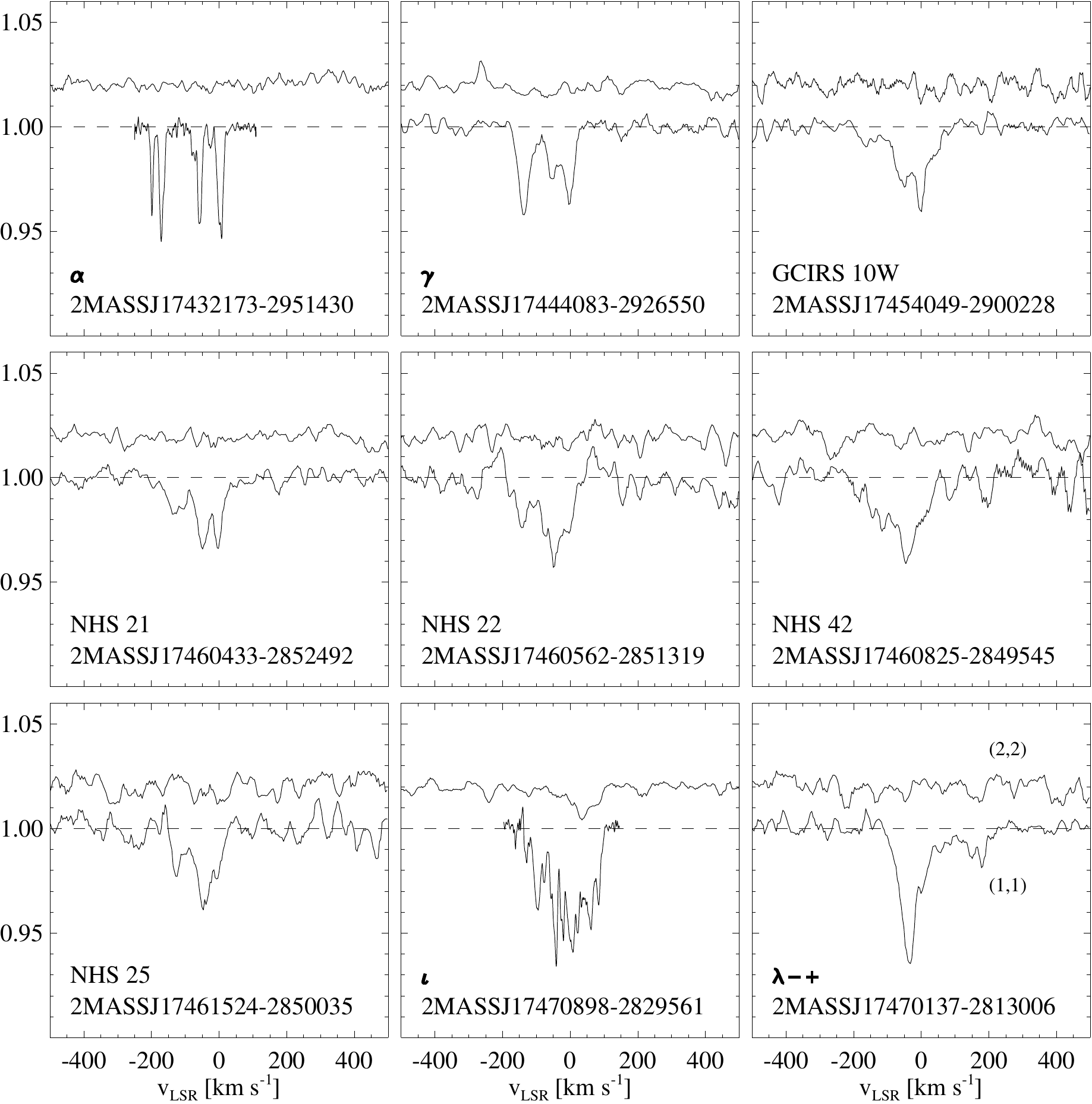}
\caption{Comparison of the $R$(1,1)$^l$ spectra from the (1,1)
  ground rotational level (lower traces) and the $R$(2,2)$^l$
  spectra from the unstable (2,2) level (upper traces) toward
  nine stars observed by the IRCS spectrometer of the Subaru
  Telescope. The $R$(2,2)$^l$ absorption is definitively
  detected only toward J17470898$-$2829561 ($\iota$) indicating
  that sightlines toward other stars cross low density gas
  only.}
\end{figure*}
%----------------------------------------------------------
%------------------------------------------------------------
% Section 4.4.2
%------------------------------------------------------------
\subsubsection{H$_3^+$ in the unstable (2,2) level: evidence of
  low density}

As discussed in Section~1.6, the column density of H$_3^+$ in
the unstable (2,2) level provides a good measure of the gas
density. In Figure~8, $R$(2,2)$^l$ spectra toward nine stars
observed by the IRCS spectrometer on the Subaru Telescope are
compared with the $R$(1,1)$^l$ spectra toward those
stars. Except for the special case of the star
J17470898$-$2829561 ($\iota$) in Sgr~B complex, absorption in
the $R$(2,2)$^l$ line is totally absent, implying that in
general the density of the diffuse gas that we have observed in
the CMZ is considerably less than the critical density,
200\,cm$^{-3}$.

%============================================================
% section 4.4.3 
%------------------------------------------------------------
\subsubsection{Observed H$_3^+$ level column densities} 
In addition to column densities of the (1,1), (3,3), and (2,2)
levels, which are needed to determine temperature and density,
the column density of the (1,0) level is needed to determine the
total H$_3^+$ column density $N$(H$_3^+$)$_{\rm total}$. For
this purpose we observed the $Q$(1,0) transition (Goto et
al. 2002) or $R$(1,0) transition which appears as doublet with
the $R$(1,1)$^u$ transition (Geballe \& Oka 1996). For the
doublet, which is blended in the CMZ due to the wide velocity
profiles of each line, we determine $N$(1,1) + $N$(1,0) and
calculate $N$(1,0) by subtracting $N$(1,1) determined from the
$R$(1,1)$^l$ line. The IRCS spectrometer of the Subaru Telescope
is especially useful for this because of its wide wavelength
coverage. This procedure gives the value of $N$(1,0)/$N$(1,1) on
the order of $\sim$0.5. A detailed thermalization analysis shows
that the excitation temperature $T_{\rm ex}(1,0/1,1)$ defined by
\begin{equation}
  \frac {N(1,0)}{N(1,1)} = 2 \exp \left(-\frac{32.9\,{\rm
      K}}{T_{\rm ex}(1,0/1,1)}\right),
\end{equation}
\noindent
is approximately 30\,K (sometimes called the ``spin
temperature''), and is a slowly varying function of the kinetic
temperature for $T_{\rm k} \sim$200\,K and densities $n \sim
50$\,cm$^{-3}$, as shown in Figure~6 of Oka \& Epp (2004).  We
use $N(1,0) = 0.7N(1,1)$ as estimated values. We believe that
the errors introduced by this in calculating $N$(H$_3^+$)$_{\rm
  diffuse}$ are less than 4\,\%, well within the uncertainties
in the measured column densities.

Finally the level column densities of H$_3^+$ in diffuse gas
$N(J,K)$ are summarized in Table~6 along with the total H$_3^+$
column densities $N({\rm H_3^+})_{\rm diffuse}$. The latter is
calculated from $N({\rm H_3^+})_{\rm diffuse} = 1.1[N(1,1) +
  N(3,3) + N(1,0)]$ where the 10\,\% increase is allowed for
populations of metastable levels higher than (3,3), such as
(4,4), (5,5), (6,6) etc. which individually are not observable
but will amount to 10\,\% of the total as discussed in Figure 5
of Oka and Epp (2004). Such metastable levels are recently
discussed by Smith et al. (2018). Table 6 also contains
population ratios and information on temperature and density. We
here list stars with relatively complete H$_3^+$ measurements
but we have observed many other stars with similar
  $N(J,K)$.

%============================================================
% Table 4.4.3 Observed H3+ level column densities by Subaru
%------------------------------------------------------------
% \setcounter{table}{4} => Table 6
\begin{deluxetable*}{l ccc ccc ccc}
  \tablewidth{0pt}
  \tablecaption{Level column densities and total column
    densities of H$_3^+$ in diffuse clouds and temperature and
    density of diffuse clouds.\label{t6}}
  \tablehead{
%   \multicolumn{3}{c}{$N$ ($J$,$K$) [10$^{15}$\,cm$^{-2}$]} \\
  \colhead{Star} &                                              %1
  \colhead{$N(1,1)$} &                                          %2 
  \colhead{$N(3,3)$} &                                          %3 
  \colhead{$N(1,0)$\tablenotemark{a}} &                         %4
  \colhead{$N(2,2)$\tablenotemark{b}} &                         %5  
  \colhead{$N({\rm H_3^+})_{\rm diffuse}$\tablenotemark{c}} &
  \colhead{$N(3,3)/N(1,1)$} &
  \colhead{$T_{\rm ex}(3,3/1,1)$\tablenotemark{d}} &
  \colhead{$T_{\rm k}$\tablenotemark{e}} &
  \colhead{$n$\tablenotemark{f}}  \\
%   }
  \colhead{} &
  \colhead{[$10^{15}$\,cm$^{-2}$]} & 
  \colhead{[$10^{15}$\,cm$^{-2}$]} & 
  \colhead{[$10^{15}$\,cm$^{-2}$]} & 
  \colhead{[$10^{15}$\,cm$^{-2}$]} & 
  \colhead{[$10^{15}$\,cm$^{-2}$]} & 
  \colhead{} &
  \colhead{[K]} &
  \colhead{[K]} &
  \colhead{[cm$^{-3}$]} 
  }

  \startdata
%      1       2                       3                        4                       5          6      7         8      9       10          
$\alpha$     & 0.73                  & 0.19\tablenotemark{h}  & 0.51                  & $<$0.03  & 1.57 & 0.26 & 125  &  154 &  $<$17  \\
$\alpha+$    & 2.10                  & 0.86\tablenotemark{h}  & 1.47                  & $<$0.29  & 4.84 & 0.41 & 148  &  182 &  $<$58  \\
$\beta$      & 0.58                  & 0.40\tablenotemark{h}  & 0.41                  & ---      & 1.53 & 0.69 & 189  &  --- &  ---    \\
$\gamma$     & 1.16                  & 0.64                   & 0.81                  & $<$0.14  & 2.87 & 0.55 & 169  &  227 &  $<$45  \\
$\delta$     & 0.85                  & ---                    & 0.60                  & ---      & ---  & ---  & ---  & ---  &  ---    \\
NHS\,21      & 0.45                  & 0.17                   & 0.32                  & $<$0.21  & 1.03 & 0.38 & 172  &  141 &  $<$651 \\
NHS\,22      & 0.55                  & ---                    & 0.39                  & $<$0.29  & ---  & ---  & ---  & ---  &  ---    \\
NHS\,42      & 1.24                  & 0.64                   & 0.87                  & $<$0.41  & 3.03 & 0.52 & 164  &  181 &  $<$178 \\
NHS\,25      & 1.43                  & ---                    & 1.00                  & $<$0.33  & ---  & ---  & ---  &  --- &  ---    \\
GCS\,3-2     & 1.45                  & 0.70                   & 1.01                  & $<$0.14  & 3.48 & 0.48 & 159  &  211 &  $<$36  \\
FMM\,362     & 1.13                  & 0.47                   & 0.79                  & ---      & 2.63 & 0.35 & 138  &  --- &  ---    \\
$\theta$     & 1.24                  & 0.95\tablenotemark{g}  & 0.87                  & ---      & 3.37 & 0.77 & 200  &  --- &  ---    \\
$\lambda-$   & 1.27\tablenotemark{g} & 0.43                   & 0.89\tablenotemark{g} & ---      & 2.85 & 0.34 & 138  & ---  & ---  \\
$\lambda-+$  & 2.94\tablenotemark{g} & ---                    & 2.16\tablenotemark{g} & $<$0.30  & ---  & ---  & ---  & ---  & ---  \\
\enddata
\tablenotetext{a}{Calculated from $N(1,0) = 0.7 N(1,1)$, see text.} 
\tablenotetext{b}{Upper limits are 1\,$\sigma$.}
\tablenotetext{c}{Sum of $N(1,1)$, $N(3,3)$, $N(1,0)$ and multiplied by 1.1, see text.}
\tablenotetext{d}{Excited temperature calculated from equation (4).}
\tablenotetext{e}{Kinetic temperature $T_{\rm k}$ of diffuse clouds calculated in Section 5.1.1.}
\tablenotetext{f}{Upper limit of densities $n = n({\rm H}) + n({\rm H_2})$ of diffuse clouds calculated in Section 5.1.1.} 
\tablenotetext{g}{$N(1,1)$ values include contributions from those in spiral arms.}
\tablenotetext{h}{$N(3,3)$ values may be compromised by atmospheric water absorption.}

% \tableline
%\enddata
\end{deluxetable*}

%============================================================

%============================================================
\section{Analysis}
%------------------------------------------------------------
\subsection{Temperature and density of the diffuse gas in the CMZ}

As shown in Table~6 the observed $N(3,3)/N(1,1)$ ratios give
excitation temperatures which average to $160\pm24$\,K. When
cooling due to symmetry-breaking spontaneous emission, discussed
in Section~1.6, is taken into account in the rotational
thermalization of H$_3^+$ (Oka \& Epp 2004), the actual kinetic
temperature $T_{\rm k}$ of the diffuse gas is significantly
higher than $T_{\rm ex}(3,3/1,1)$. On the other hand, a low
number density of the diffuse gas $n$ is suggested by the
non-detection of absorption in the $R$(2,2)$^l$ line, which sets
an upper limit on $N$(2,2).

%============================================================
% Figure 9 NT diagram 5.1.1 Determining Tk and setting upper limit
%------------------------------------------------------------
\setcounter{figure}{8}
\begin{figure}
\includegraphics[width=0.5\textwidth]{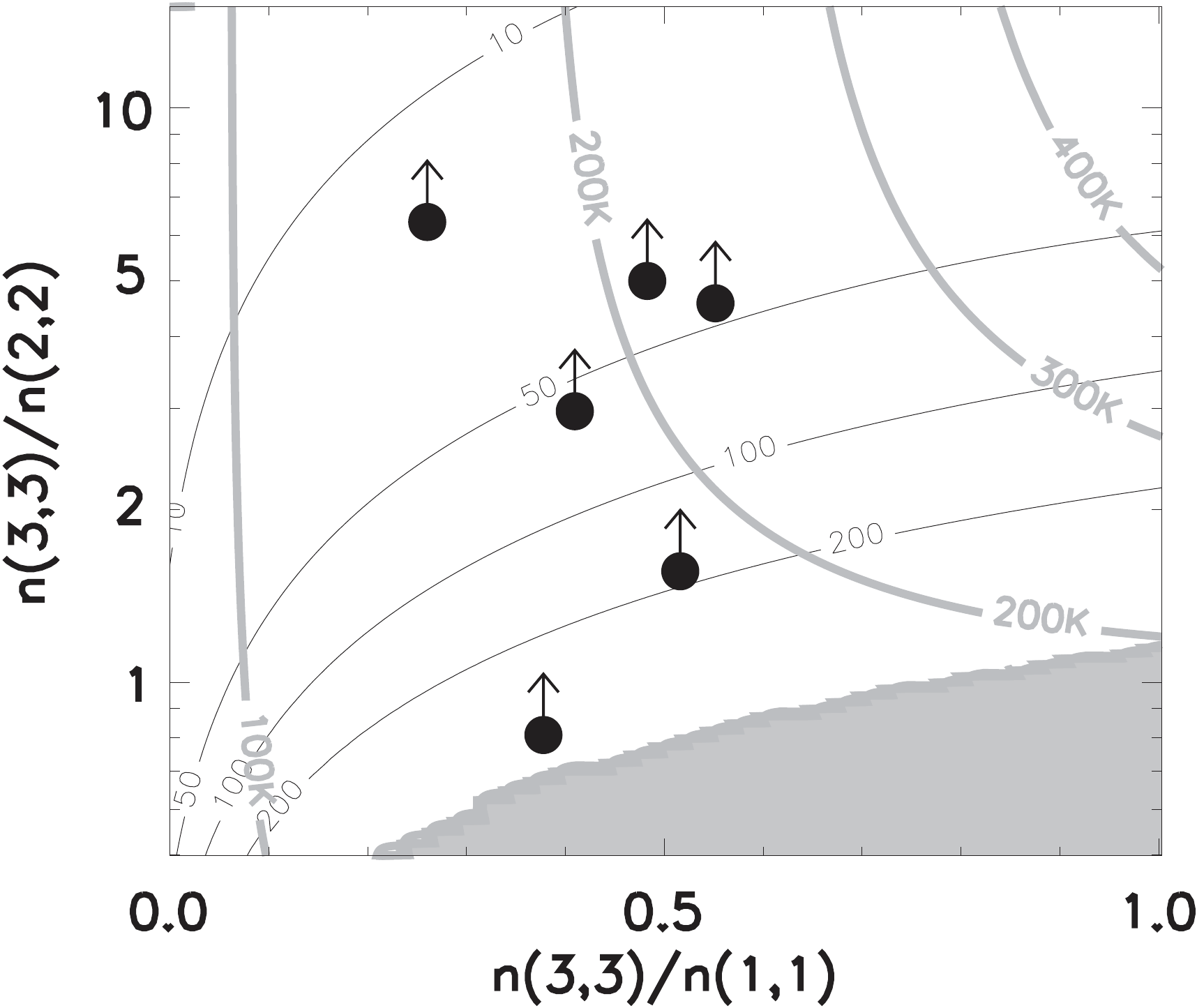}
\caption{Temperature and density plotted in filled circles as
  functions of population ratios $n$(3,3)/$n$(1,1) and
  $n$(3,3)/$n$(2,2) by inverting the diagram of Oka and Epp
  (2004). The $n$(3,3)/$n$(2,2) are all lower limits.}  %
\end{figure}
%------------------------------------------------------------
%------------------------------------------------------------
% 5.1.1
%------------------------------------------------------------
\subsubsection{ Determining $T_{\rm k}$ and setting upper limits on the density}

Figure~4 of Oka \& Epp (2004) and Figure~6 of Oka et al. (2005)
plot the ratios $N(3,3)/N(1,1)$ and $N(3,3)/N(2,2)$ as functions
of kinetic temperature $T_{\rm k}$ and number density $n$. Here
we inverted the plots to generate ones in which $T_{\rm k}$ and
$n$ are shown as functions of $N(3,3)/N(1,1)$ and
$N(3,3)/N(2,2)$ (as in Figure~4 of Goto et al. 2008). Using the
observed values of $N(3,3)/N(1,1)$ and the observed lower limits
to $N(3,3)/N(2,2)$, $T_{\rm k}$ and $n$ are determined as shown
in Figure~9.  This gives $T_{\rm k} = 140-230$\,K and $n <
20-650$\,cm$^{-3}$, with higher temperatures generally
associated with lower upper limits on density. The values of
$T_{\rm k}$ and upper limits to $n$ are listed in Table~6 for
sightlines toward 6 stars from $\sim$140\,pc west to
$\sim$120\,pc east of Sgr~A$^\ast$. Readers are referred to
Fig.~4 of Goto et al. (2008) where measurements had higher
uncertainties but give similar density and temperature.

%------------------------------------------------------------
% 5.1.2
\subsubsection{Comparison with the Meudon analysis}

While the treatment of thermalization by Oka \& Epp (2004)
(hereafter called the Chicago model) uses the Einstein
coefficients calculated by Neale et al. (1996), which are
accurate perhaps to within 5\,\%, it uses roughly estimated
collision rate constants based only on the Boltzmannian
principle of detailed balancing and their symmetric division
(their equations 1--3). This treatment also ignores the nuclear
spin selection rules in the H$_3^+$ - H$_2$ collisions (Quack
1977; Oka 2004). A more detailed state-to-state calculation of
collisional rate constants including the spin selection rules
has since been given by Park \& Light (2007a; 2007b) using a
statistical method. A more recent paper by G\'omez-Carrasco et
al. (2012) which is based on the statistical treatment of Park
\& Light but uses the full potential energy surface of the
H$_5^+$ complex (Aguado et al. 2010) and the quasi-classical
trajectory method is used in the Meudon model analysis of Le
Petit et al. (2016).

A comparison between the Chicago model and the Meudon model
calculations of the two-dimensional $N(3,3)/N(1,1)$ and
$N(3,3)/N(2,2)$ plots in terms of $n$ and $T_{\rm k}$ is given
in the top and the bottom panels, respectively, of Figure~10 of
Le Petit et al. (2016). Their study indicates that the crudely
estimated collision rates of the Chicago model provide the
essential features of the Meudon model in the region of interest
of $n$ and $T_{\rm k}$, and that the difference between the two
approaches is due mainly to a scaling factor for the number
density $n$; the two diagrams agree approximately if the scale
of the number density $n$ in the Chicago models is multiplied by
a factor of $\sim 8$. This suggests that the rate constant for
the H$_3^+$ (2,2) $\rightarrow$ (1,1) collision-induced
transition used in the Chicago model is 8 times higher than in
the Meudon model. Since the total collision rate constant is
normalized approximately to the Langevin rate constant of $2
\times 10^{-9}$\,cm$^{-3}$ in Equation (3) of Oka \& Epp (2004),
in Section 3F of Park \& Light (2007a), and in Fig.~7 of
G\'omez-Carrasco et al. (2012), the difference must be due to
the quantum number dependence of the rate constant, which is not
considered in the Chicago model. Whether a quantum number
dependence as significant as that shown in Fig.~7 of
G\'omez-Carrasco et al. exists for chemical collisions remains
to be seen.

In spite of this difference, Le Petit et al. (2016) obtain a
roughly similar, although somewhat higher kinetic temperature
range, $T_{\rm k} = 212-505$\,K and a similar upper density
limit, $n \leq 100$\,cm$^{-3}$, as the Chicago model.

%------------------------------------------------------------
\subsection{Cosmic ray H$_2$ ionization rate $\zeta$ and dimension of diffuse clouds}

In this section, we estimate the cosmic ray ionization rate in
the CMZ, using $T_{\rm k} = 200$\,K and $n \leq 100$\,cm$^{-3}$
as representative values for the diffuse interstellar medium
there. We start from the original Chicago model, in which the
quantum number dependence of collision rate and the nuclear spin
selection rules are neglected; otherwise the calculation becomes
extremely complicated. For example, the quantum number
dependences of rate constants themselves depend on
temperature. Also, in order to use nuclear spin selection rules,
it is necessary to include an additional parameter, the H$_2$
spin temperature.
% Although less detailed, our approach will be useful for
% approximate calculations, for example in the cooling of
% primordial gas (Glover \& Savin 2009) or plasmas in planetary
% ionospheres (Miller et al. 20 13).

%------------------------------------------------------------
% 5.2.1
\subsubsection{Reconsidering the relation between $\zeta L$ and $N$(H$_3^+$)}

In deriving the extremely simple equation (3) relating the
observed total H$_3^+$ column density $N$(H$_3^+$) and the
cosmic ray H$_2$ ionization rate $\zeta$, Oka et al. (2005)
neglected two effects: (1) the charge exchange reaction,

\begin{equation}
{\rm H_2^+ + H \rightarrow H_2 + H^+},
\end{equation}
\noindent
which is exothermic by 1.831\,eV and reduces the H$_3^+$
production rate if the fraction of molecular hydrogen $f({\rm
  H_2})$ is significantly lower than unity; and (2) the
production of free electrons by cosmic-ray ionization of H$_2$
and H, which increases the H$_3^+$ destruction rate if $\zeta$
is higher than $\sim 10^{-15}$\,s$^{-1}$. (Previously, only
electrons produced by vacuum ultraviolet first photoionization
of C atoms had been included.)  Both of these effects reduce
$N({\rm H_3^+})$ from values calculated from equation (3), so
that a higher value of $\zeta$ is required to obtain the same
value of $N({\rm H_3^+})$.

Values of $\zeta \sim (2-7) \times 10^{-15}$\,s$^{-1}$ obtained
for the diffuse gas in the CMZ by Oka et al. (2005) and a
slightly lower value estimated by Goto et al. (2008), using
equation (3), are more than 100 times higher than the early
canonical values of $\sim 10^{-17}$\,s$^{-1}$ (Spitzer \&
Tomasko 1968; Webber 1998: see Dalgarno 2006 for a review). We
did not consider $\zeta$ values  higher orders of magnitude
  than 10$^{-15}$\,s$^{-1}$ in those papers.   In
  the meantime McCall et al. (2003) obtained a value of $\zeta$
on the order of $10^{-16}$\,s$^{-1}$ in the diffuse cloud toward
$\zeta$~Persei, which led subsequently to the universally high
$\zeta$ values on the order of $10^{-16}$\,s$^{-1}$ in diffuse
clouds in the Galactic disk (Indriolo et al. 2007; Indriolo \&
McCall 2012).

Initially the high values of $\zeta$ in the Galactic center were
viewed with skepticism (e.g. Amo-Baladr\,on et
al. 2011). However, soon after they were first reported, even
higher values of $2 \times 10^{-14}$ and $5 \times
10^{-13}$\,s$^{-1}$ were reported toward Sgr~B1/B2 and Sgr~C,
respectively from multi-wavelength observations and analysis of
$^{13}$CO radio emission, X-rays, and $\gamma$-rays (Yusef-Zadeh
et al. 2007). Yusef-Zadeh et al. (2013) further reported $\zeta
= (1-10) \times 10^{-15}$\,s$^{-1}$ for a wider region of the
central 300$\times$120\,pc. Also, Indriolo et al. (2015)
reported even higher values of $\zeta$, approaching $4 \times
10^{-14}$\,s$^{-1}$, for some velocity components of four far
infrared sources in the CMZ [M-0.13-0.08, M-0.02-0.07,
  Sgr~B2(M), and Sgr~B2(N)] using the extensive Herschel
observations of OH$^+$, H$_2$O$^+$, and H$_3$O$^+$. Finally, Le
Petit et al. (2016) reported values of $(1-11) \times
10^{-14}$\,s$^{-1}$ from an analysis of the H$_3^+$ column
densities reported by Oka et al. (2005) and Goto et al. (2008;
2011).

All of these results imply that to neglect the production of
free electrons by ionization of H and H$_2$, assumed in deriving
equation (3), is not justified. Here we develop a formalism in
which the effects (1) and (2) mentioned at the start of this
subsection, are included. We use large-scale chemistry taking
into account only hydrogenic species and electrons. The
chemistry of helium, which may introduce uncertainties on the
order of 10\,\%, is neglected. We believe our treatment is
meaningful, even after the highly-detailed analysis of Le Petit
et al. (2016) using the Meudon code, in order to have an overall
view, in particular because the Meudon results underestimate the
observed H$_3^+$ column density of $3 \times 10^{15}$\,cm$^{-2}$
in obtaining their reported value of $\zeta = (1-11) \times
10^{-14}$\,s$^{-1}$ (see their Figure~1) for $n <
10^2$\,cm$^{-3}$, which is a reasonable choice of density as
discussed above.

%------------------------------------------------------------
\subsubsection{Reduction of H$_3^+$production rate due to charge exchange reaction}

If $f({\rm H_2})$ is significantly less than 1, the charge
exchange reaction between H$_2^+$ and H, Equation (6), competes
with the H$_3^+$ production reaction of equation (1) and reduces
the production rate of H$_3^+$ by
\begin{equation}
  \Phi\left( f({\rm H_2}), \frac{k_6}{k_1} \right) = \left[ 1 +
    \frac{2k_6}{k_1} \left( \frac{1}{f({\rm H_2})} -1 \right)
    \right]^{-1}
\end{equation}  
\noindent
(Indriolo \& McCall 2012; equation (6) of Oka 2013), where $k_1$
and $k_6$ are the rate constants of equation (1) and (6),
respectively.  The reduction of the production rate due to
dissociative recombination of H$_2^+$ is negligible due to the
small number density of electrons. While $k_1$ in equation (7)
is a well-established Langevin rate constant $k_1 = (2.08 \pm
0.21) \times 10^{-9}$\,cm$^3$\,s$^{-1}$ (Anicich \& Huntress
1986), the value of $k_6$ is not well known. Based on the
arguments given in Appendix A4, we assume $k_6 = k_1/2$, partly
because it gives a simple expression
\begin{equation}
\Phi = f({\rm H_2})
\end{equation} % equation (8)
\noindent
Compared with the previous H$_3^+$ production rate $\zeta f({\rm
  H_2})$ used by Oka et al. (2005), in which $f({\rm H_2})$ was
assumed to be 1, the revised H$_3^+$ production rate $\zeta
[f({\rm H_2})]^2$ is approximately 3 times lower for the value
$f({\rm H_2}) = 0.6$ reported by Le Petit et al. (2016) (see
Appendix A3).

%------------------------------------------------------------
\subsubsection{Increase of H$_3^+$destruction rate due to ionization of
H$_2$ and H}

Considering only pure hydrogenic species and electrons, the
production rate of electrons from H$_2$ and H due to cosmic ray
ionization is
\begin{equation}
  \left( \frac{dn_{\rm e}}{dt} \right)_{\rm prod} = \zeta n({\rm
    H_2}) + \zeta_{\rm H} n({\rm H}) \sim \frac{\zeta n_{\rm
      H}}{2} % equation (9)
\end{equation}
\noindent
where $\zeta_{\rm H} \sim \zeta/2$ (Glassgold \& Langer 1974;
Rudd et al. 1985; Dalgarno 2006) is the ionization rate of H and
$n_{\rm H} = 2n({\rm H_2}) + n({\rm H})$ is the total number of
H. The destruction rate of free electrons is
\begin{equation}
  \left( \frac{dn_{\rm e}}{dt} \right)_{\rm dest} = [k_{\rm e}
    n({\rm H_3^+}) + k_{\rm r} n({\rm H^+})]n_{\rm e}^\ast,
\end{equation} % equation (10)
\noindent where $n_{\rm e}^\ast$ is the number density of
electrons produced from cosmic ray ionization of H$_2$ and H,
and $k_{\rm r}$ is the radiative recombination rate constant for
H$^+$. The destruction of electrons due to dissociative
recombination of H$_2^+$ is negligible because, unlike H$_3^+$
and H$^+$, H$_2^+$ is rapidly destroyed both by H$_2$ and H and
thus has an extremely small number density. While the production
rates of H$_3^+$ and H$^+$ are comparable, their destruction
rates are widely different because the dissociative
recombination rate constant, $k_{\rm e} \sim
10^{-7}$\,cm$^3$\,s$^{-1}$ (Appendix A1), is more than 4 orders
of magnitude higher than the radiative recombination rate
constant $k_{\rm r} \sim 5 \times 10^{-12}$\,cm$^3$\,s$^{-1}$
(Appendix A5). This means $n({\rm H^+})$ is more than 4 orders
of magnitude higher than $n({\rm H_3^+})$. Thus, from the
condition of neutrality $n({\rm H^+}) + n({\rm H_3^+}) = n_{\rm
  e}^ \ast$, $n({\rm H^+}) = n_{\rm e}^\ast$ to a good
approximation. We therefore obtain the steady state equation for
electrons, $\zeta n_{\rm H}/2 = [k_{\rm e} n({\rm H_3^+}) +
  k_{\rm r}n_{\rm e}^\ast ] n_{\rm e}^\ast$, which gives,
\begin{equation} % equation (11) 
 n_{\rm e}^\ast = \frac{k_{\rm e}}{2k_{\rm r}} n({\rm H_3^+})
 \left[ \sqrt{1 + \frac{2 \zeta k_{\rm r} n_{\rm H}}
     {\left[k_{\rm e} n({\rm H_3^+}) \right]^2}} -1 \right]
\end{equation} % equation (11) 
\noindent
The total electron number density is given by
\begin{equation} % equation (12) 
  n_{\rm e} = \left( \frac{n_{\rm C}}{n_{\rm H}} \right)_{SV} R
  n_{\rm H} + n_{\rm e}^\ast
\end{equation} % equation (12) 
\noindent
where the first and second term represent electrons from the
photoionization of carbon atoms and the cosmic ray ionization of
H$_2$ and H, respectively. Using the carbon to hydrogen ratio
after depletion of $(n_{\rm C}/n_{\rm H})_{\rm SV} = 1.6 \times
10^{-4}$ (Sofia et al. 2004) and the increase of metallicity
from the solar vicinity to the GC of $R >3$ (Appendix A2), we
find the electron number density from photoionization of carbon
to be higher than $5 \times 10^{-4} n_{\rm H}$.

For a low value of $\zeta$, $n_{\rm e}^\ast$ is negligible as
given in our earlier analysis (Oka et al. 2005). Equation (11)
indicates that as $\zeta$ increases, $n_{\rm e}^\ast$ increases
approximately as $\zeta n_{\rm H}/2k_{\rm e} n({\rm
  H_3^+})$. Numerical estimates indicate that the fraction of
$n_{\rm e}^\ast$ arising from ionization becomes significant for
$\zeta = 10^{-15}$\,s$^{-1}$, in approximate agreement with the
result in Figure~2 of Le Petit et al. (2016). The above
simplified treatment, in which the number density from carbon
atoms is set as a constant and their participation neglected in
the analysis, introduces some error in the region where $n_{\rm
  e}^\ast$ is comparable to $(n_{\rm C}/n_{\rm H})_{\rm SV} R
n_{\rm H}$, but Equation (8) and equations thereafter are
accurate in regions with $\zeta \gg 10^{-15}$\,s$^{-1}$ where
$n_{\rm e}^\ast \gg (n_{\rm C}/n_{\rm H})_{\rm SV} R n_{\rm H}$,
which applies to the CMZ, as shown below.

%------------------------------------------------------------
% 5.2.4
\subsubsection{Self-consistent solution for $\zeta$}

With the reduced H$_3^+$ production rate discussed in
Section~5.2.2. and the increased electron number density
discussed in Section~5.2.3, steady state H$_3^+$ chemical
equilibrium is described by

\begin{equation} % equation (13)
  \zeta n_{\rm H}\left[f({\rm H_2})\right]^2 = k_{\rm e} \left [
    (n_{\rm C}/n_{\rm H})_{SV} R n_{\rm H} + n_{\rm e}^\ast
    \right] n({\rm H_3^+}).
\end{equation}  % equation (13)
\noindent
Since $n_{\rm e}$ is a function of $n({\rm H_3^+})$ [Equation
  (8)], which in turn is a complicated function of $\zeta$, it
is a hopeless task to solve this equation directly. Instead, we
make use of the observed total H$_3^+$ column density $N({\rm
  H_3^+})$ and calculate $n({\rm H_3^+}) = N({\rm H_3^+})/L$ for
assumed column lengths $L$. Equations (8) and (9) then have only
two unknowns, $\zeta$ and $n_{\rm H}$. For each value of $n_{\rm
  H}$ we obtain $\zeta$ by simply solving a quadratic equation
as shown in Appendix B.

We use $N({\rm H_3^+}) = 3 \times 10^{15}$\,cm$^{-2}$ which is
typical of observed H$_3^+$ column densities toward deeply
embedded stars that are located near the center of the CMZ such
as GCS\,3-2, GCIRS\,3, GCIRS\,1W, etc. Plots of $\zeta$ as a
function of $n_{\rm H}$ for assumed column lengths $L$ are given
in Figure~10.

%============================================================
% Figure 10 % 5.2.4
%------------------------------------------------------------
\setcounter{figure}{9}
\begin{figure}
\includegraphics[width=0.5\textwidth]{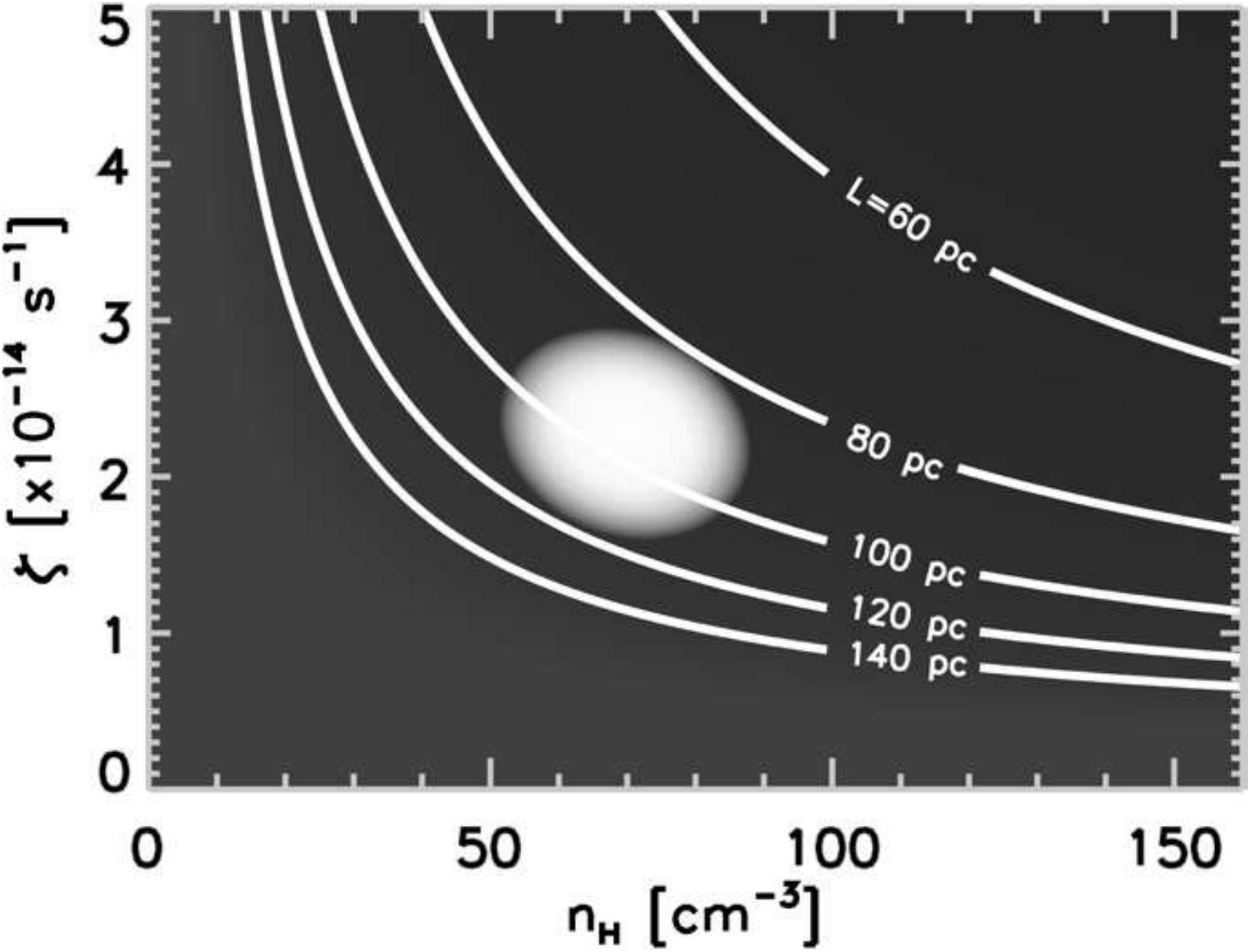}
% \hspace{4mm}
\caption{Cosmic ray ionization rate $\zeta$ versus hydrogen atom
  number density $n_{\rm H}$ for different values of diffuse
  cloud pathlength, $L$. This is a plot of Equation (B5) in
  which $x \equiv n_{\rm H}$ and $H \equiv N({\rm H_3^+})/L$,
  where $N({\rm H_3^+})$, the total column density of H$_3^+$,
  is assumed to be $3 \times 10^{15}$\,cm$^{-2}$.  The white
  ellipse represents the most plausible values of $\zeta$,
  $n_{\rm H}$, and $L$. \label{f10}}
\end{figure}
%------------------------------------------------------------
% 5.2.5
\subsubsection{Estimating $\zeta$ and $L$}

Figure~10 indicates that the observed H$_3^+$ column densities
can be obtained by a wide range of $\zeta = (1-5) \times
10^{-14}$ \,s$^{-1}$, $n_{\rm H} = (20-140)$\,cm$^{-3}$, and $L
= (60-140)$\,pc. This situation is parallel to Oka et al. (2005)
in which $\zeta L$ was obtained from observed $N({\rm H_3^+})$,
but $\zeta$ and $L$ could not be uniquely separated. In this
section, we attempt to separate them using various
considerations to arrive at the most likely approximate mean
values of $\zeta$, $n_{\rm H}$, and $L$ in the front half of the
CMZ.

First, the analysis of observed data in Section~4.4. gave $n <
100$\,cm$^{-3}$. Since $n = n({\rm H}) + n({\rm H_2})$ and
$n_{\rm H} = n({\rm H}) + 2n({\rm H_2})$ are related as $n = [1
  - f({\rm H_2})/2] n_{\rm H}$, the constraint on $n$ sets the
upper limit of $n_{\rm H}$ in Figure~10 to be $n_{\rm H} <
140$\,cm$^{-3}$ for $f({\rm H_2}) = 0.6$ (Appendix A3). Although
observational data from which to determine the value of $n$ is
lacking, $n$ is unlikely to be much lower than the above upper
limit in view of the high column density of H$_3^+$.  We adopt
$n_{\rm H}\sim(60-70)$~cm$^{-3}$, and $L\sim(60-140)$~pc with
$n({\rm H_3^+})/n_{\rm H}\sim10^{-7}$ as the most plausible
values.

As for the most likely value of the cloud length $L$, we regard
the large velocity spreads of the H$_3^+$ absorption troughs of
many of the spectra of centrally located stars (those shown in
Figures~3 and 4) as indicating long absorption path lengths
covering most of the distance from the center of the CMZ to its
edge. As the sightline moves away from the center the velocity
dispersion decreases with relatively sharp absorptions found
toward stars located near the eastern and western edges. This
behavior is consistent with the warm and diffuse gas occupying a
large portion of the CMZ and therefore we assume $L \sim
100$\,pc corresponding to a volume filling factor $f \sim 2/3$
for it.

From these considerations we propose the following parameters to
be the most likely average values of the warm and diffuse
gaseous environment: $\zeta \sim 2 \times 10^{-14}$\,s$^{-1}$,
$L \sim 100$\,pc, $f = 2/3$, $n_{\rm H} \sim 70$\,cm$^{-3}$, $n
\sim 50$\,cm$^{-3}$, $n({\rm H_2}) \sim 20$\,cm$^{-3}$, $n({\rm
  H}) \sim 30$\,cm$^{-3}$, $n({\rm H_3^+}) \sim 1 \times
10^{-5}$\,cm$^{-3}$, $n({\rm H^+}) \sim n_{\rm e}^\ast
\sim 0.30$\,cm$^{-3}$, $n_{\rm e} \sim 0.33$\,cm$^{-3}$.
These values of $\zeta$, $L$, and $n_{\rm H}$ are shown in
Figure~10 with approximate error limits. Assuming the warm
diffuse gas is in the shape of a disk with of radius 150\,pc and
scale height 30\,pc, its total mass is estimated to be $\sim 6
\times 10^6$\,$M_\odot$. A summary is given in Table~7.

%============================================================
% Table 5 % 5.2.5 Estimating zeta and L
%------------------------------------------------------------
%\setcounter{table}{4}
% \setcounter{table}{4}
\begin{deluxetable}{l | l l l }
\tablecaption{Most likely parameters for the diffuse molecular cloud in the CMZ\label{t7}}
\tablehead{\colhead{Physical Parameter} &  \colhead{Symbol} &  \colhead{Value} &  \colhead{Unit}
}
\startdata
% \tableline \tableline
% \multicolumn{1}{c}{parameter}&\multicolumn{1}{c}{symbols}&\multicolumn{1}{c}{ }&\multicolumn{1}{c}{units}\\
% density of electron from ionization of hydrogen   & $\underline{n(\rm{e})}$  & 0.30        & [cm$^{-3}$]\\
% \tableline
% \tableline
comic ray ionization rate      & $\zeta$         & $2 \times 10^{-14}$  & [s$^{-1}$] \\
pathlength                     & $L$             & 100                  & [pc]       \\ 
volume filling factor          & $f$             & $\frac{2}{3}$        &            \\ 
density of hydrogen nuclei     & $n_{\rm H}$     & 70                   & [cm$^{-3}$]\\
gas density                    & $n$             & 50                   & [cm$^{-3}$]\\
density of molecular hydrogen  & $n(\rm{H_2})$   & 20                   & [cm$^{-3}$]\\
density of atomic hydrogen     & $n(\rm{H})$     & 30                   & [cm$^{-3}$]\\
density of ${\rm H_3^+}$       & $n(\rm{H_3^+})$ & $1 \times 10^{-5}$   & [cm$^{-3}$]\\
density of ionized hydrogen    & $n(\rm{H^+})$   & 0.30                 & [cm$^{-3}$]\\
total density of electron      & $n(\rm{e})$     & 0.33                 & [cm$^{-3}$]\\
mass of diffuse molecular cloud&                 & $6 \times 10^6$      & [$M_\odot$]\\ 
% \tableline
\enddata
\end{deluxetable}

%============================================================
%------------------------------------------------------------
\subsubsection{Comparison with the Meudon analysis}

The Meudon analysis of Le Petit et al. (2016) uses a highly
sophisticated code taking into account 165 species and 2850
chemical reactions. In contrast, here we consider only hydrogen
and electrons. Le Petit et al. also consider the effects of the
radiation scaling factor ($G_0$), radiative transfer, grains,
PAHs etc. which are all ignored in our treatment. We regard
these differences of the treatments as minor in view of the
dominant abundance of hydrogen and the smallness of the latter
effects reported by Le Petit et al.

There, however, are two essential differences between the two
methods. First, the Meudon analysis is for ``small'' diffuse
clouds with maximum visual extinction $A_V{\rm max} = 1$\,mag
with stars located outside of the cloud, while our calculation
is for a much larger diffuse cloud with $A_V \sim 10-30$\,mag
and with the stars inside. Le Petit et al. thus explain our
measured H$_3^+$ column densities as due to a pile-up of
absorptions from many small diffuse clouds, while we regard them
as due to a single or a few large diffuse clouds covering large
areas of the CMZ.  Perhaps this is the reason why the
pathlengths $L$ shown in their Figures~13 and 14 are so much
smaller than those shown in Figure~10 of this paper.

Second, the Meudon code calculates the H$_3^+$ abundance ``{\it
  ab initio}'' from assumed chemical abundances and
environmental parameters. Thus, none of the H$_3^+$ column
densities shown in Figure~1 of Le Petit et al. matches the
observed $N({\rm H_3^+})$ of $3 \times 10^{15}$\,cm$^{-2}$
unless $n_{\rm H}$ is considerably larger than 100\,cm$^{-3}$,
contradictory to their limit of $n_{\rm H} \leq
100$\,cm$^{-3}$. On the contrary, all points in Figure~10 of
this paper give the observed $N({\rm H_3^+})$.

%============================================================
\section{Discussion}
%------------------------------------------------------------
\subsection{Predominance of the warm and diffuse gas in the CMZ}

The analysis given above strongly suggests that the central
300\,pc of the GC is dominated by warm ($T \sim 200$\,K) and
diffuse ($n < 100$\,cm$^{-3}$) gas. This gas is of a different
category than the diffuse component associated with dense gas,
with $n\sim 10^{2.5}$\,cm$^{-3}$ and with lower temperature
reported by Oka et al. (1998a) from observations of the $J = 2
\rightarrow 1$ CO emission and by Dahmen et al. (1998) from $J =
1 \rightarrow 0$ C$^{18}$O emission. The new category of gas
reported here, with lower density and higher temperature, is not
probed efficiently by molecular rotational radio emission since,
even for CO with its abnormally small permanent dipole moment of
$\mu = 0.1098$\,D (Muenter 1975), the critical density for the
lowest $J = 1 \rightarrow 0$ emission is much higher, on the
order of 700\,cm$^{-3}$. Nevertheless, there have been several
previous observations of the diffuse gas, albeit not as clearly
stated as in this paper.

%------------------------------------------------------------
% Section 6.1.1 
\subsubsection{Previous observations of diffuse ($n < 100$\,cm$^{-3}$) gas in the CMZ}

The highly blue-shifted broad OH radio absorptions at 1.6\,GHz
toward Sgr~A, observed in early days of radio astronomy at
velocities near $-$120\,km\,s$^{-1}$ by Goldstein et al. (1964)
and more accurately with higher spatial resolution at
$-$135\,km\,s$^{-1}$ by Bolton et al. (1964b), originate in the
diffuse gas discussed in this paper. Robinson \& McGee (1970)
observed the 1667\,MHz absorption continuously over a wide range
of Galactic longitudes and noted that the blue-shift decreases
with longitude. Their measurements along with the morphological
studies by McGee (1970) led to the claimed existence by Kaifu et
al. (1972) of the ``Expanding Molecular Ring'' (EMR) with an
expansion velocity of 130\,km\,s$^{-1}$.  The variation of
blue-shift with the Galactic longitude in Figure~2 of Robinson
\& McGee is similar to our observations of H$_3^+$ shown in
Figure~7.

Likewise, the 4830\,MHz absorption line of H$_2$CO, whose
distribution is ``very similar'' to that of OH as observed (with
higher angular resolution) by Scoville et al. (1972), must also
be in the diffuse gas, although the blue-shifted absorption is
not as clearly noted in their Figure~2 as in Figure~2 of
Robinson \& McGee (1970). Nevertheless, Scoville (1972) also
interpreted their observations as due to the ``Expanding
Molecular Ring'', with an expansion velocity of 145\,km\,s$^{-1}$
similar to that of Kaifu et al. (1972). The existence of H$_2$CO
in diffuse clouds is well documented by Liszt \& Lucas (1995)
and Liszt et al. (2006).

While the expansion velocity of the rings, 130\,km\,s$^{-1}$ and
145\,km\,s$^{-1}$ observed in OH and H$_2$CO, respectively, are
close to the maximum radial velocity, $\sim$140\,km\,s$^{-1}$,
of the diffuse clouds observed in H$_3^+$, as shown in Figure~4,
the velocity dispersion is much higher for the H$_3^+$
lines. Therefore, the morphology of the gas observed in OH and
H$_2$CO appears more like a ring. This suggests that H$_3^+$ is
pervasive in the diffuse gas while OH and H$_2$CO are more
localized at higher radius, closer to the front of the expanding
diffuse gas, and perhaps have a lower temperature (Goto et
al. 2011). More detailed discussions on the morphology of this
gas will be given in Part II of this series of papers.

Radio emission can also probe low-density gas if the critical
density of the observed transition is low. The 3335\,MHz CH
emission line, with a critical density on the order of
3\,cm$^{-3}$, observed at high velocity dispersion by Magnani et
al. (2006), must originate in the diffuse gas discussed in this
paper. We note, though, that line emission has disadvantages
compared with absorption in that (1) it does not
straightforwardly discriminate between expansion and
contraction, and (2) determinations of column densities are
complicated since collisions need to be taken into
account. However, we expect that the general observability of CH
makes its emission a powerful future probe for studying the
diffuse gas in the CMZ.

In a recent noteworthy paper Corby et al. (2018) observed low
frequency (1--50\,GHz) rotational {\it absorption} lines of 10
molecules, OH, c-C$_3$H$_2$, H$_2$CO, SiO, CS, CCS, HCS$^+$,
H$_2$CS, $l$-C$_3$H and $l$-C$_3$H$^+$ in the sightline toward
Sgr~B2(N) and interpreted all of them to be in diffuse or
translucent clouds. While not all of them are likely in the $n <
100$\,cm$^{-3}$ gas in which H$_3^+$ reside, their large column
densities demonstrate the richness of chemistry in the diffuse
interstellar medium in the CMZ.

%------------------------------------------------------------
% Section 6.1.2 
\subsubsection{The CMZ is not as opaque as previously proposed}

As pointed out in Section 1.2 and demonstrated by the relatively
low extinctions we derive to stars in the Central Cluster, the
Quintuplet Cluster, and other stars stretching across the CMZ,
the CMZ is not as dominated by dense molecular clouds ($f \geq
0.1$), and therefore not as opaque, as originally proposed by
Morris \& Serabyn (1996). The intensity of the millimeter wave
dust continuum across the CMZ as mapped by Bally et al. (2010)
also yields much lower extinctions over the vast majority of the
CMZ than predicted by the above estimate (Appendix D). The most
direct evidence for the much higher transparency of the CMZ
based on data reported here comes from the absorption spectra of
CO overtone lines, which are shown in Figure~6, whose
characteristics are given in Table~3, and which are discussed in
Section~4.3. All 18 stars listed in Table~3 are deeply embedded
in the CMZ, as indicated by the large H$_3^+$ column densities
toward them, and yet for most of them the CO absorptions are
dominated by those in the foreground three spiral arm, whose
total CO column densities correspond to visual extinctions
ranging from 10 to 30 mag. The CO column densities arising in
the CMZ correspond to visual extinctions of 10\,mag or less for
these stars. For the stars physically associated with the
Sagittarius complexes A and E, the total CO column densities in
the CMZ are higher, but still only comparable to those in the
foreground arms rather than vastly larger.  On all but one of
these sightlines the visual extinction is at least an order of
magnitude less than the mean value associated with such a large
filling factor of dense gas.

It is only toward star $\iota$ in the Sagittarius B complex that
overtone CO absorption is dominated by CO in the CMZ, as
described in Section 4.2.3. The CO column density measured
toward this star corresponds to $A_V \sim 300$\,mag, which is
unrealistically large as discussed in Section~4.3.2. The nature
of this star, how deeply it is buried in Sgr~B, the extinction
law in Sgr~B, and the dust-to-gas ratio in that dense molecular
cloud complex are intriguing questions.

Radiation from stars located behind giant molecular clouds will
be largely blocked. Therefore, there is some bias in
interpreting column densities determined from CO infrared
absorption spectroscopy as characterizing the entire CMZ.
  However, this effect must be minor in view of the very much
  smaller values of $A_V$ due to CMZ gas compared to the
  $A_V$ in the spiral arms toward almost all of the
  sightlines listed in Table~3. The only three cases in Table~3
  for which $A_V$ in the CMZ is large and comparable to
  $A_V$ from the spiral arms are star $\alpha$,
  which lies behind two localized dense clouds, and GCIRS~3 and
  GCIRS~1W, which probably lie behind the CND. The cases of
  $\alpha$ and GCIRS~3 were discussed earlier in the paper
  (Sections 4.2.2. and 4.2.3.). Moreover, although absorption
spectroscopy probes only the columns of CMZ gas in front of the
stars, it seems highly unlikely that, whatever determines the
current distribution of gas in the CMZ, the asymmetry between
gas abundances in the front and rear halves of the CMZ would be
so extreme as having virtually all of the $n \geq
10^4$\,cm$^{-3}$ molecular gas located in the rear half of the
CMZ, in order to explain the claimed filling factor of 0.1 or
higher.

We therefore maintain that the combination of $n \geq
10^4$\,cm$^{-3}$ and $f \geq 0.1$ in the CMZ is an overestimate
by a factor of more than 30. On the other hand, if the observed
CO is mostly located in gas with number densities of $\sim
10^{2.5}$\,cm$^{-3}$ as reported by Oka et al. (1998a) and
Dahmen et al. (1998) the volume filling factor of $f\geq 0.1$
may be approximately right. The filling factor of 0.1 for
  gas of density ${10^4}$\,cm$^{-3}$ corresponds to a
  mass of $\sim 1 \times 10^9\,M_\odot$, one-thirtieth of
  which is comparable to the masses given by Oka et al. (1998a),
  (2-6)$\times 10^7\,M_\odot$, and Dahmen et al (1998),
  (2-5)$\times 10^7\,M_\odot$.

The discovery of the predominance of diffuse molecular gas
  reported in this paper makes the term Central Molecular Zone
  even more fitting.

%------------------------------------------------------------
% 6.1.3
\subsubsection{The highly spatially extended ultra-hot (${\it 10^8}$K) X-ray-emitting plasma does not exist}

Intense apparently diffuse emission in the 6.7\,keV iron line
was discovered from the Galactic Ridge (Koyama et al. 1986) and
from the GC (Koyama et al. 1989) and was interpreted by those
authors as thermal X-ray emission from ultra-hot plasmas with
temperatures of $\sim 10$\,keV$\sim 10^8$\,K. The emission
extends over an angular size of 1.8\degr, which suggested its
volume filling factor in the GC is large, possibly approaching
unity. On the other hand, others (e.g. Pavlinsky et al. 1992),
who observed X-rays in the same energy range, interpreted its
origin differently because of the difficulty in finding a
mechanism to create and maintain such spatially extended high
temperature gas (Sunyaev et al. 1993). Nevertheless, the concept
of an ultra-hot plasma dominating interstellar medium in the GC
was accepted by many authors (e.g. Figure~9 of Lazio \& Cordes
1998).

The Chandra X-ray Observatory, with a high spatial resolution
(0\farcs5), resolved part of the diffuse X-ray emission
into point sources (Muno et al. 2003), although it was reported
that the dominant part of observed X-rays was still diffuse
(Muno et al. 2004) in the GC, as well as in the Galactic Ridge
(Ebisawa et al. 2001, 2005). On the other hand, Wang et
al. (2002) suggested that the presence of a large amount of
$10^8$\,K gas was no longer required. Belmont et al. (2005)
maintained that an ultra-hot helium plasma (after the ultra-hot
hydrogen escaped the GC region) extends over a few hundred
parsecs in the Galactic center region and proposed a viscous
heating mechanism to heat and maintain it (Belmont \& Tagger
2006). In contrast, Revnivtsev et al. (2006) interpreted the
diffuse X-rays from Galactic ridge as due to accumulation of
point sources, mostly cataclysmic variables and coronally active
binaries. Revnivtsev et al. (2009) resolved 80\,\% of the
seemingly diffuse X-ray emission into discrete sources and
Warwick (2014) and Warwick et al. (2014) associated virtually
all X-ray sources with coronally active late-type stars. Others,
however, have continued to maintain the presence of a large
amount of pervasive ultrahot gas in the CMZ (Yamauchi et
al. 2016; Nobukawa et al. 2016). Thus, whether such an ultra-hot
plasma occupies a large fraction of the volume of the CMZ has
been a long-standing debate.

The predominance of the warm and diffuse gas claimed in this
paper argues against the presence of a pervasive ultra-hot
plasma. The mechanism for production of such gas and maintaining
an ultrahigh temperature is not known. Temperatures as high as
$10^8$\,K are possible only in the vicinities of stars and
supernovae. The high energy ultra-hot gas cannot coexist with
extensive low energy gas, such as the warm, diffuse gas reported
here, because it is quickly cooled by the latter. Thus, we
suspect that diffuse X-rays must be due to stars yet to be
resolved and to the scattering of stellar X-rays by interstellar
matter.

%------------------------------------------------------------
\subsection{High H$_2$ ionization rate and cosmic-ray flux}

The estimated cosmic ray H$_2$ ionization rate of $\zeta \sim 2
\times 10^{-14}$\,s$^{-1}$ in the CMZ derived in Section~5.2 is
an order of magnitude higher that that reported by Oka et
al. (2005). The value derived here is near the low end of the
range of values calculated by Le Petit et al. (2016), $\zeta =
(1-11) \times 10^{-14}$\,s$^{-1}$, but is the highest value
determined for an extended region apart from that of Le Petit et
al. From X-ray observations, Yusef-Zadeh et al. (2007) give much
higher values of $\zeta = 5 \times 10^{-13}$\,s$^{-1}$ for Sgr~C
and a similar value as ours for Sgr~B1, Sgr~B2 and the radio
arc. Yusef-Zadeh et al. (2013) give a much lower value, $\zeta =
(1-10) \times 10^{-15}$\,s$^{-1}$, for a much more extended
region ($\sim 300 \times 120$\,pc) of the GC. The highest local
value of $\zeta$ reported so far is $\zeta \sim 4.5 \times
10^{-12}$\,s$^{-1}$ by Becker et al. (2011) for
$\gamma$-ray-emitting supernova remnants.

%------------------------------------------------------------
\subsubsection{Magnetic fields in the CMZ}

This work establishes that the cosmic ray flux in the CMZ is
$\sim$1000 times higher than that in the solar vicinity (Webber
1998). If the equipartition assumption (e.g. Yoast-Hull et
al. 2016) applies to the CMZ, as it does in the solar vicinity,
this indicates an average magnetic field in the CMZ on the order
of 100\,$\mu$G. This is an order of magnitude lower than the
estimates of $\sim$1\,mG by Yusef-Zadeh \& Morris (1987) along
the giant Radio Arc (Yusef-Zadeh et al. 1984) and by Morris \&
Yusef-Zadeh (1989), who argued that mG magnetic fields pervade
much of the volume of the CMZ. On the other hand, Sofue et
al. (1987) suggested a smaller magnetic field of 10--100\,$\mu$G
in the region of the Radio Arc. More recently a mG magnetic
field was reported by Chuss et al. (2003) from submillimeter
polarimetric observations based on the model calculations of
Uchida et al. (1985).

However, as a result of observations of smaller scale
non-thermal filaments which look randomly oriented, and of
magnetic field related phenomena such as the Zeeman effect and
Faraday rotation, as well as other astrophysical measurements,
smaller pervading magnetic fields of less than 100\,$\mu$G and
even in the range 1--10\,$\mu$G have become favored (Boldyrev \&
Yusef-Zadeh 2006). For example, LaRosa et al. (2005) estimate
the mean magnetic field in the GC to be on the order of
10\,$\mu$G, an order of magnitude less than our estimate based
on equipartition. Thus, there remain huge uncertainties in the
mean value of the magnetic field in the GC, and the
applicability or non-applicability of equipartition law to the
GC (Morris 2006) remains to be determined.

%------------------------------------------------------------
% 6.2.2
\subsubsection{Heating and cooling of the gas in the CMZ}

The high ionization rate on the order of $10^{-14}$\,s$^{-1}$
indicates that the CMZ belongs to the category of giant cosmic
ray-dominated regions (CRDRs) (Papadopoulos 2010) where cosmic
rays are the ultimate regulator of temperature and
ionization. The  efficiency of X-rays in heating the gas
was calculated by Ao et al. (2013) to be far too small. The
ionization of the gas by X-rays is more than 3 orders of
magnitude lower (Notani and Oka, to be published). Other dynamic
heating mechanisms such as turbulence claimed for heating of
dense clouds in the CMZ (e.g. Ginsburg et al. 2016) will not
work for the large scale diffuse gas reported in this paper.

Heating the gas to temperature much higher than the dust
temperature, which is on the order of 20\,K (e.g. Lis et
al. 2001) in giant molecular clouds in the GC and 20\,K + 50\,K
in diffuse clouds of the Arches and Quintuplet Cluster in the
two temperature model of Kaneda et al. (2012), must be due to
cosmic rays. On the other hand, the temperature of the diffuse
gas, 200\,K, is far below that expected from the equipartition
hypothesis. This is at least partly due to the efficient cooling
of the gas by the spontaneous emission of the H$_3^+$
symmetry-breaking rotational transitions (Pan \& Oka 1986)
discussed in Section~1.6. H$_3^+$ is known to be an efficient
coolant in planetary ionospheres (Miller et al. 2010,
2013). There, the role of H$_3^+$ is so decisive that the term
``H$_3^+$ thermostat'' was coined by Miller et
al. (2000). H$_3^+$ as a coolant in the chemistry of primordial
star formation was considered by Glover \& Savin (2006, 2009).

%============================================================
\section{Summary and Conclusions}

In this paper, we have presented and analyzed velocity-resolved
spectra, obtained at several telescopes, of selected lines of
the fundamental band of H$_3^+$ in the 3.5--4.0\,$\mu$m interval
and of the first overtone band of CO near 2.34\,$\mu$m, on
sightlines to stars stretching across the Central Molecular Zone
of the Galaxy, a region of radius 150\,pc and thickness of
$\sim$30\,pc. Previous observations of these lines, obtained
toward stars within a few tens of pc of Sgr~A$^\ast$ and
therefore near the very center of the CMZ, strongly suggested
that on those sightlines most of the CMZ's H$_3^+$ resides in
warm ($T\sim 200$\,K) diffuse ($n < 100$\,cm$^{-3}$) gas
covering a large fraction of radius of the CMZ, and that the
cosmic ray ionization rate in the CMZ is considerably higher
than in either dense or diffuse molecular clouds in the Galactic
disk. We have since identified additional stars suitable for
spectroscopy of interstellar H$_3^+$ and CO that cover a much
wider range of Galactic longitudes within the CMZ. The present
study demonstrates that the above gaseous environment and the
high cosmic ray ionization rate exist essentially throughout the
front half of the CMZ. We believe that it is likely to exist in
the rear half of the CMZ as well.

The key to understanding the physical conditions in this warm,
diffuse gaseous environment is the H$_3^+$ molecule
(trihydrium), first detected in the interstellar medium in 1996
(Geballe \& Oka 1996) and first observed toward the Galactic
center in 1997 (Geballe et al. 1999). Although the number
density of H$_3^+$ in the CMZ is only $\sim 10^{-6}$ times that
of H$_2$ due to its high chemical reactivity, the fully
dipole-allowed infrared vibration-rotation spectrum of H$_3^+$
is $\sim 10^9$ times stronger than the quadrupole spectrum of
H$_2$ (Oka 1981). Thus, the infrared absorption spectrum of
H$_3^+$ is much more readily observable than that of
H$_2$. H$_3^+$ also has a number of characteristics that make
spectroscopy of it a highly sensitive probe of physical
conditions in interstellar molecular gas, the main ones being
its suitability as both a thermometer and (at low densities) a
densitometer, as discussed in Section~1 of this paper. Being a
charged molecule and being created in a simple and
straightforward manner following cosmic ray ionization of H$_2$,
it also is a powerful in situ probe to measure the cosmic ray
flux.

The spectra of H$_3^+$ lines in the CMZ reveal that the warm
diffuse gas has high velocity dispersion, often producing
line widths exceeding 100\,km\,s$^{-1}$. For such a large range
of velocities to exist on many widely-separated sightlines, the
gas must occupy a large volume within the CMZ. In contrast, in
almost all cases the spectra of CO on these sightlines show
little absorption attributable to CO in the CMZ, and in cases
where such CO is evident, it exists in mostly narrow absorption
features, suggesting that those absorptions arise in compact
dense clouds. Because of the ubiquity of cosmic rays, H$_3^+$
exists wherever H$_2$ abounds, be it dense clouds or diffuse
molecular clouds. CO, on the other hand, resides mostly in dense
clouds. Thus, by comparing spectra of H$_3^+$and CO on the same
sightline one usually can discriminate between H$_3^+$
absorption in dense and diffuse clouds. This is particularly
useful for separating H$_3^+$ in the CMZ from H$_3^+$ in the
foreground Galactic spiral arms.

Prior to our studies, the CMZ was widely considered to contain a
large amount of dense ($n \geq 3\times 10^3$\,cm$^{-3}$) gas with a
significant volume filling factor ($f \geq 0.1$).  The mean
visual extinction of $A_V \geq 500$\,mag over the $\sim$150\,pc
radius of the CMZ would make it impossible to observe deeply
into the CMZ on most sightlines. This is clearly far from the
situation for the stars in the Central Cluster, which surround
the supermassive central black hole, Sgr~A$^\ast$ and are easily
observed, as well as for stars in the Quintuplet Cluster. The
depths of the other stars that we have observed in this study
are less well known or not known at all, but observations of
high H$_3^+$ column densities comparable to those found toward
stars in the Central Cluster, as well as similar extinctions to
many of them as to the stars in the Central Cluster, indicate
that most of them are also deeply embedded in the CMZ. Indeed,
the ubiquity of the warm diffuse clouds containing copious
H$_3^+$ allows one to adopt $N({\rm H_3^+})$ as an approximate
depth meter for the stars. We find that the above extinction is
an overestimate, at least by a factor of 30.

That finding appears to be in accord with the large scale
infrared photometric surveys of stars in the GC by Cotera et
al. (2000) and Schultheis et al. (2009), as well as with the
millimeter-wave dust continuum mapping of Bally et al. (2010)
(Appendix~D). Undoubtedly there are some selection effects: our
study is brightness-limited; we cannot observe stars deeply
embedded in or behind giant molecular clouds; and our
observations are limited to only detecting absorption by gas in
front of our background stars. But the statistics seem
overwhelming. Whatever the mechanism of molecular production and
dynamics, it is highly unlikely that average gas densities
differ by an order of magnitude or more between the front and
the back halves of the CMZ.

Two effects not included in our previous analyses of the diffuse
cloud chemistry of H$_3^+$ have been added in the analysis
presented here (Section~5). The exothermic charge exchange
reaction H$_2^+$ + H $\rightarrow$ H$_2$ + H$^+$, which reduces
the H$_3^+$ production rate significantly for clouds with low
H$_2$ fractions is now included. Also, unlike in Oka et
al. (2005) in which only electrons resulting from the
photoionization of neutral carbon were considered, electrons
produced from cosmic ray ionization of H and H$_2$ are now taken
into account (following Le Petit et al. 2016). We have developed
a chemical model calculation in which only hydrogen and
electrons are considered. Instead of the simple linear equation
connecting the cosmic ray ionization rate $\zeta$ and the
observed $N({\rm H_3^+})$ (Oka et al 2005), a quadratic equation
results from the present analysis. As in Oka et al. (2005),
$\zeta$ and the cloud length $L$ cannot be separated uniquely,
but various considerations (chemistry, velocity dispersion,
morphology, etc.) point to $\zeta \sim 2 \times
10^{-14}$\,s$^{-1}$ and $L \sim 100$\,pc being the most likely
average values, the latter corresponding to a volume filling
factor of $\sim$2/3 for the warm diffuse gas We find the most
likely chemical parameters to be $n({\rm H_2}) \sim
20$\,cm$^{-3}$, $n({\rm H}) \sim 30$\,cm$^{-3}$, $n({\rm H_3^+})
\sim 1 \times 10^{-5}$\,cm$^{-3}$, and $n_{\rm e} \sim
0.33$\,cm$^{-3}$.

These observations and analyses lead to the following conclusions.

\begin{itemize}
\item[(1)] Warm ($T \sim 200$\,K) diffuse ($n \sim
  50$\,cm$^{-3}$) gas dominates the volume of the CMZ. We have
  already pointed out above that the filling factor for dense
  gas must be much less than 0.1. The ultra-hot ($T \sim
  10^8$\,K) X-ray emitting plasma, which some thought to
  dominate the CMZ, cannot coexist with the warm diffuse gas and
  thus does not exist over extended regions. The observed
  diffuse X-ray emission in the CMZ must be due to point sources
  yet to be resolved and to scattering by interstellar atoms and
  molecules.

\item[(2)] The average cosmic-ray H$_2$ ionization rate in the
  CMZ that we deduce, $\zeta$, $\sim 2 \times
  10^{-14}$\,s$^{-1}$, is about 1000 times higher than in
  Galactic dense clouds and 10--100 higher than in Galactic
  diffuse clouds. If the equipartition law stands, this suggests
  a pervading magnetic field in the CMZ on the order of $\sim
  100$\,$\mu$G.
\vspace*{3.0cm}
\end{itemize}

%============================================================
\acknowledgments

The first author (T.O.) acknowledges many years of discussions
on the Galactic center with Harvey S. Liszt, Tomoharu Oka, and
Farhad Yusef-Zadeh. Discussions with Franck Le Petit, Evelyne
Roueff and other members of LERMA at the Meudon Observatory have
been crucial for our analyses of H$_3^+$ chemistry.  We
especially thank the referee, E. A. C. Mills, for many
illuminating and helpful comments and suggestions.  T.O. also
thanks, C. Westrick and Y-S. M. Chen for their help in producing
Figures~2 and 10, respectively. T. O. has been supported by a
generation of NSF grants, the latest being NSF grant AST
1109014. M.G. is supported by the German Research Foundation
(DFG) grant GO 1927/6-1.  This paper is based in large part on
observations obtained at the Gemini Observatory (Programs
GS-2003A-Q-33, GS-2008A-C-2, GS-2009A-C-6, GN-2010A-Q-92,
GN-2011A-Q-105, GN-2011B-Q-12, GN-2011B-Q-90, GN-2012A-Q-75,
GN-2012A-Q-121, GN-2013A-Q-114, GN-2014A-Q-108, GS-2014A-Q-95,
GN-2015A-Q-402, GS-2015A-Q-96, GN-2016A-Q-96, GS-2016A-Q-102,
GS-2017A-Q-95), which is operated by the Association of
Universities for Research in Astronomy, Inc., under a
cooperative agreement with the NSF on behalf of the Gemini
partnership: the National Science Foundation (United States),
the National Research Council (Canada), CONICYT (Chile),
Ministerio de Ciencia, Tecnolog\'ia e Innovaci\'on Productiva
(Argentina), Minist\'erio da Ci\^encia, Tecnologia e
Inova\c{c}\~ao (Brazil), and Korea Astronomy and Space Science
Institute (Republic of Korea).

\facilities{Gemini:Gillett, Gemini:South, Subaru, UKIRT, VLT:Antu}

%============================================================
\clearpage
\appendix
\section{Critical Evaluation of Parameters} 

\subsection{Rate constant of the H$_3^+$ dissociative recombination, $k_{\rm e}$}

In our previous analyses (Oka et al. 2005; Goto et al. 2008),
the temperature-dependent experimental formula for $k_{\rm e}$
determined using an ion storage ring by McCall et al. (2004),
their equation (7), was used. It was found subsequently that the
rotational temperature $T_{\rm rot}$ of H$_3^+$ which had been
measured to be low (22--37\,K) at the inlet of the ring, was
much higher in the body of the ring and $\sim$380\,K (Petrignani
et al. 2011; Kreckel et al. 2012).  The dependence of $k_{\rm
  e}$ on $T_{\rm rot}$ of H$_3^+$ is a complicated issue.

Here we continue to use the value of McCall et al. (2004) based
on the following two justifications: (i) Figure~2 of Forenseca
dos Santos et al (2007) gives their theoretical calculations of
$k_{\rm e}$ as a function of electron energy $E_{\parallel}({\rm
  eV})$ for $T_{\rm rot} = 13$\,K and 300\,K. The two curves
cross at $E_{\rm e} \sim 21.5$\,meV $ \sim250$\,K, suggesting
nearly equal $k_{\rm e}$ for $T_{\rm rot} = 30$\,K and
360\,K. (ii) It has been reported that the reaction rates for C
+ H$_3^+$ (O'Connor et al. 2015) and O + H$_3^+$ (de Ruette et
al. 2016) are not significantly affected by the H$_3^+$ internal
temperature. Argument (ii) is not as direct as (i) but the
temperature range is very high ($\sim3000$\,K).

%------------------------------------------------------------
\subsection{The C/H ratio in the CMZ, $(n_{\rm C}/n_{\rm H})_{\rm SV} R$}

Along diffuse and translucent sightlines in the solar vicinity
the carbon to hydrogen ratio in the gas phase is $(n_{\rm
  C}/n_{\rm H})_{\rm SV} = 1.6 \times 10^{-4}$ (Sofia et
al. 2004). In the GC the ratio is multiplied by a factor $R$ to
take into account the increase of metallicity from solar
vicinity to the GC. In the past we (Oka et al. 2005; Goto et
al. 2008) have used $R = 3-10$ on the assumption that the
increase of metallicity is close to that of $1/X = I_{\rm
  CO}/N({\rm H_2})$ (Sodroski et al. 1995; Arimoto et al. 1996).

There is considerable literature on the measurements of the
elemental abundance gradient $g$ (in dex\,kpc$^{-1}$) in the
Galactic disk which gives $R = 10^{gD}$ where $D$ (in\,kpc) is
the distance to the GC. The summary of measured abundance
gradients in Table~1 of Chiappini et al. (2001) indicates that
interstellar measurements in the H II regions and planetary
nebulae give fairly consistent abundance gradients ($g \sim
0.07$\,dex\,kpc$^{-1}$) with low measurement uncertainties,
while those of stars give smaller gradients with high
uncertainties and values scattered from 0 to 0.1
dex\,kpc$^{-1}$, perhaps reflecting individual characteristics
of the stellar atmospheres. Here we use interstellar
measurements because of their consistency and also because we
are applying them to interstellar gas.

Most measurements do not extend all the way to the GC. Among
those few who do, Simpson et al. (1995) give an [N/H] gradient
of $-(0.10 \pm 0.02)$\,dex\,kpc$^{-1}$ corresponding to $R = 7.1
\pm 1.4$. Afflerbach et al. (1997) find an [N/H] gradient of
$-(0.072\pm0.006)$\,dex\,kpc$^{-1}$ and an [O/H] gradient of
$-(0.064\pm0.009)$\,dex\,kpc$^{-1}$, corresponding to $R = 4.1
\pm0.4$ and $R = 3.5 \pm0.5$, respectively. Rudolph et
al. (1997) give an [N/H] gradient of $-(0.111\pm
0.012)$\,dex\,kpc$^{-1}$ corresponding to $R = 8.8\pm 0.9$. All
those authors assumed GC distance of 8.5\,kpc. Other
observations reported in many other papers which do not reach
the GC tend to give somewhat lower gradient and $R$ values. Most
measurements are for O, N, and S and very few are for C although
Rolleston et al. (2000) (GC distance 8.5\,kpc) give a [C/H]
gradient of $-(0.07\pm 0.01)$\,dex\,kpc$^{-1}$ corresponding to
$R = 3.9\pm 0.5$. A higher [C/H] gradient of $-(0.103\pm
0.018)$\,dex\,kpc$^{-1}$ has been reported by Esteban et
al. (2005) (GC distance 8.0\,kpc) which corresponds to $R = 6.7
\pm 1.3$.

Based on the above values we continue to use $R = 3$ (Oka et
al. 2005; Goto et al. 2008) as the lower limit for the increase
of C/H ratio from the solar vicinity to the GC. The accuracy of
$R$ is less essential in the chemical analysis of this paper
than in Oka et al. (2005) since the electrons from cosmic ray
ionization of H$_2$ and H, rather than from photoionization of
C, are dominant.

%------------------------------------------------------------
\subsection{Fraction of molecular hydrogen, $f(H_2)$}

This parameter affects the value of $\zeta$ seriously because
its square appears in the master Equation (13). In our previous
papers (Oka et al. 2005; Goto et al. 2008) where the master
equation was linear in $f({\rm H_2})$, we assumed $f({\rm H_2})
= 1$. This seriously underestimates the value of $\zeta$ if
$f({\rm H_2})$ is significantly smaller than 1. Values of
$f({\rm H_2})$ are given for diffuse clouds in the Galactic disk
toward the star $\zeta$~Per and X~Per as 0.60 and 0.76,
respectively, in Table~4 of Indriolo et al. (2007) based on the
direct measurements of $N({\rm H_2})$ and $N({\rm H})$ by the
Far Ultraviolet Spectroscopic Explorer (Rachford et
al. 2002). For other stars, they used $f({\rm H_2}) = 2/3 =
0.67$ which approximates those values and corresponds to the
special case of $n({\rm H}) = n({\rm H_2})$.

For the CMZ, Le Petit et al. (2016) find $f({\rm H_2}) = 0.6$,
from their analysis using the Meudon PDR code of the OH$^+$,
H$_2$O$^+$, and H$_3$O$^+$ column densities toward Sgr~B2(N)
reported by Indriolo et al (2015). Le Petit et al. claim that
this value agrees with the value calculated directly from
$N({\rm H})$ reported by Indriolo et al. (2015) and $N({\rm
  H_2})$ obtained from the Meudon analysis of HF (Godard et
al. 2012) listed in their Table~6. The value of 0.6 is also
consistent with their Figure~3 in which $f({\rm H_2})$ is
plotted as a function of $\zeta$.

The Meudon value of $f({\rm H_2}) = 0.6$ is drastically
different from $f({\rm H_2}) = 0.08 \pm 0.02$ of the analysis
toward Sgr~B2(N) by Indriolo et al (2015). Le Petit et
al. pointed out that this discrepancy was caused by the small
electron fraction $x_{\rm e}\sim1.5 \times 10^{-4}$ based on the
assumption that electrons are all from photodissociation of C
atoms, the same assumption Oka et al. (2005) made. This is
reasonable for ordinary diffuse clouds in the Galactic disk
where $\zeta$ is less than $10^{-15}$\,s$^{-1}$ but not for the
GC where $\zeta$ and thus $x_{\rm e}$ is much higher.

Based on these considerations, here we use $f({\rm H_2}) = 0.6$.

%------------------------------------------------------------
\subsection{Rate constant for charge exchange reaction ${H_2^+ + H \rightarrow H_2 + H^+}$}

The experimental value of $k_6 = (6.4\pm1.2) \times 10^{-10}$ by
Karpas et al. (1979) may not be accurate because the quoted low
uncertainty does not include the systematic error in the ratio
of atomic and molecular hydrogen [H]/[H$_2$] used in its
derivation. The theoretical cross sections of the reaction given
in Table~I of Last et al. (1997) are ordinary Langevin cross
sections suggesting that the experimental $k_6$ may be too
low. Also, Table~I of Karpas et al. (1979) indicates that their
method tends to give a low rate constant even for ordinary ion
neutral reactions. In view of this uncertainty we here assume
$k_6 = k_1/2$ partly because it gives a simple expression of the
H$_3^+$ production rate $\zeta [f({\rm H_2})]^2$. If the
experimental value is correct, this assumption underestimates
the H$_3^+$ production rate by 14\,\% for $f({\rm H_2}) = 0.6$
reported by Le Petit et al. (2016). If $k_6 \sim k_1$, this
overestimates the H$_3^+$ production rate by 40\,\%. Since
charge exchange reactions often are slower than the Langevin
rate (Huntress private communication), the assumption of $k_6 =
k_1/2$ may be reasonable. Determination of $k_6$ using modern
experimental techniques (e.g. Allmendinger et al. 2016) and more
theoretical study of this important reaction are highly
desirable.

%------------------------------------------------------------
\subsection{Radiative recombination rate constant of H$^+$}

Since the groundbreaking work by Menzel (1937), radiative
recombination of protons has been studied by many
astrophysicists (e.g. Spitzer 1948; Seaton 1959). Hollenbach and
McKee (1989) used
\begin{equation}
  k_{\rm r} = 3.60 \times 10^{-12} (T/300)^{-0.75}
       ~~{\rm cm^3\,s^{-1}},
\end{equation}
\noindent
by Aldovandi and P\'equignot (1973) which gives $k_{\rm r} = 4.1
\times 10^{-12}$\,cm$^3$\,s$^{-1}$ for $T = 200$\,K. A more
recent calculation by Badnell (2006) using
\begin{equation}
  k_{\rm r}= A \left[ \sqrt{\frac{T}{T_0}}
    \left( 1 + \sqrt{\frac{T}{T_0}} \right)^{1-B}
    \left( 1 + \sqrt{\frac{T}{T_1}} \right)^{1+B} \right]^{-1}
\end{equation}
\noindent
with $A = 8.32 \times 10^{-11}$\,cm$^3$\,s$^{-1}$, $T_0 =
2.97$\,K, and $T_1 = 7.0 \times 10^5$\,K gives
$k_{\rm r} = 5.7 \times 10^{-12}$\,cm$^3$\,s$^{-1}$ for $T =
200$\,K. Thus, the radiative recombination rate constant $k_{\rm
  r}$ of H$^+$ is $2 \times 10^4$ times smaller than the
dissociative recombination of H$_3^+$, $k_{\rm e}$.

%============================================================
\section{Solution of Equation (9)}

From Equations (8) and (9), the steady state equation for
H$_3^+$ is written explicitly as
\begin{equation}
  \zeta n_{\rm H} \left[ f({\rm H_2}) \right]^2
  = k_{\rm e} n({\rm H_3^+})
  \left[ 
    \left(\frac{n_{\rm C}}{n_{\rm H}} \right)_{\rm SV} Rn_{\rm H} 
    + \frac{k_{\rm e}}{2 k_{\rm r}} n({\rm H_3^+})
    \left(\sqrt{1 + \frac{2 \zeta k_{\rm r} n_{\rm H}}
      {\left[ k_{\rm e} n({\rm H_3^+}) \right]^2}}
      -1 
    \right)
  \right], 
\end{equation}  
\noindent
which gives the solution,
\begin{equation}
  \zeta = \frac{2 \rho K}{f^2}
  + \frac{K^2 (1-f^2)}{kf^4}
  + \frac{1}{f^4}
  \sqrt{\frac{4 \rho K^3 f^2}{k}
       + \frac{K^4 (1-f^2)^2}{k^2}}, 
\end{equation}  
\noindent
where $\rho \equiv (n_{\rm C}/n_{\rm H})_{\rm SV} R$, $K \equiv
k_{\rm e} n({\rm H_3^+})$, $f \equiv f({\rm H_2})$, and $k =
n_{\rm H} k_{\rm r}$. Using numerical values $k_{\rm e} = 8.68
\times 10^{-8}$\,cm$^3$\,s$^{-1}$, calculated from Eq.(7) of
McCall et al. (2004) for $T = 200$\,K, $\rho = 4.8 \times
10^{-4}$, $f({\rm H_2}) = 0.6$, and $k_{\rm r} = 5.7 \times
10^{-12}$\,cm$^3$\,s$^{-1}$ discussed in A1, A2, A3, and A5,
respectively, we obtain a numerical expression for $\zeta$ in
terms of $n_{\rm H} \equiv x$ and $n({\rm H_3^+}) = N({\rm
  H_3^+})/L \equiv H$ as,

\begin{equation}
  \zeta = \left(1.50\times 10^{-11} H x^2
  + 4.23 \times 10^{-4} H^2 x
  + \sqrt{2.00 \times 10^{-14} H^3 x^3
          + 1.79 \times 10^{-7} H^4 x^2}
  \right)
  /0.0648 x^2 .
\end{equation}  
\noindent
In Figure~10, $\zeta$ is plotted as a function of $x$ for assumed
values of $L$.
%============================================================
% C
\section{Similarities and Differences between H$_3^+$ and NH$_3$}

H$_3^+$ is similar to NH$_3$ in many ways. They have identical
symmetry of $D_{3h}$ and hence the same ortho ($I = 3/2$) and
para ($I = 1/2$) spin modifications and parity $(-1)^K$. H$_3^+$
and NH$_3$ (and isoelectronic H$_3$O$^+$) are the only oblate
symmetric tops observed in the CMZ and their rotational level
structures are similar (Figure~1); both have $J = K$ metastable
levels that are lower than $J > K$ levels. However, they have a
few differences that result in profound differences in their
astrophysical behaviors as observed in the CMZ.

\subsection{Planar non-polar H$_3^+$versus quasi-planar polar NH$_3$}

While the equilibrium structure of H$_3^+$ is planar, NH$_3$ is
pyramidal with high permanent dipole moment $\mu = 1.468$\,D
(Townes \& Schawlow 1995) and its plane symmetry is caused by
the inversion motion of the molecule. The motion produces many
strong inversion lines in the centimeter region whose
frequencies are only mildly dependent on the rotational quantum
numbers $J$ and $K$ and thus very high rotational levels can be
observed. NH$_3$ and its isoelectronic H$_3$O$^+$ (Liu \& Oka
1985; Lis et al. 2014) are unique in this respect. On the other
hand H$_3^+$ is non-polar and its observation is limited to the
infrared.

Initially the lifetimes of the $J = K$ metastable levels were
thought be longer than that of the Universe (Cheung et
al. 1969). Theoretical work indicated that because of SBS due to
vibration-rotation interaction they are many orders of magnitude
shorter (Oka et al. 1971). The lifetime of the spontaneous
emission is approximately proportional to $B^{-6}$, where $B$ is
the rotational constant, and is much shorter for H$_3^+$ ($B
\sim44$\,cm$^{-1}$) than for NH$_3$ ($B \sim9.9$\,cm$^{-1}$)
(Pan \& Oka 1986). While for NH$_3$ the life time of the (2,2)
$\rightarrow$ (1,1) spontaneous emission is $\sim230$ years, it
is 27 days for H$_3^+$ with the critical density on the order of
200\,cm$^{-3}$.

Although the lowest (0,0) rotational level is allowed for the
asymmetric inversion (-) level of NH$_3$, it is not allowed for
H$_3^+$ making the (1,1) level the ground level (Landau \&
Lifshitz 1977). NH$_3$ in the (3,3)$^{\pm}$ levels may reach the
ground (0,0)$^-$ level via 3 spontaneous emissions while H$_3^+$
in the (3,3) level cannot make transitions to lower levels
without violating the ortho-para rule which takes longer than
the lifetime of the Universe.

\subsection{Chemical (H$_3^+$) versus physical (NH$_3$) collisions}

Collisions between H$_3^+$ and H$_2$ are qualitatively different
from those between NH$_3$ and H$_2$. The former are chemical
reactions in which protons scramble in the (H$_5^+$)$^\ast$
intermediate complex, making ortho-para transitions possible
both for H$_3^+$ and H$_2$ (Quack 1977; Oka 2004). Thus,
collision-induced transitions can occur between any pair of
H$_3^+$ levels. On the other hand, a collision between NH$_3$
and H$_2$ is a physical collision in which protons do not
scramble. The collisional processes follow near rigorous
ortho-ortho and para-para selection rules (Oka 1968,
1973). While H$_3^+$ thermalizes quickly to any level at the
high Langevin rate, NH$_3$ thermalize slowly following the spin
isomer rule to limited levels. An NH$_3$ molecule in a high
metastable $J = K$ ($J > 2$) level changes its quantum state
only to ($J^\prime$, $K-3$) with $J^\prime \geq K-3$.

\subsection{Sub-thermal H$_3^+$ versus supra-thermal NH$_3$}

Because of the rapid symmetry breaking spontaneous emission, the
rotational distribution of H$_3^+$ is sub-thermal, as
demonstrated by the observed low population in the (2,2)
level. Along with the rapid vibrational spontaneous emission,
this makes H$_3^+$ an efficient coolant in planetary ionospheres
(Miller et al. 2000; see Section~5.2.2.). On the other hand, for
NH$_3$, the ($J$, $J$) metastable level is higher than the
($J-1$,$J-3$) level for $J \leq 10$ but their order reverses at
$J = 10$. Therefor the symmetry breaking spontaneous emissions
($J$,$J$) $\rightarrow$ ($J-1$,$J-3$) are quenched beyond $J >
10$; the only way for them to cool is via collisions. But the
${\it \Delta} K = 3$ collisions are slow (Oka 1973).  NH$_3$ in
those metastable levels cannot cool easily. The rotational
distribution of NH$_3$ is supra-thermal. The observation of
NH$_3$ at high energy levels, 1930\,K for the (14,14) metastable
level reported by H\"uttemeister et al. (1995) and 3130\,K for
the (18,18) metastable level reported by Wilson et al. (2006),
are due to this effect. The presence of metastable levels may
also explain the higher kinetic temperatures $T_{\rm k} >
600$\,K toward Sgr~B2 reported by H\"uttemeister et al. (1995),
$T_{\rm k} = (700\pm 100)$\,K toward Sgr~B2 by Ceccarelli et
al. (2002) and $T_{\rm k} \sim 600$\,K toward many giant
molecular clouds by Mills \& Morris (2013) using NH$_3$,
compared with lower temperatures of $T_{\rm k} \sim300-500$\,K
toward Sgr~B2 reported by Cernicharo et al. (2006) and $T_{\rm
  k}\geq 400$\,K toward Sgr~A complex reported by Sonnentrucker
et al. (2013) using H$_2$O, which does not have the
metastability.

%============================================================
\section{Agreement between Visual Extinction from our CO absorption and from Dust Continuum Emission by Bally et al. (2010)}

The visual extinctions $A_V$ obtained from the infrared
absorption of CO listed in Table~3 of this paper and those
determined from millimeter dust continuum by Bally et al. (2010)
in their Table~3 are remarkably similar in view of the entirely
different methods of the observations and analyses.
Both observations indicate that the previous estimate of volume
filling $f \geq 0.1$ for dense clouds with density $n \geq
10^4$\,cm$^{-3}$, is an overestimate by a factor of at least 30.
Out of the 1429 Bolocat clumps listed in the paper by Bally et
al., 616 lie within the CMZ. In both tables, there is a region
of minimum optical thickness in the direction of the Quintuplet
Cluster. Their Bolocam Galactic Survey (BGPS) \#63 ($G_{\rm
  lon}=0.164\degr, G_{\rm lat}=0.079\degr$) gives equivalent
column density $N_{40} = 6 \times 10^{21}$\,cm$^{-2}$ which
corresponds to $A_V \sim 6$\,mag in their analysis. Our CO
measurement in Table~3 gives $A_V({\rm CMZ}) \sim 7$\,mag toward
GCS\,3-2 (0.164\degr, $-$0.061\degr). In this comparison we
assume that the emission observed by Bally et al. is from
relatively high temperature dust in the CMZ and that the
emission from low temperature spiral arms is negligible.  In
both tables the region of highest optical thickness is toward
Sgr~B2; BGPS\,\#227 (0.680\degr, $-$0.029\degr) gives $N_{40} =
1.9 \times 10^{24}$\,cm$^{-2}$ corresponding to $A_V
\sim1900$\,mag while our CO observation toward J17470898
($\iota$) (0.549\degr, $-$0.059\degr) between Sgr~B1 and B2
gives $A_V \sim270$\,mag. Out of the 616 BGPS data BGPS\,\#227
is the only one which shows $A_V > 500$\,mag. The second highest
is BGPS\,\#223 (0.656\degr, $-$0.045\degr) with $N_{40} = 4.6
\times 10^{23}$\,cm$^{-2}$ and $A_V \sim460$\,mag.
The values of $A_V$ of all other BGPS are lower than 120\,mag.

These and other agreements between the results of two different
measurements must be somewhat fortuitous but this justifies the
following statistical argument using Table~3 of Bally et
al. (2010). The majority (80\,\%) of the BGPS data give more
than 30 times lower values of $A_V$ than would be produced by
dust in dense ($n \geq 10^4$\,cm$^{-3}$) clouds with a CMZ
filling factor $f \geq 0.1$.

Although the 1.1\,mm and 350\,$\mu$m images and contours given
in figures of Bally et al. may give an impression that the CMZ
is covered by giant molecular clouds, only 9 BGPS points have
column densities higher than $N_{40} = 1 \times
10^{23}$\,cm$^{-2}$. They are \#227 ($1.9 \times
10^{24}$\,cm$^{-2}$), \#223 ($4.6 \times 10^{23}$\,cm$^{-2}$)
and \#170 ($1.2\times 10^{23}$\,cm$^{-2}$) near Sgr~B, \#1411
($1.6\times 10^{23}$\,cm$^{-2}$) near Sgr~A, \#1388 ($1.4\times
10^{23}$\,cm$^{-2}$) and \#1391 ($1.1\times 10^{23}$\,cm$^{-2}$)
near star $\epsilon-$, {\#96 ($1.1 \times 10^{23}$\,cm$^{-2}$)
  and} \#99 ($1.0\times 10^{23}$\,cm$^{-2}$) near star $\theta$,
and \#1314 ($1.0 \times 10^{23}$\,cm$^{-2}$) near star
$\delta$.
If H$_2$ column densities $N_{40} \geq 1 \times
10^{23}$\,cm$^{-2}$ define giant molecular clouds, their areal
filling factor in the CMZ is 1--2\,\%.

% ===========================================================

\bibliographystyle{aasjournal}
%============================================================
\end{document}